\documentclass[manuscript]{emulateapj}
\usepackage{color}
\usepackage{gensymb}

\shorttitle{kinematics of galaxies and the role of group dynamics}
\shortauthors{M. Raouf et al.}

\begin{document}
	
	\title{THE SAMI GALAXY SURVEY: Kinematics of stars and gas in brightest group galaxies; the role of group dynamics}
	
	\author{Mojtaba Raouf,\altaffilmark{1}
		Rory Smith,\altaffilmark{1} 
		Habib G. Khosroshahi,\altaffilmark{2} 
		Jesse van de Sande,\altaffilmark{4,3} 
		Julia J. Bryant,\altaffilmark{4,5,3}  
		Luca Cortese,\altaffilmark{6,3}  
		S. Brough,\altaffilmark{7,3} 	
		Scott M. Croom,\altaffilmark{4,3} 
		Ho Seong Hwang,\altaffilmark{1} 
		Simon Driver,\altaffilmark{6} 	
		\'Angel R. L\'opez-S\'anchez, \altaffilmark{8,9,3}    
		Jongwan Ko ,\altaffilmark{1} 
		Jae-Woo Kim ,\altaffilmark{1} 
		Jihye Shin ,\altaffilmark{1} 
		Nicholas Scott,\altaffilmark{4,3} 
		Joss Bland-Hawthorn, \altaffilmark{4} 	
		Samuel N. Richards,\altaffilmark{10} 
		Matt Owers,\altaffilmark{8,12} 
		J.S. Lawrence,\altaffilmark{11} 
		Iraklis S. Konstantopoulos,\altaffilmark{13}
	}    
	\affil{$^1$Korea Astronomy and Space Science Institute, 776 Daedeokdae-ro Yuseong-gu, Daejeon 305-348, Korea}
	\affil{$^2$School of Astronomy, Institute for Research in Fundamental Sciences (IPM), Tehran, 19395-5746, Iran}
	\affil{$^3$    ARC Centre of Excellence for All Sky Astrophysics in 3 Dimensions (ASTRO 3D)}
	\affil{$^4$Sydney Institute for Astronomy (SIfA), School of Physics, \\University of Sydney, NSW 2006, Australia}
	\affil{$^5$Australian Astronomical Optics, AAO-USydney, School of Physics, \\University of Sydney, NSW 2006, Australia}
	\affil{$^6$International Centre for Radio Astronomy Research (ICRAR), The University of Western Australia, 35 Stirling Highway, Crawley, WA 6009, Australia}
	\affil{$^7$School of Physics, University of New South Wales, NSW 2052, Australia}
	\affil{$^8$Department of Physics and Astronomy, Macquarie University, NSW 2109, Australia}
	\affil{$^9$ Australian Astronomical Optics, Macquarie University, 105 Delhi Rd, North Ryde, NSW 2113, Australia}
	\affil{$^{10}$SOFIA Science Center, USRA, NASA Ames Research Center, \\Building N232, M/S 232-12, P.O. Box 1, Moffett Field, CA 94035-0001, USA}
	\affil{$^{11}$ Australian Astronomical Optics - Macquarie, Macquarie University, NSW 2109, Australia}
	\affil{$^{12}$ Astronomy, Astrophysics and Astrophotonics Research Centre, Macquarie University, Sydney, NSW 2109, Australia}
	\affil{$^{13}$ Contact Energy 29 Brandon Street Wellington 6011 New Zealand}
	\email{mojtaba.raouf@gmail.com}
	
	\begin{abstract}    
		We study the stellar and gas kinematics of the brightest group galaxies (BGGs) in dynamically relaxed and unrelaxed galaxy groups for a sample of 154 galaxies in the SAMI galaxy survey. 
		We characterize the dynamical state of the groups using the luminosity gap between the two most luminous galaxies and the BGG offset from the luminosity centroid of the group. We find that the misalignment between the rotation axis of gas and stellar components is more frequent in the BGGs in unrelaxed groups, although with quite low statistical significance. Meanwhile galaxies whose stellar dynamics would be classified as `regular rotators' based on their kinemetry are more common in relaxed groups.  We confirm that this dependency on group dynamical state remains valid at fixed stellar mass and Sersic index. 
		The observed trend could potentially originate from a differing BGG accretion history in virialised and evolving groups.  
		Amongst the halo relaxation probes, the group BGG offset appears to play a stronger role than the luminosity gap on the stellar kinematic differences of the BGGs. However, both the group BGG offset and luminosity gap appear to roughly equally drive  the misalignment between the gas and stellar component of the BGGs in one direction.
		This study offers the first evidence that the dynamical state of galaxy groups may influence the BGG's stellar and gas kinematics and calls for further studies using a larger sample with higher signal-to-noise.  
		
	\end{abstract}
	\section{Introduction}
	Galaxies are rarely found in isolation - they tend to \mbox{aggregate} together into small collections of galaxies known as galaxy groups, consisting of a few up to several hundred members, depending on the mass of the group. In most cases, the low-velocity dispersion of groups makes them ideal \mbox{environments} for efficient galaxy interactions and mergers. Furthermore, environmental mechanisms such as tidal stripping \citep{Han2018}, starvation \citep{Larson1980}, and ram pressure stripping \citep{gunngott1972,Park2009} may impact galaxy evolution, long before galaxies reach the denser environments of clusters  \citep[so called `group pre-processing';][]{Mihos2004}.
	The impact of mergers on a galaxy may differ depending on the type of merger. Minor mergers are less disruptive but more frequent, while major mergers can significantly transform a galaxy's morphology and dynamics. The mergers may also be dry (i.e., the merger does not bring new gas) or wet (the secondary galaxy brings new gas), and this also impacts on the final formation result.
	
	The impact of a merger on the gas-kinematics is greater than on the stellar dynamics of galaxies \citep{Barrera-Ballesteros2015}. 
	Simulations of galaxy mergers suggest timescales of $\sim$ 2 Gyr for the gas disk to realign with the stellar component following the merger in the formation of massive early-type galaxies \citep{vandeVoort2015}.
	
	\cite{Bryant2019} used 618 galaxies with fitted gas and stellar position angle (hereafter: PA) from the SAMI galaxy survey and found that $\sim$ 45\% of early-type galaxies and $\sim$5\% of late-type galaxies have a gas PA offset of more than 30 degrees with respect to the stars. They found a stronger correlation of the gas-star misalignment with morphology compared to the stellar mass, color, or local environment of galaxies.  
	Furthermore, \cite{Davis2016} showed that the gas-star misalignment in early-type galaxies could also originate from a continuous external gas accretion and so major mergers are not the only possible origin. \cite{Jin2016} studied the fraction of gas-star misalignment in different stellar mass ranges and demonstrated that the fraction peaks at $10^{10.5} M_{\odot}$. \cite{Bassett2017} showed that the dust in the majority of early-type galaxies with an existing kinematical misalignment between the gas and stars, mostly has its origin in external dust brought in by the merger and is less likely to be produced internally by the AGB stars \cite[see for e.g][]{Saremi2020}.

	\begin{figure}
		\centering
		\includegraphics[width=1.01\linewidth]{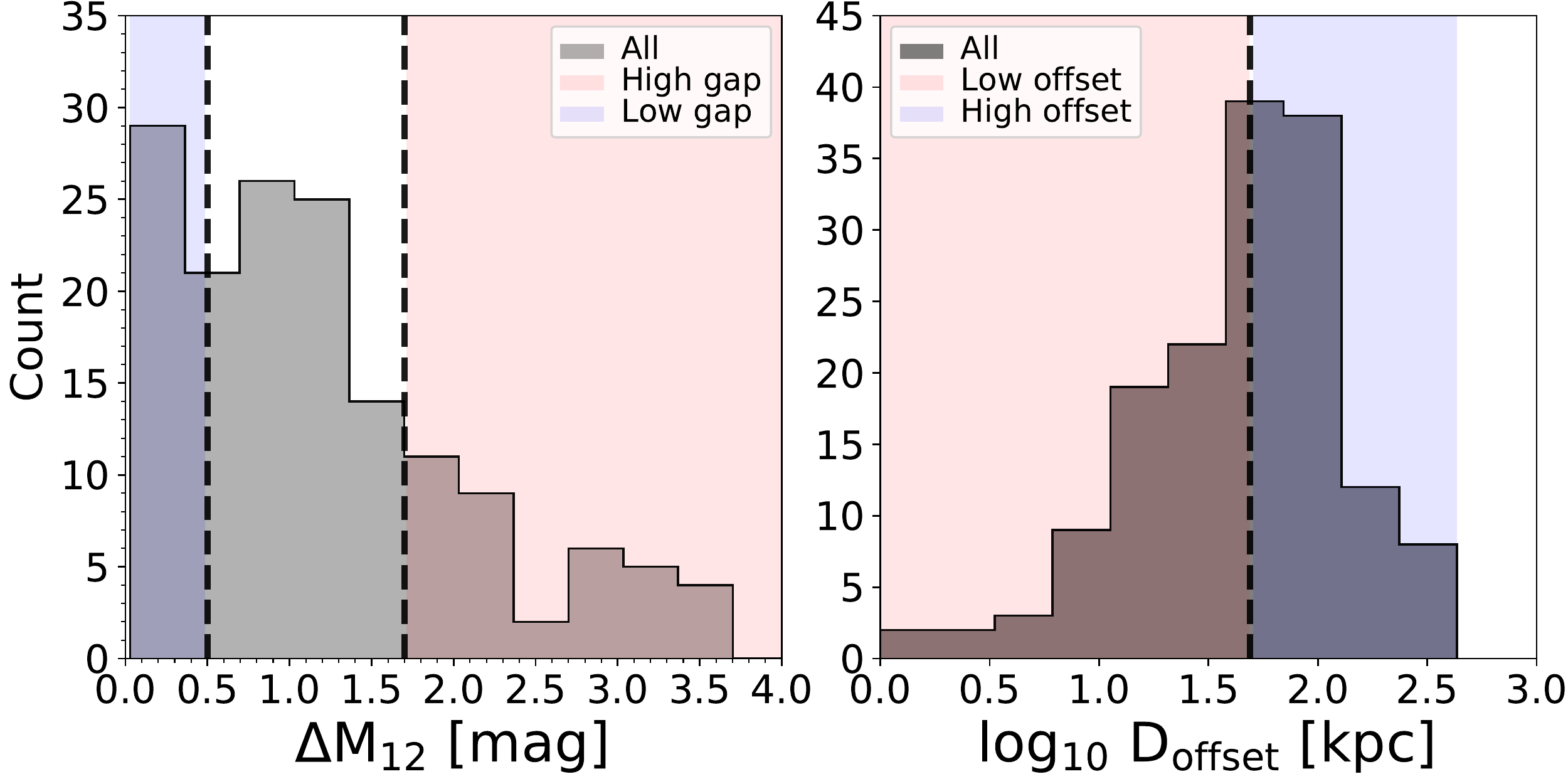}
		\caption{Histograms of the luminosity gap, $\Delta M_{12}$ (left), and BGG offset, $D_{offset}$ (right), for all galaxies in the sample. Superimposed are limits for selecting high/low gap and low/high offset groups as shown by the shaded red/blue regions, respectively, separated by dashed-lines.}
		\label{fig:sami-hist-gap-off}
	\end{figure}

	\begin{figure*}
		\centering
		\includegraphics[width=1.\linewidth]{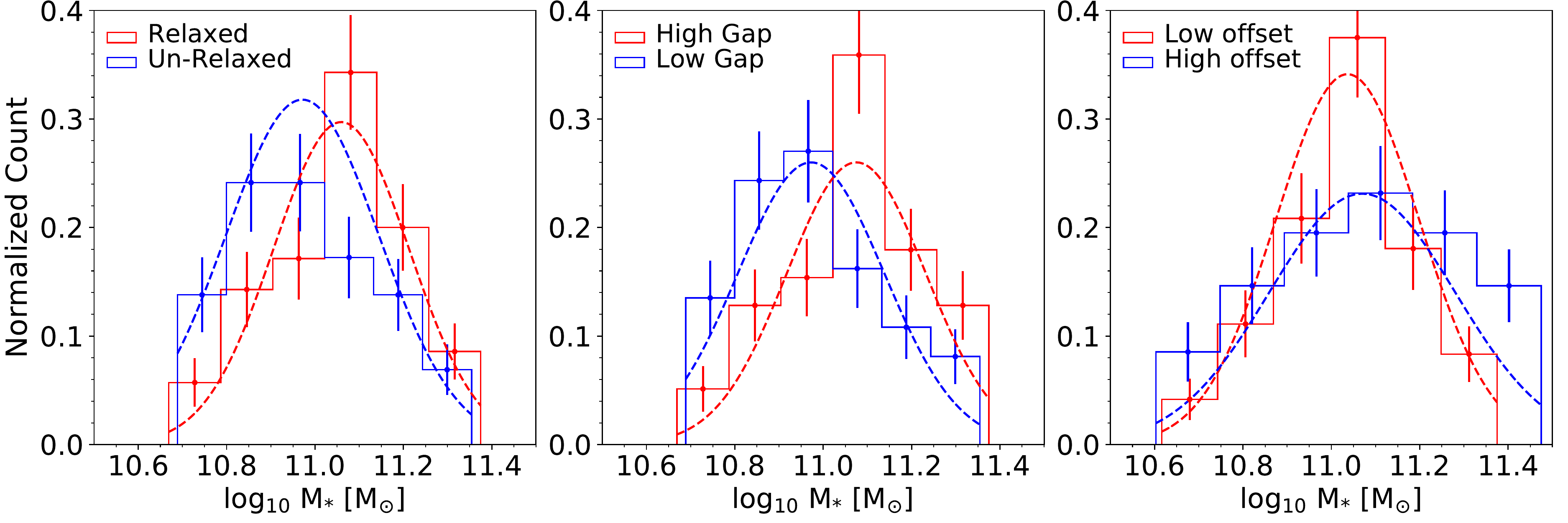}
		\caption{Distribution of stellar mass with Poisson errors for all sub-samples considered in this study including; relaxed/unrelaxed, high/low gap, low/high offset as shown by red/blue color in each panel from left to right. Dashed lines show the Gaussian fit to each distribution.}
		\label{fig:SAMI-Hist-mh_ms}
	\end{figure*}
	
	\cite{Krajnovic2011} classified galaxies by kinematic asymmetry to analyze the mean velocity map labeled as regular and non-regular rotators using a sample of 260 early-type galaxies in the ATLA$S^{3D}$ survey.  They reported around 82\% of galaxies are regular rotators compared to just 17\% non-regular rotators, where non-regular rotators are typically found in dense regions and are massive \citep[e.g;][]{Brough2017}. The dynamical properties of early-type galaxies are related to a measure of their specific angular momentum, such that the merger remnant absorbs the orbital angular momentum of the merged galaxies in its internal dynamics \citep{vandeVoort2015,Lagos2015}. Furthermore, using the projected angular momentum/spin and ellipticity, 
	some studies show that the vast majority of early-type galaxies are actually fast rotating and have a regular stellar rotation \citep{Cappellari2016,Emsellem2011}.
	
	\begin{figure}
		\centering
		\includegraphics[width=0.95\linewidth]{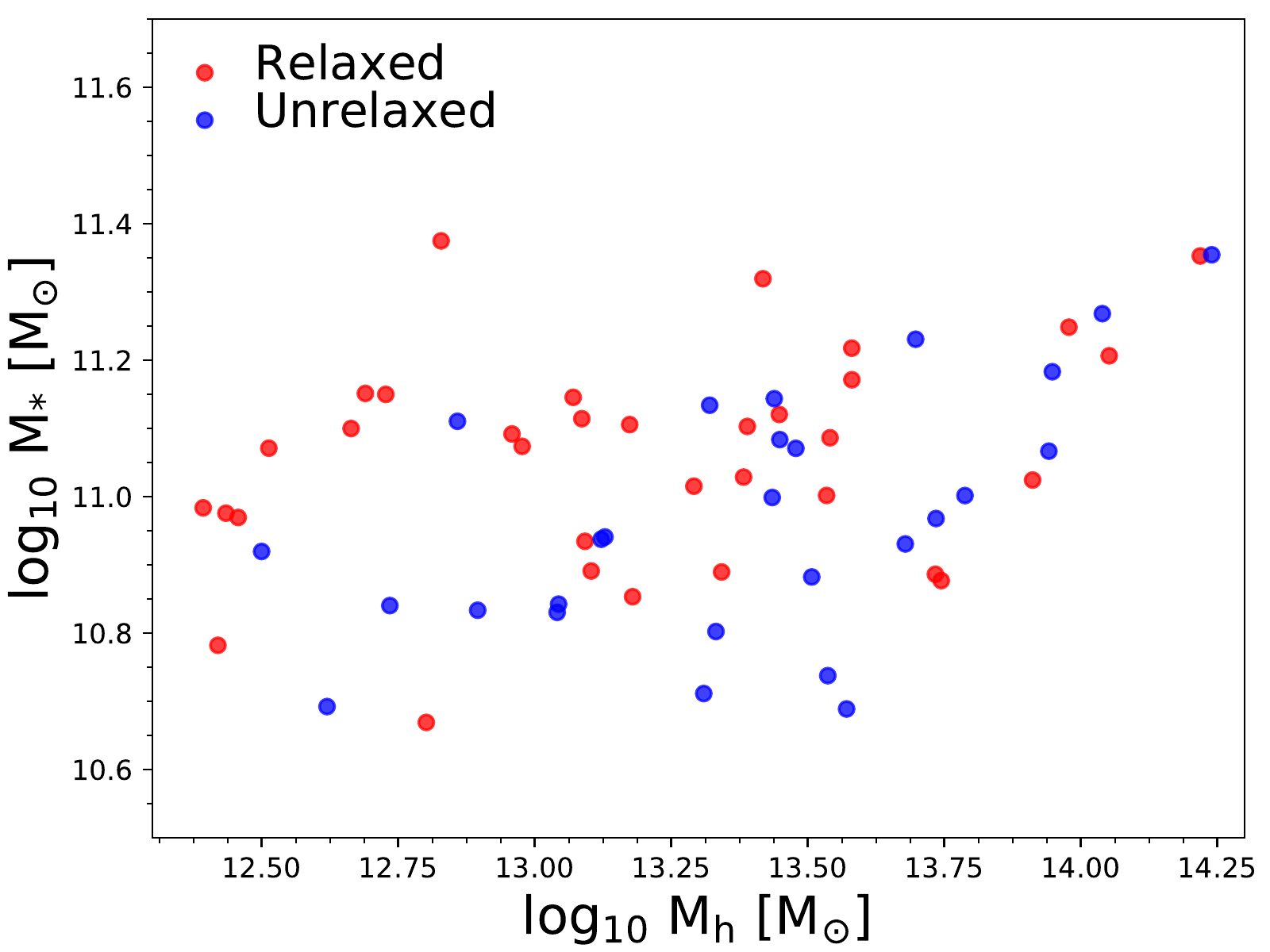}
		\caption{The distribution of stellar mass as a function of halo mass for the relaxed (red) and unrelaxed (blue) samples.}
		\label{fig:sami-ms-mh}
	\end{figure}
	
	While the time difference between the peak of the BGG major merger in relaxed and unrelaxed groups is typically around $\sim 2$ Gyrs, most of the BGG  last major mergers occur over the past one Gyr in unrelaxed groups \citep{Raouf2018}. This time-scale for the last major merger of the BGG in unrelaxed group is below the relaxation time-scale in which galaxy gas-star misalignment PA persists following an episode of re-accretion of cold gas \citep{vandeVoort2015} post merger. 
	
	Using the Galaxy And Mass Assembly(GAMA) galaxy survey, we showed that the BGGs of unrelaxed groups are significantly bluer in NUV-r colors and tend to have higher stellar metallicity and star formation rate compared to BGGs of relaxed group at a given stellar mass \citep{Raouf2019b}. Moreover, \citet{Khosroshahi2017} used the same sample of BGGs and found that the radio luminosity of the BGGs strongly depends on their dynamical state, such that the BGGs in dynamically unrelaxed groups are an order of magnitude more luminous in the radio than those with a similar stellar mass but residing in dynamically relaxed groups. They suggested that the presence of a high r-band magnitude gap between the two most luminous galaxies (luminosity gap) points to a scenario in which an earlier major merger could have triggered cold mode accretion, consistent with the associated 1.4 GHz AGN radio emission predicted by our semi-analytical galaxy formation model \citep[Radio-SAGE;][]{Raouf2017,Raouf2019}. In our earlier study, using hydrodynamical simulations, we showed that the black hole accretion in BGGs of dynamically relaxed groups is lower, for a given stellar mass than that in unrelaxed groups \citep{Raouf2016}. These pieces of evidence highlight clear differences in the observed properties of BGGs hosted by groups with different dynamical states. Here, we investigate the gas and star kinematics of BGGs to evaluate \textit{the role of group dynamical states}. 
	
	We define the group dynamical state using a combination of two  parameters, the luminosity centroid deviation of the BGG (i.e., the BGG offset) and the luminosity gap, which provide valuable information on the host halo formation history of BGGs \citep[as demonstrated in cosmological simulations;][]{Raouf2014}.  In this study, we determine whether the group halo formation history has any impact on the kinematic properties of the BGGs.  
	We attempt to answer the above question by studying data from the Galaxy And Mass Assembly survey \citep[GAMA;][]{Driver2011} and Sydney-AAO Multi-object Integral field (SAMI) galaxy survey \citep{Croom2012}. Our relaxedness state indicators are best applied to statistically large samples of galaxies, rather than on a one-by-one basis, and these surveys provide us with sufficient numbers of galaxies ($\sim$ 154) to make our study possible for the first time. Throughout this paper, we adopt $H_0 = 100h\ km\ s^{-1}Mpc^{-1}$ for the Hubble constant with $h = 0.7$.
	\begin{figure*}
		\centering
		\includegraphics[width=0.99\linewidth]{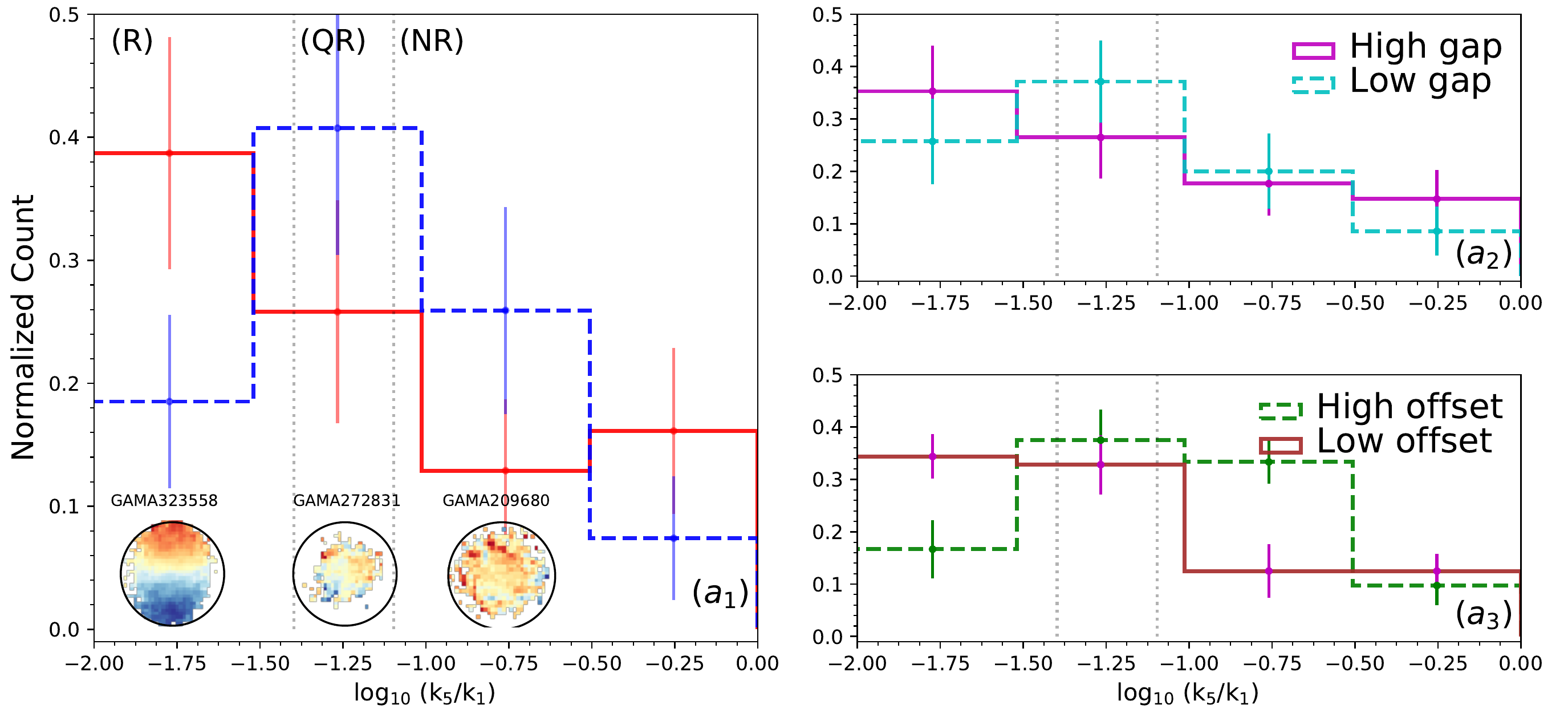}
		\includegraphics[width=0.99\linewidth]{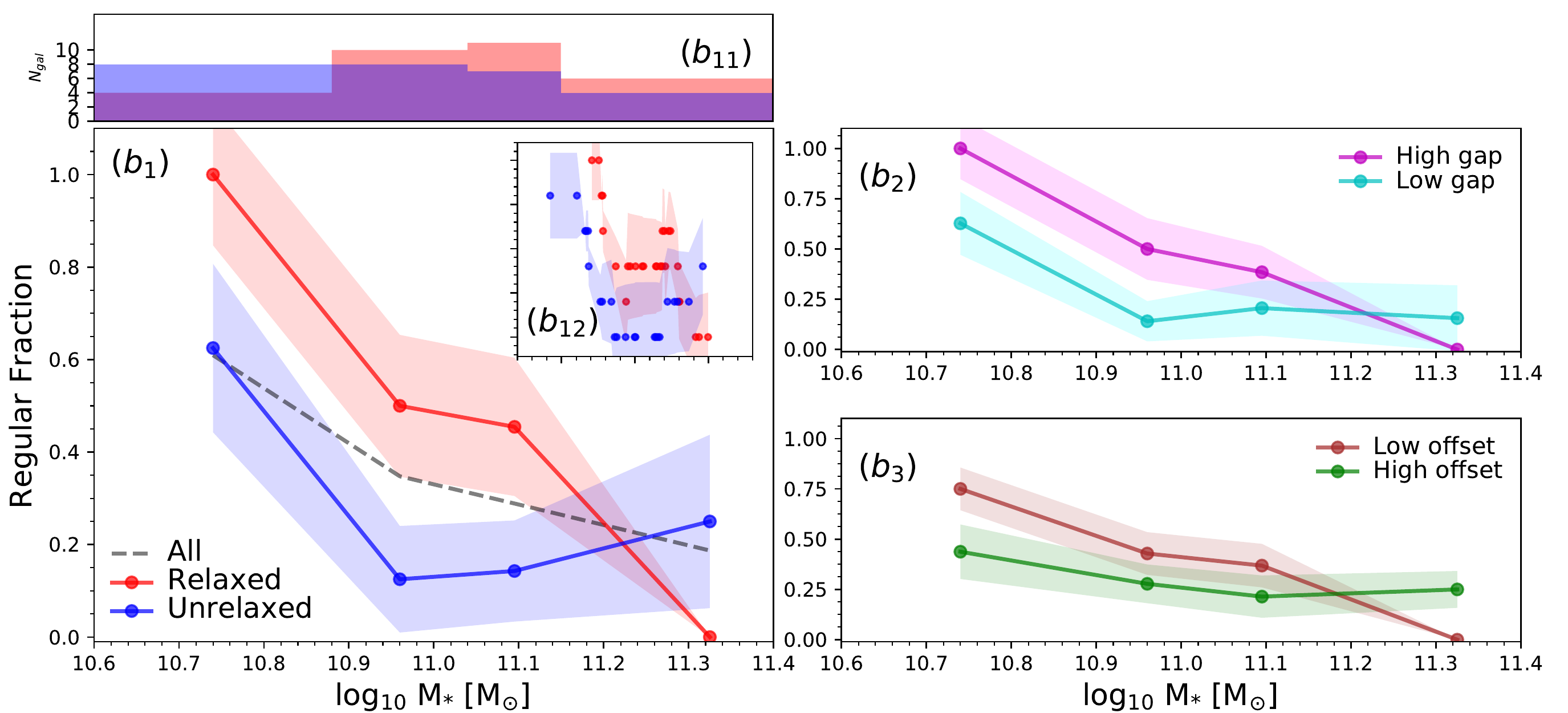}
		\caption{ Top: Distribution of the kinematic asymmetry ($k_5/k_1$) with standard deviation ($\sigma$) errors calculated by the bootstrap method for BGGs in relaxed (red) and unrelaxed (blue) group in the left panel ($a_{1}$), and  High/Low gap and Low/High offset  groups on the top ($a_{2}$) and bottom ($a_{3}$) right panels, respectively. The fraction of non-regular(NR) rotators ($k_5 /k_1 >$ 0.08), quasi-regular(QR) (0.04 $< k_5/k_1 <$ 0.08) and regular(R) rotating ($k_5/k_1 \leq$ 0.04) \citep[dotted lines;][]{vandeSande2017}  are presented in Table \ref{tab:value} for the BGGs reside in sub-sample groups. Representative stellar kinematic maps of BGGs defined as NR, QR, and R rotators are shown in the corresponding regions. Bottom: The fraction of regular BGGs as a function of stellar mass, separated by the subsample indicated in the legend (panels $b_{1}$, $b_{2}$, $b_{3}$). The dashed-line shows the median trend for all BGGs in our sample. In each panel, the errors (color shade-lines) are $\sigma$ confidence intervals on the fractions calculated using the bootstrap method. In the histogram along the upper x-axis of the panel ($b_{11}$), we show the number of galaxies ($N_{gal}$) in each stellar mass data point for the relaxed and unrelaxed samples. The inset figure in the left panel ($b_{12}$) shows the median fraction of regular rotation BGGs as a function of stellar mass using the moving average method where the axes range is the same as in the main plot.
		}
		\label{fig:sami-hist-k51}
	\end{figure*}
	\begin{figure}
		\includegraphics[width=1.05\linewidth]{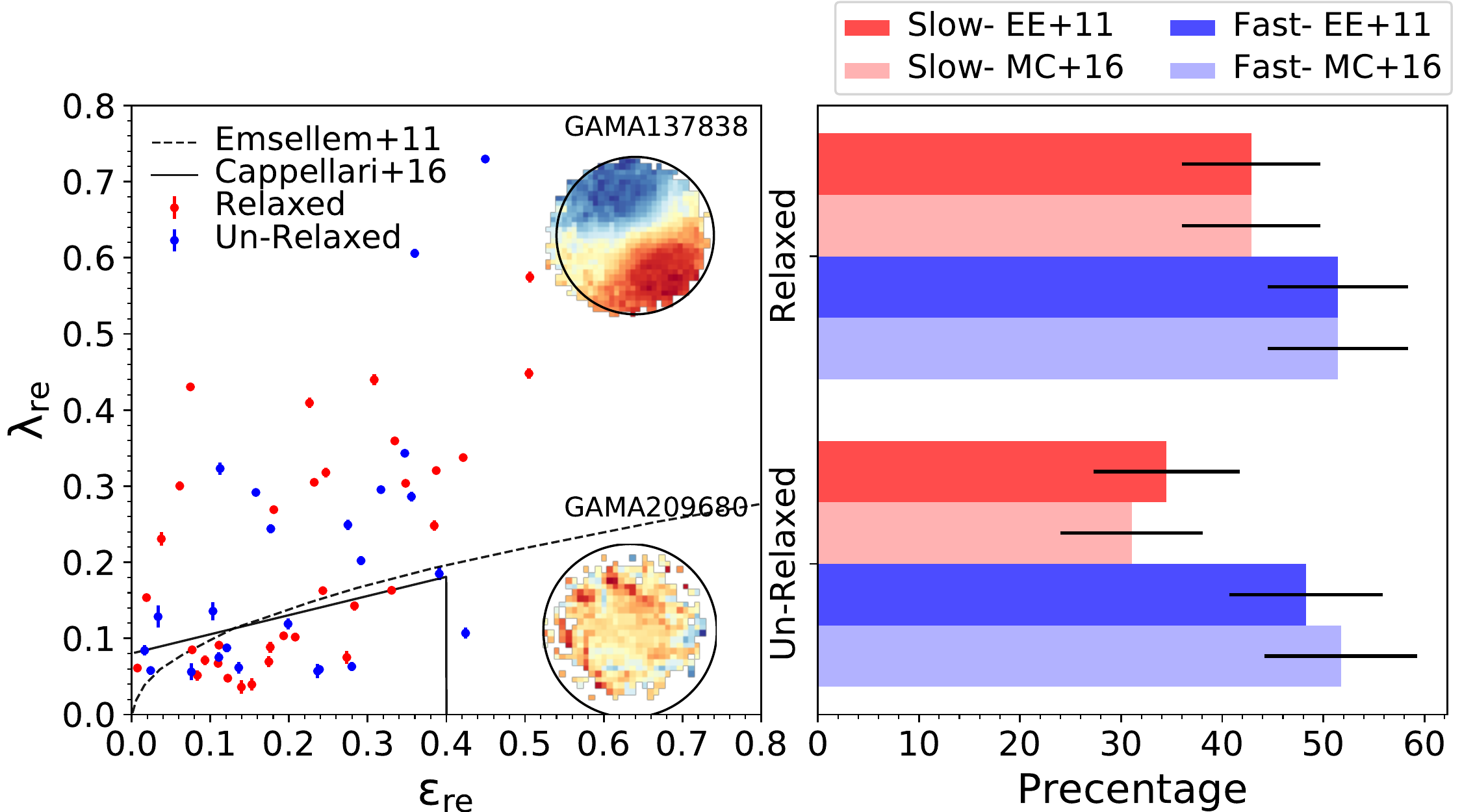}
		\caption{In the left panel, the spin parameter proxy within the effective radius ellipse, $\lambda_{Re}$ including measurement errors,  is plotted as a function of galaxy ellipticity measured at the effective radius for BGGs in relaxed and unrelaxed groups. The various curves/lines separate galaxies into slow and fast-rotators following the two separate schemes proposed in Emsellem et al. (2011, EE+11), and Cappellari (2016, MC+16). In the right panel, the fraction of slow and fast rotators is compared using horizontal bars with binomial errors for relaxed and unrelaxed groups. Representative examples of stellar kinematic maps are shown inside the figure for the fast(top) and slow(bottom) rotator BGGs. }
		\label{fig:sami-spin-ru}
	\end{figure}
	
	\section{Data and sample selection}
	The main sources of data for this study are the Galaxy And Mass Assembly third data release \citep[GAMA-DRIII;][]{Driver2011,Baldry2018}  and Sydney AAO Multi-object Integral field galaxy \citep[SAMI;][]{Croom2012,Bryant2015} surveys.
	
	GAMA is a multi-wavelength spectroscopic data set covering an area of 180 deg$^2$ as described in \citep{Baldry2010,Robotham2010,Driver2011,Hopkins2013}. We use the third data release, GAMA-DRIII,  group catalog generated for a spectroscopic component using a friends-of-friends (FoF) based grouping algorithm as described in \cite{Robotham2011}. The group catalog contains 23,838 galaxy groups which reduce to about 4,000 galaxy groups (and about 19,000 group members) with a multiplicity of at least four spectroscopically confirmed members and within the limit of our redshift range  0.02$<z<$0.22, and in terms of having a complete set of groups in which the first and second most luminous galaxy is detectable above the GAMA luminosity limit of $r$=19.8 mag. The luminosity centroid\footnote{Defined as the center of light derived from the $r$-band luminosity of all the galaxies identified to be within the group \citep{Robotham2011}}, stellar mass \citep{Taylor2011} and ellipticity from an r-band Sersic fit \citep{Kelvin2012} of the group members are taken from the GAMA catalogs.
	
	The SAMI instrument \citep{Croom2012} is mounted on the 3.9m Anglo-Australian Telescope (AAT) and fed into the AAOmega spectrograph \citep{Sharp2006, Bland-Hawthorn2011,Sharp2015,Allen2015,Green2018,Scott2018} which gives a median resolution of $\mathrm{FWHM}_{blue}$ = 2.65  \AA~from 3700-5700 \AA~and $\mathrm{FWHM}_{red}$ = 1.61 \AA~from 6300-7400 \AA~\citep{vandeSande2017}.
	
	From all galaxies in the GAMA survey, there are $\sim$ 2200 galaxies that overlap with SAMI data within the redshift limit z $<$ 0.115 from the  G09, G12, and G15 regions \citep{Driver2011}. We focus on a sample that includes both stellar and gas kinematics in order to measure the PA offset. We include only those brightest group galaxies (BGGs) with $M_{star} \gtrsim10^{10.6} M_{\odot}$ in order to have early-type galaxies in the group 
	\citep[see fig. 2;][for the mass distributions of the main sub-samples]{Raouf2019b}. Statistically, the final sample reduces to 154 galaxies, and we use the luminosity gap and BGG offset to categorize them into four different group sub-samples (see Figure \ref{fig:sami-hist-gap-off}). In this study, we use kinematic properties from the internal data release Version 0.11.   
	
	The detailed description of the kinematic asymmetry, galactic spin, and misalignment angle measurements are given in the following subsections. Note that we use the Spearman correlation coefficient for the linear regression in our analysis. 	
	
	\subsection{Method to estimate kinematic asymmetry} \label{Method-K51}   
	Similar to the method outlined in \cite{vandeSande2017}, the kinematic asymmetry of the galaxy stellar velocity fields is determined by the amplitude of the Fourier harmonics on all velocity data that pass the velocity quality criteria measured using the kinemetry  routine \citep{Krajnovic2006,Krajnovic2011}. The kinemetry routine determines a best-fitting amplitude for $k_1$, $k_3$, and $k_5$ for each ellipse. The first-order decomposition $k_1$ is equivalent to the rotational velocity, whereas the higher-order terms ($k_3$ , $k_5$ ) describe the kinematic anomalies. 
	In this study the kinematic asymmetry\footnote{There is another definition of kinematic asymmetry, $(k_3+k_5)/2k_1$ as described in \cite{Shapiro2008},  which is consistent with the $k_5/k_1$ measurement of  \cite{vandeSande2017}} is defined as the luminosity-weighted average ratio of $k_5/k_1$ within one effective radius, $R_e$, determined by the flux in SAMI images. The uncertainty on $k_5/k_1$ for each measurement is estimated from Monte Carlo simulations. Note that we require a filling factor of 85 percent of the spaxels that pass the velocity quality criteria \citep[for details see;][]{vandeSande2017,Scott2018} within the aperture for producing $R_e$ measurements. We now classify 
	BGGs into regular (low $k_5/k_1$) and non-regular (high $k_5/k_1$) rotators, hosted by groups with differing dynamical states.
	
	\begin{figure*}
		\centering
		\includegraphics[width=1\linewidth]{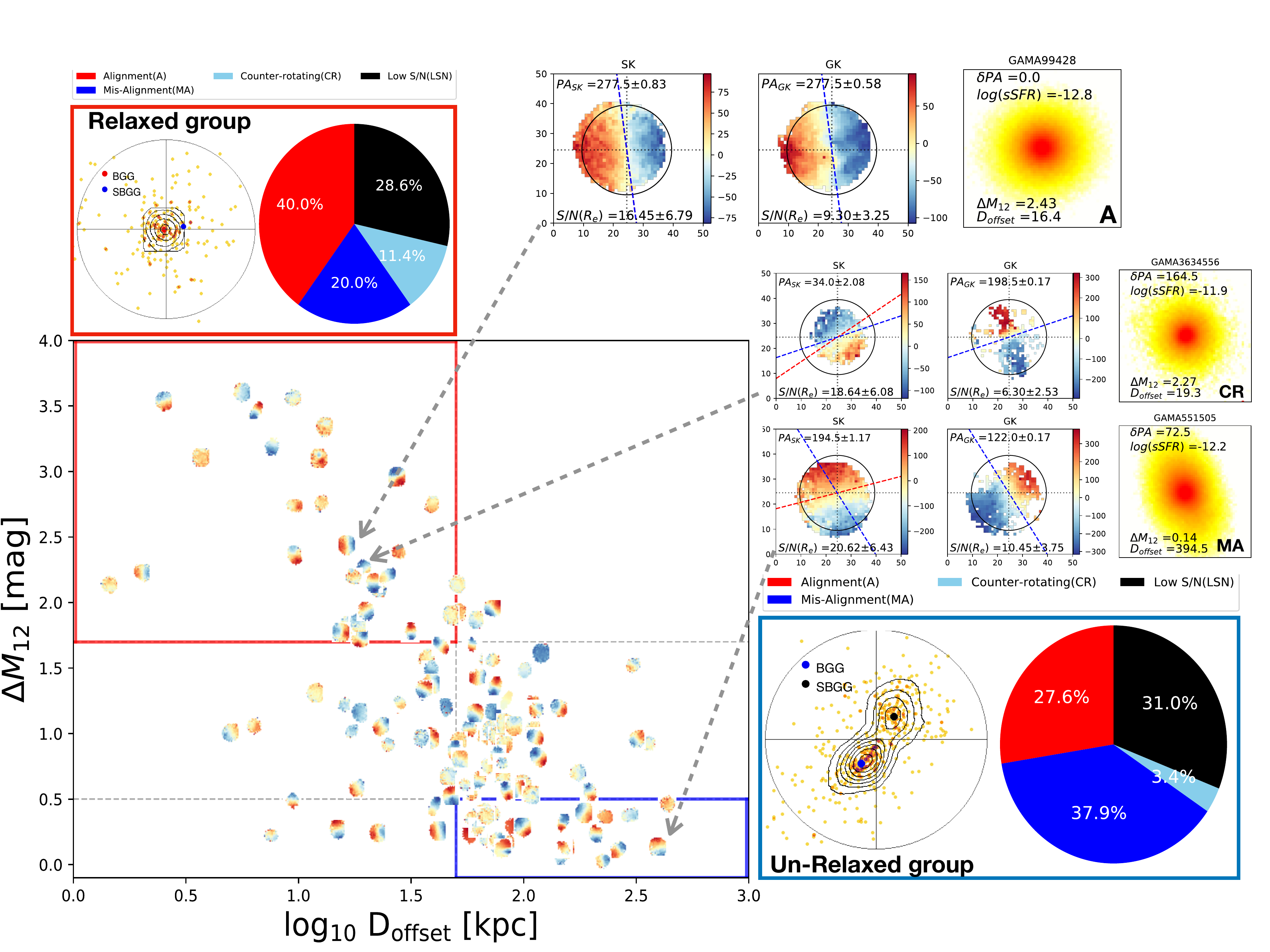}
		\caption{A map of luminosity gap, $\Delta M_{12}$ as a function of BGG luminosity offset, $D_{offset}$ symbolized by the BGGs stellar kinematic. The red and blue bordered squares highlight the limits of the regions inhabited by the relaxed and unrelaxed groups, respectively. The pie chart shows the fraction of aligned(A: red), misaligned(MA: blue) and counter-rotating (CR: sky blue) BGGs, and BGGs where there is too much uncertainty in their misalignment PA to classify them, due to low signal-to-noise (LSN: black) in the $H_\alpha$ velocity maps, within the effective radius(Re). Besides the pie charts, we present a representative example of a member of the relaxed and unrelaxed group, as found in the semi-analytic model from \citep{Raouf2017}. The location of the brightest group galaxies (BGGs) and the second brightest group galaxies (SBGGs) are highlighted with a filled circle symbol for each group. In the upper-right panels, we provide representative examples of the stellar kinematics (SK), gas kinematics (GK) velocity map, and a postage stamp in $r$-band from the GAMA Panchromatic Swarp Imager (PSI), for a BGG in the relaxed and unrelaxed group. The major axes of rotation are shown by red and blue dashed lines in the SK and GK kinematic maps, respectively, along with the median signal to noise (S/N) in $Re$ with STD error. Each kinematic map inside the panels has a 15" side length and the SAMI bundle size is shown by the circle.  In the r-band optical postage stamp (right panels), we also show the $\delta$PA, sSFR, $\Delta M_{12}$ and $D_{offset}$.
		}
		\label{fig:samiimage}
	\end{figure*}
	
	\subsection{Spin of galaxies}        
	For each galaxy, the parameter $\lambda_{R}$ is used as a proxy for the spin parameter and following \cite{Emsellem2007} is calculated using:  
	\begin{equation}
	\lambda_{R}  = \sum_i ( F_i*R_i* |V_i| ) /  \sum_i ( F_i*R_i* (V_i^2+ \sigma_i^2)^{1/2} ),
	\end{equation}
	where, the subscript $i$ refers to the $i$th spaxel within the ellipse, $F_i$ is the flux of the $i$th spaxel, $V_i$ is the stellar velocity in $km/s$, and $\sigma_i$ is the velocity dispersion in $km/s$.
	For $\lambda_R$, $R_i$ is the semi-major axis of the ellipse on which the spaxel $i$ lies, where we use an ellipse at the effective radius, $Re$, in this study. We also estimate the ellipticity at the effective radius, $\epsilon_{Re}$, from the best-fit MGE model (F. D'Eugenio in Prep.). 
	\begin{figure*}
		\centering
		\includegraphics[width=0.8\linewidth]{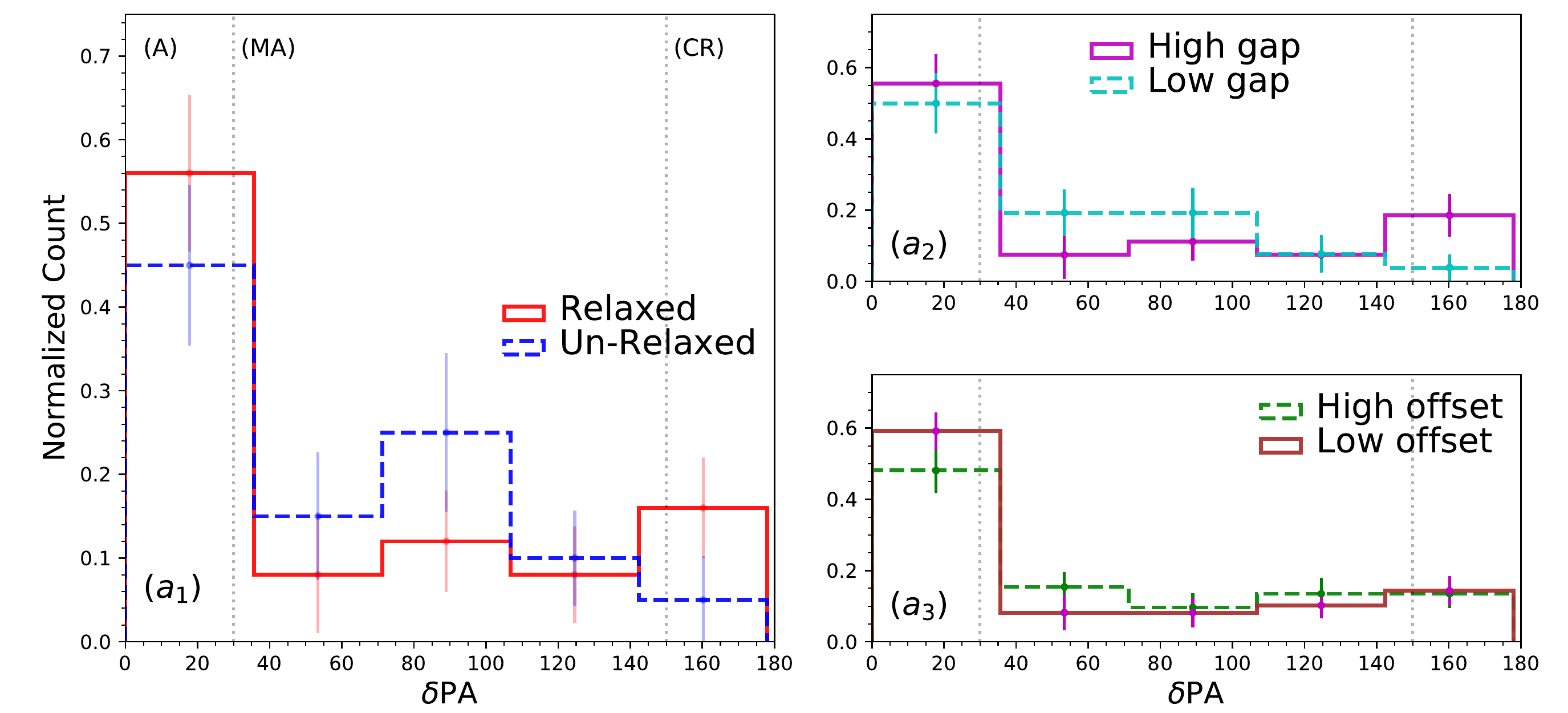}
		\includegraphics[width=0.8\linewidth]{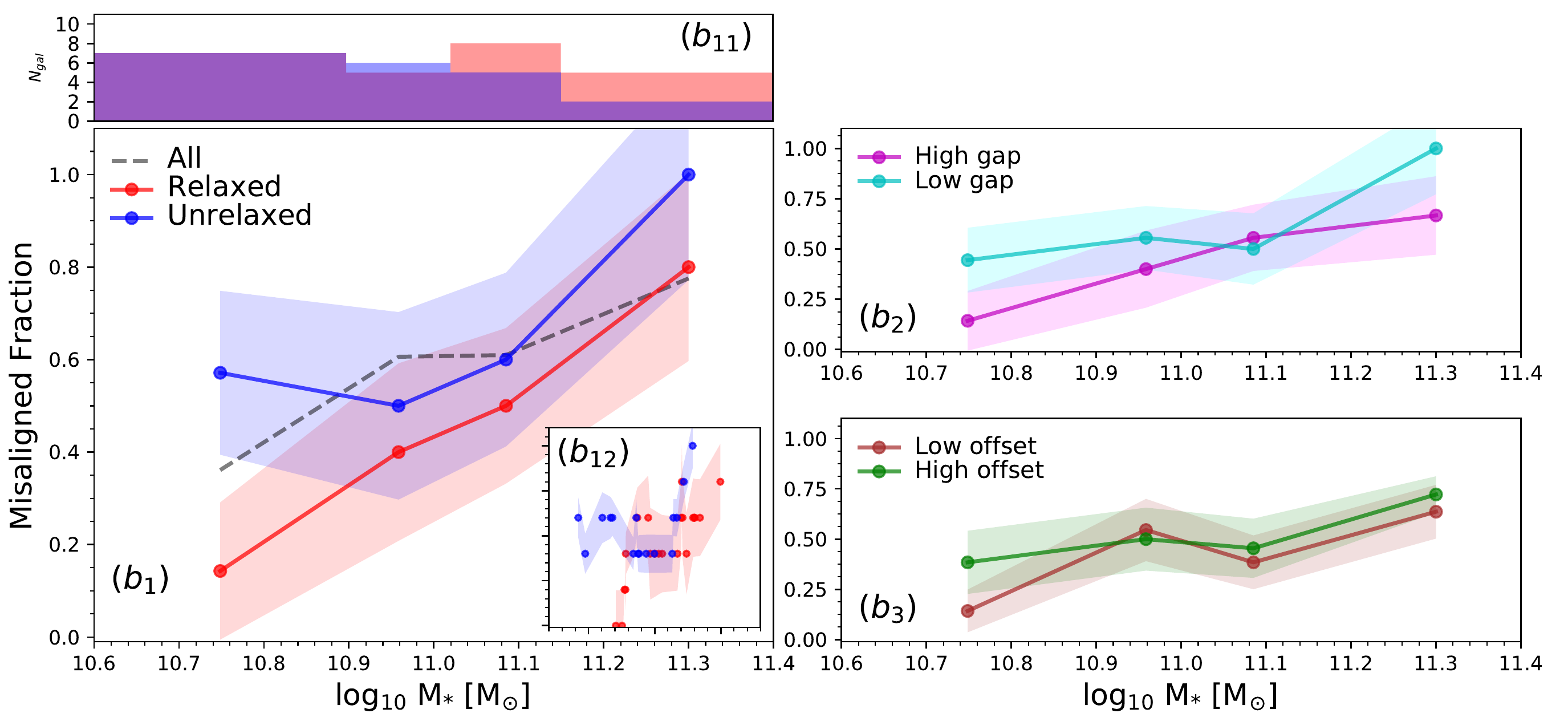}
		\includegraphics[width=0.8\linewidth]{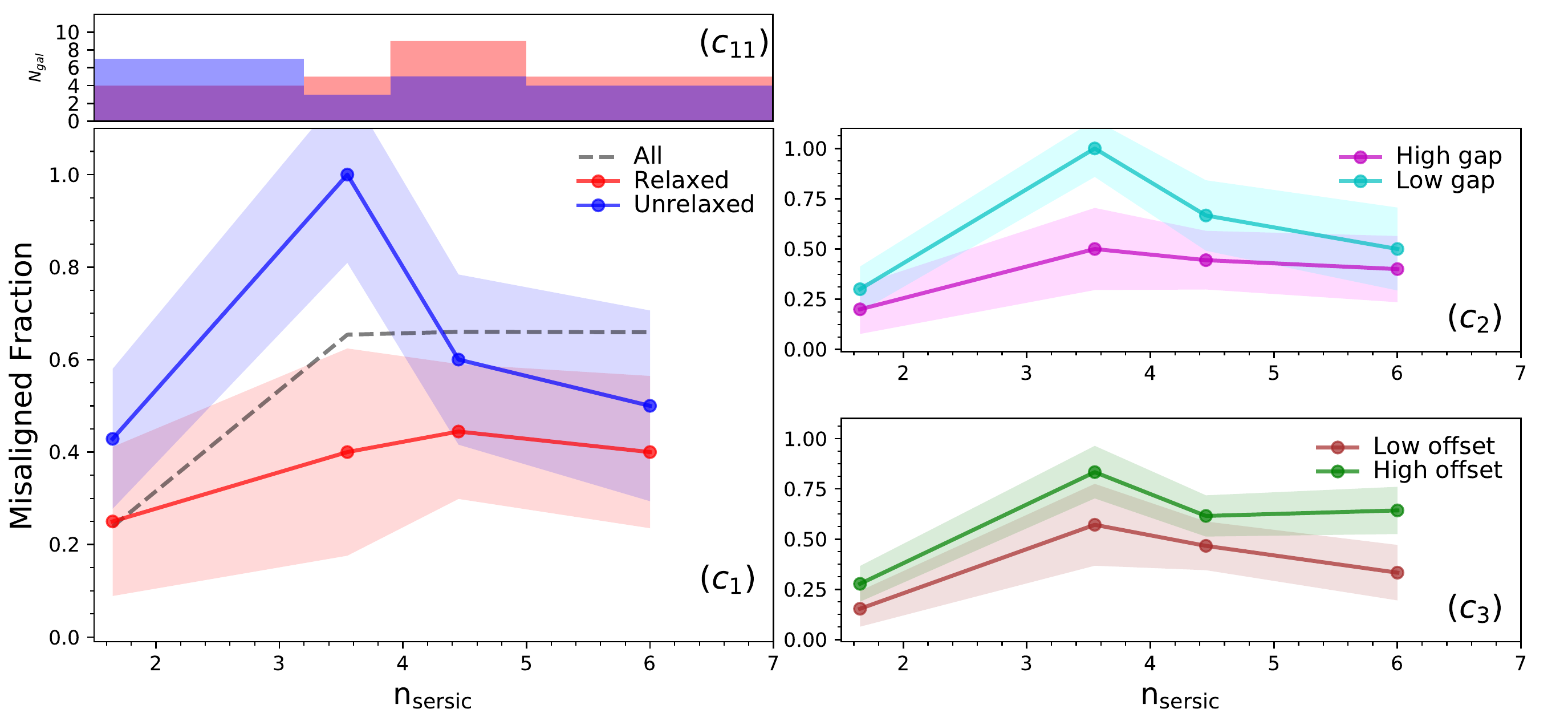}
		\caption{ Top: Distribution of the star-gas misalignment PA, $\delta$PA, for BGGs  for relaxed/unrelaxed (panel ($a_{1}$)), High/Low gap (top right panel ($a_{2}$)) and Low/High offset (bottom right panel ($a_{3}$)) groups. Errors are standard deviation (one-$\sigma$) errors calculated using the bootstrap method. The vertical dotted lines show $\delta$PA of 30 and 150 degrees that we used for discriminating between alignment/misalignment(A/MA) and contour rotating(CR) BGGs, respectively. 
			Middle: The fraction of misaligned $\delta$PA as a function of stellar mass for BGGs, separated into the sub-samples ($b_{1}$, $b_{2}$, $b_{3}$) given in the legend. The dashed-line show the median trend for all BGGs in our sample. In each panel, the errors (color shade-lines) are one-$\sigma$ confidence intervals on the fractions calculated using the bootstrap method. Along the upper x-axis of the panel ($b_{11}$), we show the stellar mass distribution of relaxed and unrelaxed in the bins ($N_{gal}$) in order to present the statistics in each data point. The inset figure in the left panel ($b_{12}$) shows the median fraction of misaligned $\delta$PA  BGGs as a function of stellar mass using the moving average method with standard deviation error, where the axes range is the same as in the main panel.
			Bottom: The fraction of misaligned $\delta$PA as a function of BGGs Sersic index ($c_{1}$, $c_{2}$, $c_{3}$), $n_{Sersic}$, with the same description as given in the middle panel. }
		\label{fig:SAMI-Hist-align_hist}
	\end{figure*}
	\begin{figure}
		\centering
		\includegraphics[width=1.0\linewidth]{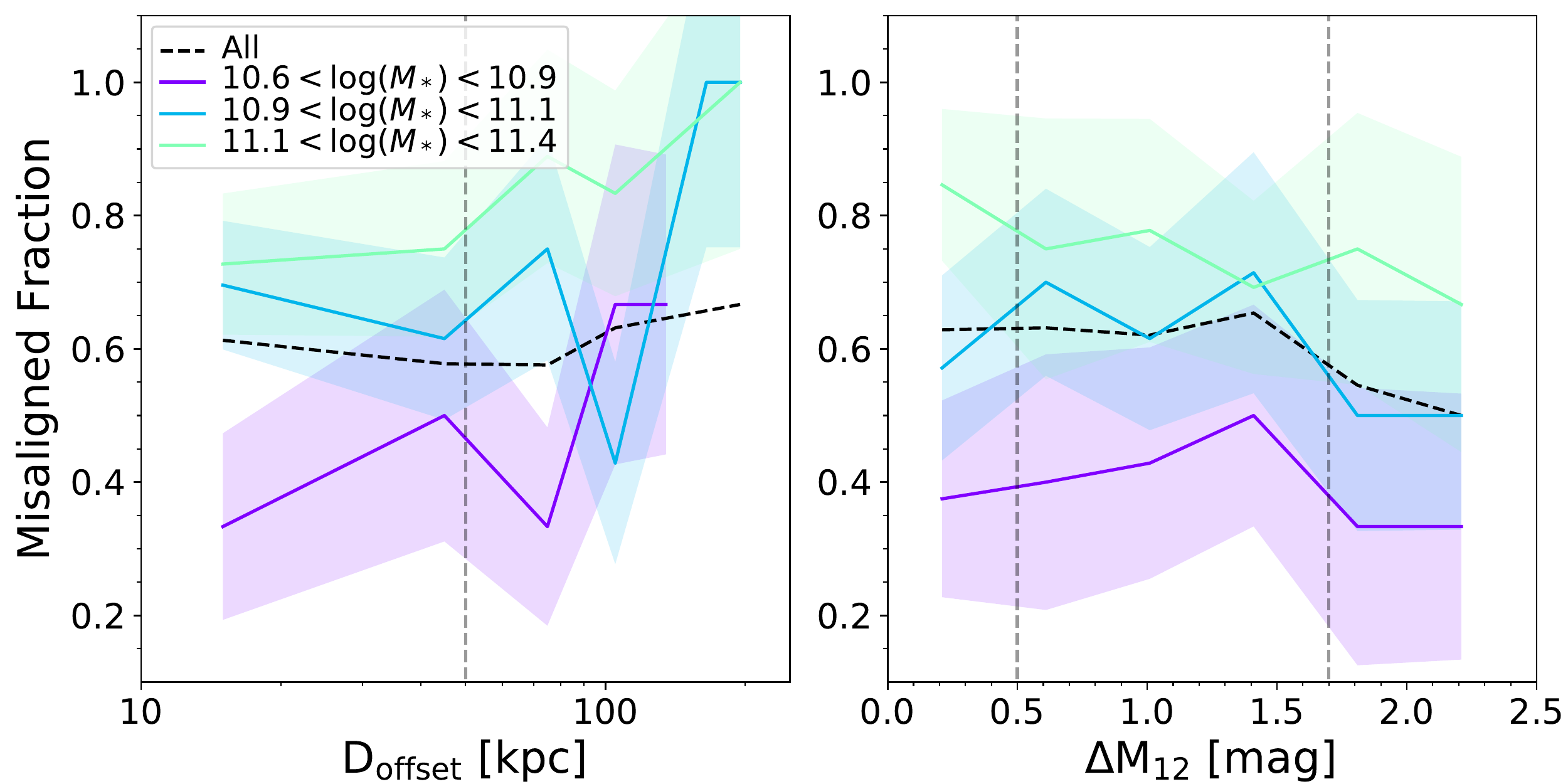}
		\caption{ The fraction of misaligned gas-star PA for BGGs residing in groups with a varying luminosity gap, $\Delta M_{12}$(right), and BGG offset, $D_{offset}$(left). The color lines show the trend for 3 stellar mass bins. The black dashed-lines show the median trend for all BGGs in our sample. The errors (color shade-lines) are one-$\sigma$ confidence intervals on the fractions calculated by the bootstrap method.  The grey vertical dashed-lines in each panel show the selection for Low/High offset and High/Low gap groups.
		}
		\label{fig:SAMI-Frac-MA}
	\end{figure}
	
	\subsection{Misalignment angle}   \label{Sec:MA} 
	Stellar and gas kinematic measurements are derived for SAMI galaxies using the penalized pixel fitting code \citep[pPXF;][]{Cappellari2004,Cappellari2017} based on the method described in \cite{Krajnovic2006}.
	We use the \textsc{FIT-KINEMATIC-PA} code to measure the PA of the 2D stellar/gas rotation as described in \cite{vandeSande2017} and \cite{Fogarty2014} on all spaxels that pass the quality cut for the velocity measurements and following the method described in Appendix C of \cite{Krajnovic2006}. 
	The kinematic PA is measured assuming the center of the map is at (24.5,24.5). The fit to the kinematic PA also includes measuring a 3-$\sigma$ uncertainty on that fit that can maximally be 90 degrees. Because we provide a one-$\sigma$ uncertainty in this catalog, the maximum error of the PAs is 30 degrees. The kinematic PA is measured anticlockwise, with 0 degrees being North ($y = 0$) in the sky. Note that, for the one-$\sigma$ errors, any PA with an uncertainty value close to 30 degrees means the fit is not very reliable. But this occurs in less than 5\% of all the SAMI samples.  
	
	Misalignment angles for SAMI galaxies are defined as the difference between the kinematic PA of the gas ($PA_{GK}$) and the stars ($PA_{SK}$) when both velocities could be fitted with a reliable PA. We refer to these misalignment angles as the PA offset($\delta$PA) and they always lie between 0 and 180 degrees. 
	Following other studies in the literature, \citep[e.g.,][]{Davis2016,Lagos2015,Bryant2019}, we assume that galaxies with a misalignment angle, $\delta PA = |PA_{SK} - PA_{GK}|$ of less than 30 degrees are aligned (A). Otherwise, we refer to them as misaligned(MA) galaxies. If $\delta PA$ is between 150\degree and 180\degree, we consider them as counter-rotating(CR) BGGs (extreme cases of MA galaxies). We also ran tests where we varied the critical angle required for a BGG to be considered misaligned from the standard value of 30 degrees. In general, when we change the angle from 30 to 15 degrees the results agree within the errors and so they are not extremely sensitive to our exact choice of this critical angle.

	The $H_{\alpha}$ emission line is used to measure the gas dynamics. The signal to noise(S/N) is estimated for each galaxy individually based on the median value of the S/N for non-zero spaxels within $R_e$ of a galaxy after removing the outliers spaxel ($\gtrsim$ 3-$\sigma$) using a $\sigma$-clipping approach. In this study, we labeled those galaxies with S/N $>$ 4 and S/N $<$ 4 as high (HSN) and low (LSN) signal-to-noise galaxies, respectively.
	We also did further cleaning of our sample by checking the trustworthiness of the $H_{\alpha}$ emission line maps by eye, in order to remove those few objects ($\lesssim$5\% of the sample) that survive that S/N cut, yet we still cannot measure a reliable $PA_{GK}$ from. 
	
	\subsection{Sample selection}    
	We focus on the BGGs ($M_{*} > 10^{10.6} M_{\odot}$) hosted by the groups with at least four members and whose halo mass peaks around $10^{13.25} -10^{13.75} M_{\odot}$\footnote{Groups with N$<$ 4 spectroscopic members have a large uncertainty on the estimation of halo mass and other group parameters \citep{Robotham2011}}. 
	Our galaxy groups are separated into dynamically relaxed and unrelaxed sub-samples using the following criteria:
	
	{\it{Criteria I}}: Galaxy groups with a large luminosity gap between the BGG and the second brightest group member, $\Delta M_{12} \ge 1.7$ (``high gap'') in $r$-band within half the virial radius of the group. Besides, we also impose that the BGG is located within a radius of 50 kpc from the luminosity/stellar-mass centroid of the group (``low offset''). This criterion reduces our sub-sample to 35 galaxy groups, labeled as ``relaxed" systems. 
	
	{\it{Criteria II}}: Galaxy groups with a small luminosity gap, $\Delta M_{12} \le 0.5$ (``low gap'') in $r$-band within half the virial radius of the group. We impose the BGG to be located outside a radius of 50 kpc centered on the luminosity/stellar-mass centroid of the group (``high offset''). This criterion reduces our sub-sample to 29 galaxy groups labeled as ``unrelaxed" systems. 
	
	This selection criterion leads us to identify groups with a similar halo mass but differing sub-structure (see Figure \ref{fig:samiimage}) and formation epoch \citep{Raouf2014,Raouf2016,Raouf2018,Farhang2017,Haghighi2020}. Note that the small difference between the adopted $\Delta m_{12}=1.7$ limit used for the selection of the relaxed and high gap groups and the one conventionally used in previous studies of optical fossil groups \citep{Jones2003,Ponman1994}, $\Delta m_{12}=2.0$, is chosen to increase the number statistics of the relaxed and high gap sub-samples (see Figure \ref{fig:sami-hist-gap-off} left panel). We also use $\Delta M_{12} \le 0.5$ as our definition of optical non-fossil groups, based on the simulation study of \citep{Dariush2010}. We choose the offset criteria based on the distribution of offset values found in our semi-analytic model \citep[see fig. 3, ][]{Raouf2014} where the BGG offset distribution of dynamically relaxed groups peaks at around $\sim$ 50 - 70 kpc. In this study, we use 50~kpc for the minimum offset criteria, in order to increase the statistics in the unrelaxed and high offset sub-samples (see right panel of Figure \ref{fig:sami-hist-gap-off}). 
	We also note that the measurements derived from the semi-analytical models are based on the average of three projections along the three Cartesian coordinates of the simulation box, to allow them to be fairly compared with the observations.	
	The normalized distributions of BGG stellar mass  for each sub-sample are shown in \mbox{Figure \ref{fig:SAMI-Hist-mh_ms}} (See Table \ref{tab:value} for the group number counts). The figure shows that our various sub-samples (high/low gap, low/high offset, and relaxed/unrelaxed groups) have a fairly similar range of stellar mass. We note that there is a small offset in the stellar mass distributions between the relaxed and unrelaxed sub-samples, that actually contributes to strengthen our conclusions later (sec. \ref{Sec:map_SAMI}). In Figure \ref{fig:sami-ms-mh} we also show the distribution of stellar mass vs. halo mass for the sample of relaxed and unrelaxed groups. As can be seen, the small stellar mass difference mentioned earlier is more significant in low mass halos. Note that the host halo mass distribution is quite similar for the various sub-samples and does not influence our results significantly overall.
	To try to remove source confusion, all group BCGs are required to have at least a 10 kpc projected separation between the most luminous galaxies and other group members. In this way, less than 5\% of the total sample has more than one galaxy in the SAMI field of view that could artificially affect the PA estimation. But we also visually check by eye the kinematic maps to confirm that our results are not affected in these few cases.
	
	\section{Results}
	Here we compare the stellar regularities and rotational classification then focus on the gas dynamics in order to compare the offset between the rotational axis of gas and stars in the BGGs of our sub-samples. In order to get a feeling for the significance of the difference between the sub-samples, we calculate one-sigma errors using the bootstrap method with iterations of N = 1000. 	
	
	\subsection{Stellar dynamics in our BGG sample}
	\subsubsection{Measurement of regularity of BGG rotation using $k_5/k_1$}
	The top panels of Figure \ref{fig:sami-hist-k51}  show the distribution of the kinematic asymmetry ($k_5/k_1$) for BGGs in relaxed/unrelaxed ($a_1$), high/low gap ($a_2$), and low/high offset ($a_3$) groups. We define a `regular rotation'(R) BGGs with $k_5/k_1 \leq$ 0.04, a `quasi-regular rotation'(QR) BGGs with 0.04 $< k_5/k_1 <$ 0.08, and a `non-regular rotation'(NR) BGGs with $k_5 /k_1 \geq$ 0.08, following the approach of \cite{Krajnovic2011} and similarly to in \cite{vandeSande2017}. A representative example of a stellar kinematic map for an R(GAMA323558), QR(GAMA272831) and NR(GAMA209680) rotator BGG are presented in the figure. 
	In the left panel of Figure \ref{fig:sami-hist-k51} ($a_1$), the BGG in relaxed groups peaks at $k_5/k_1$ values corresponding to those defined as regular rotators
	and decreases dramatically towards higher values of $k_5/k_1$. Considering the statistical error bars, the fraction of regular rotators for BGGs in unrelaxed groups is significantly lower at the one-$\sigma$ level. 
	We also conduct a Kolmogorov-Smirnov (KS) test of the difference in the distribution of relaxed and unrelaxed groups. Note that a resulting low KS statistic and/or smaller p-value indicates the evidence for rejecting the null hypothesis of the same distribution is stronger. The resulting KS statistic is 0.17, however the large p-value (0.68) indicates the difference between the distributions is low significance (Table \ref{tab:ks}).
	
	In the right panels ($a_2$, $a_3$), we show the distribution of $k_5/k_1$ for the corresponding BGGs residing in low/high gap  and high/low offset groups. While the BGG offset shows lower uncertainty compared to the luminosity gap, the BGGs in high offset groups have a higher fraction of non-regular rotators compare to the other sub-samples.
	Once again, we conduct a KS test and find that in the majority of cases, the BGGs in high offset groups tend to be non-regular with a higher statistical difference (0.2) compared to the distribution of low offset groups (p-value $=$ 0.07). This shows the higher significance of difference compared to the high/low gap groups (statistic = 0.14 , p-value = 0.8). However, we note that these histograms of the total distributions might not be expected to be as sensitive to the differences between the sub-samples, because we do not control for any other parameter (e.g. mass).
	In general, around $\sim$45\% of the relaxed groups consist of regular rotator BGGs, while the fraction is less than $\sim$29\% for BGGs in unrelaxed systems. We also find the highest fraction of non-regular BGGs in high offset groups($\sim$49\%). All the above fractions, including their binomial errors, are reported in Table \ref{tab:value}.
	
	The bottom panels of Figure \ref{fig:sami-hist-k51} show the fraction of regular BGGs at a given stellar mass for all sub-samples, including errors based on the one-$\sigma$ confidence intervals on the fractions, calculated using the bootstrap method. We try to get similar numbers in each stellar mass bins as indicated with $N_{gal}$ in the histogram panel on the upper x-axis ($b_{11}$). This leaves us with low $N_{gal}$ in each bin, which is inevitable because we are sharing a small sample out over several bins.
	In the bottom left panel ($b_1$), the dashed line shows that the regular fraction decreases with stellar mass. The Spearman correlation coefficient for the linear regression between $k_5/k_1$ and stellar mass for all SAMI galaxies is $r_s$ = 0.33, p-value = 0.0001. The correlation increases when we focus on BGGs in relaxed ($r_s$ = 0.56, p-value = 0.002) and unrelaxed ($r_s$ = 0.45, p-value = 0.02) groups. We also note that the unrelaxed groups are less regular in the upper panel ($k_5/k_1$ distributions), despite having slightly lower masses which could have resulted in a higher regular fraction. Hence, this further strengthens the differences seen in the top-left panel where we show the histogram of $k_5/k_1$.
	
	The red line is generally above the blue line, except at the highest stellar masses, which shows that the unrelaxed groups have a less regular rotation at a fixed stellar mass, considering the 68\% confidence interval errors (one-$\sigma$). In addition, we also tried making a moving average plot of the regular fraction versus stellar mass (see the upper-right inset Figure in the bottom left panel ($b_{12}$)). In this way, we ensure that there are equal numbers (5 galaxies) of galaxies in each data point on the plot. As can be seen, the trend of regular BGG fraction in relaxed groups tends to be systematically offset above the BGGs in unrelaxed groups. Although the scatter is large, we believe this confirms the significance of the difference between the relaxed and unrelaxed sub-samples. Overall, we see similar results as are found in the main panel, which confirms our conclusions that the relaxed groups tend to have BGGs whose stellar dynamics are systematically more regular. 
	
	The regular fractions of the corresponding low/high gap (panel $b_2$) and high/low offset (panel $b_3$) groups, as a function of stellar mass, are shown in the right panels. 
	Both parameters have a similar trend in which the combination of the two parameters tend to complement each other in order to disentangle between the fraction of regular rotator BGGs in the relaxed and unrelaxed groups. 
	The result shows that there is a difference in the fraction of regular rotator BGGs in the relaxed groups (i.e., those formed earlier) at a roughly one-$\sigma$ confidence level compared to unrelaxed groups. However, the statistics in the individual points are inevitably quite poor, as we are spreading an already quite small sample over multiple stellar bins. Although in some individual points, the significance is above one-$\sigma$. When we split the sample by low and high offset we get larger statistics in each data point, and the significance of the difference between the sub-samples seems to increase to slightly greater than one-$\sigma$. 
	
	\subsubsection{The spin parameter of our sample}
	In Figure \ref{fig:sami-spin-ru}, we show the distribution of the spin parameter ($\lambda_{Re}$) 
	as a function of ellipticity ($\epsilon_{Re}$)  measured within the effective radius/ellipse as defined in \citep{vandeSande2017}. This is the conventional diagram for separating galaxies into slow and fast-rotating subcategories \citep{Emsellem2011,Cappellari2016}. By eye, it is not easy to see a clear difference between the relaxed and unrelaxed samples, but the horizontal bar plots in the right-hand panel reveal the statistics more clearly. The fraction of slow rotator BGGs is a bit higher ($\sim 10\%$) in the relaxed system compared to the unrelaxed groups, while both samples have the same fraction of fast rotators BGGs as expected for the most early-type galaxies \citep{vandeSande2017}. A representative stellar kinematic map for fast (GAMA209680) and slow (GAMA137838) rotator BGGs is shown in this figure. 
	In general, over 50\% of our sub-samples are fast-rotators BGGs with the highest fraction for BGGs in low offset groups($\sim$55\%). 
	\textit{We find that the BGGs of unrelaxed groups frequently rotate faster.} However, this result is of low significance due to low number statistics.
	
	\begin{table*}
		\centering
		\scriptsize
		\caption{The group number counts and percentages with one-$\sigma$ binomial confidence interval errors; low S/N(LSN), various kinematic $\delta$PA (A, MA and CR), kinematic asymmetry (NR, QR and R) and fast and slow rotators of BGGs residing in; High gap, Low gap,  High off-set, Low off-set, Relaxed and Unrelaxed groups. }
		\label{tab:value}
		\begin{tabular}{lcccccc}            
			\hline \\
			Shape         &    High Gap &   Low Gap    & High off-set    & Low off-set  & Relaxed & Unrelaxed   \\
			\hline \hline
			Total number    & 39               & 37         & 82         & 72         & 35        & 29          \\
			\hline
			Alignment(A) & 38.5$\pm$6.9 \% & 32.4$\pm$6.6 \% & 29.3$\pm$6.5 \% & 40.8$\pm$5.4 \% & 40.0$\pm$6.8 \% & 27.6$\pm$5.9 \%\\
			Misalignment(MA) & 20.5$\pm$5.9 \% & 35.1$\pm$6.7 \% & 28.0$\pm$6.6 \% & 18.3$\pm$6.7 \% & 20.0$\pm$5.7 \% & 37.9$\pm$6.5 \%\\
			Counter rotating(CR) & 10.3$\pm$5.9 \% & 2.7$\pm$6.7 \% & 6.1$\pm$6.6 \% & 9.9$\pm$6.7 \% & 11.4$\pm$5.7 \% & 3.4$\pm$6.5 \%\\
			Low S/N(LSN) & 30.8$\pm$6.6 \% & 29.7$\pm$6.5 \% & 36.6$\pm$5.0 \% & 31.0$\pm$6.7 \% & 28.6$\pm$6.3 \% & 31.0$\pm$6.1 \%\\
			\hline
			non-regular & 41.2$\pm$6.9 \% & 40.0$\pm$6.8 \% & 48.6$\pm$4.9 \% & 37.5$\pm$5.0 \% & 38.7$\pm$7.2 \% & 44.4$\pm$7.8 \%\\
			quasi-regular& 17.6$\pm$5.5 \% & 25.7$\pm$6.1 \% & 22.2$\pm$4.1 \% & 23.4$\pm$4.4 \% & 16.1$\pm$5.6 \% & 25.9$\pm$7.0 \%\\
			regular & 41.2$\pm$6.9 \% & 34.3$\pm$6.6 \% & 29.2$\pm$4.5 \% & 39.1$\pm$5.1 \% & 45.2$\pm$7.3 \% & 29.6$\pm$7.2 \%\\
			\hline
			Slow Rotator(EE+11)& 41.0$\pm$6.5 \% & 29.7$\pm$6.2 \% & 37.5$\pm$4.7 \% & 36.6$\pm$4.4 \% & 42.9$\pm$6.9 \% & 34.5$\pm$7.2 \%\\
			Slow Rotator(MC+16)& 43.6$\pm$6.5 \% & 29.7$\pm$6.2 \% & 40.3$\pm$4.8 \% & 37.8$\pm$4.5 \% & 42.9$\pm$6.9 \% & 31.0$\pm$7.1 \%\\
			Fast Rotator(EE+11)& 53.8$\pm$6.6 \% & 54.1$\pm$6.7 \% & 55.6$\pm$4.9 \% & 56.1$\pm$4.6 \% & 51.4$\pm$6.9 \% & 48.3$\pm$7.6 \%\\
			Fast Rotator(MC+16)& 51.3$\pm$6.6 \% & 54.1$\pm$6.7 \% & 52.8$\pm$4.9 \% & 54.9$\pm$4.6 \% & 51.4$\pm$6.9 \% & 51.7$\pm$7.6 \%\\			
			\hline \\		
		\end{tabular}
	\end{table*}

	\subsection{Kinematic misalignment between the gas and stellar dynamics}\label{Sec:map_SAMI}
	To measure the kinematic misalignment between the gas and stellar dynamics, we need to be able to trust both the stellar and gas dynamics individually to measure the difference in their PA. However, in practice it is normally the gas which is the limiting factor, hence it was necessary to do some additional cleaning of the samples. 
	We removed all objects whose gas PA maps were measured to be too low signal to noise. This was accomplished in two steps. Firstly, we removed all objects whose S/N criteria $<$ 4 (LSN objects) and then we did some additional cleaning by eye (as described in sec. \ref{Sec:MA}). 
	Concerning any bias in the population that can be the LSN galaxies compare to our full sample, we find the fraction of LSN galaxies is similar between relaxed and unrelaxed groups.
	
	Figure \ref{fig:samiimage} shows the $\Delta M_{12} - D_{offset}$ map with symbols that show the stellar kinematic of each individual galaxy in our sample. 
	Representative examples of an aligned(A), misaligned(MA), and counter-rotating(CR) galaxy are shown
	for the gas and stellar velocity map and photometric postage stamp (see Appendix \ref{Sec:Apendix1} for the full sample of relaxed and unrelaxed galaxies, respectively). The statistical percentage of A, MA, CR, and LSN for BGGs residing in relaxed and unrelaxed groups are shown in the pie-charts.  Note that CR is an extreme subcategory of MA but with  very low statistics in our sample, so here we merely report the percentage without using them as part of our main conclusion. The sub-structure of an ideal relaxed and unrelaxed group is shown in the corresponding box with density contours, extracted from the semi-analytic model \cite[Radio-SAGE;][]{Raouf2017}. 
	
	In general, regarding the blue and red regions of galaxies in the map of $\Delta M_{12} - D_{offset}$, galaxies in the relaxed groups tend to be more aligned by over 40\%  and in general higher than the fraction of total misaligned BGGs including MA($\sim$20\%) and CR ($\sim$11.4\%). Around $\sim$29\% of galaxies in the relaxed group are LSN BGGs. On the other hand, almost $\sim$41\% of the BGG in the unrelaxed group are misaligned (MA + CR). Such groups have $\sim$27.6\% aligned and $\sim$31\% LSN BGGs with around 3.4\% fraction of CR galaxies. The percentage of other sub-samples, including their binomial errors, are reported in Table \ref{tab:value}. As might be expected, most of the LSN BGGs in our sample have uncertain $\delta$PA. We find that the removal of these objects has a similar impact on both the relaxed and unrelaxed groups. 
	
	Further, the top left panel of Figure \ref{fig:SAMI-Hist-align_hist} ($a_1$) shows the distribution of the kinematic misalignment angles of gas and stellar velocity map for BGGs in relaxed versus unrelaxed groups. Both relaxed and unrelaxed group have a peak in the aligned region($< 30\degree$), but with a more significant peak in the relaxed groups with respect to the unrelaxed groups. We conduct a KS test of the difference in the distributions of relaxed and unrelaxed groups and get a KS statistics of $\sim$0.2, with a large p-value (0.78), which indicates the low significance of the difference in their distribution (Table \ref{tab:ks}). On the right, we show the same plot but for high/low gap and low/high offset groups. While overall the shapes of the histograms are quite similar, a KS test gives the statistical difference between high/low gap groups is equal to 0.25 which seems slightly larger and more significant (p-value = 0.3) than in the low/high offset groups (statistic = 0.1, p-value = 0.8). This suggests that the luminosity gap is slightly more important for determining the amount of misalignment than the offset parameter. However, as we will show, it is important to take into account the other parameters such as the stellar mass and Sersic index.
	
	So far, overall, the difference in the gas-star $\delta$PA BGGs residing in groups with different dynamical states (and so different formation histories) has not been very significant. 	
	But this is partly because we have combined all the galaxies in each subsample, regardless of their stellar mass or morphological shape. So now, in the middle/bottom panel of Figure \ref{fig:SAMI-Hist-align_hist} ($b_1/c_1$) we show the fraction of misaligned BGGs at a given stellar mass/Sersic index for the same sub-samples (relaxed vs unrelaxed). Errors in the fractions are one-$\sigma$ confidence intervals, calculated using the bootstrap method. Given the small sample size, there is no fully robust manner to calculate the errors. Therefore, we also try calculating our errors using the binomial confidence levels, separately.
	Both methods of measuring the errors give similar results which gives us additional confidence in the significance of the differences that we quote.  Note that the BGGs of unrelaxed groups are slightly lower mass in the left-middle-panel ($b_1$) which actually strengthens the significance of the difference seen in the first panel (i.e., the $\delta$PA distributions).

	In general, the gas-star kinematic misalignment angle has a stronger dependence on the Sersic index in comparison to the stellar mass, as shown by comparing the Spearman coefficient correlation for linear regression ($r_s$) and the p-value for all the sample ($M_{star}$ : $r_s$ = 0.26, p-value = 0.005; $n_{Sersic}$: $r_s$ = 0.38, p-value = 0.00005). This is the same result as was found in \cite{Bryant2019} although the sample selection is very different in their study.
	
	In the middle panel, along the upper x-axis (panel $b_{11}$), we plot a histogram of the number of galaxies in each data point and this leaves us with low $N_{gal}$ in each bin, which is inevitable because we are sharing a small sample out over several bins. Note that we choose the bins to try and get similar numbers in each data point and the error bars are clearly larger when there are less galaxies in a data point.
	In the panel below the histogram ($b_{1}$), we show that, at a given stellar mass, the misaligned fraction of BGGs is higher for unrelaxed groups. The fraction for relaxed systems is consistently below the unrelaxed sample across the plot. But the significance of the difference is weak for some points, perhaps due to the low statistics in each stellar mass bins. Note that we lose around 30\% of the BGGs in the sample that do not have sufficient signal-to-noise to measure the gas dynamics position angle.
	The dashed line (panel $b_{1}$) shows that the misaligned fraction is greater in more massive galaxies. Despite the fact that our unrelaxed sample has slightly less massive galaxies (Figure \ref{fig:SAMI-Hist-mh_ms}), we still see that the unrelaxed groups have more misalignment in the upper-left panel ($b_{1}$). This strengthens the results seen in the distribution of galaxies further, as the stellar mass difference could be partially cancelling out the true difference. This also provides a motivation to split the sample by stellar mass.
	The corresponding low/high gap and high/low offset groups misaligned fraction as a function of stellar mass are shown in the centre right panels ($b_2, b_3$).
	The results show that the low gap groups tend to have a higher misaligned fraction relative to the high gap systems. The KS test results confirm that there is a greater statistical difference and significance between low/high gap groups than in the low/high offset groups (see Table \ref{tab:ks}). The sub-samples have better number statistics for high/low offset groups, which leads to lower uncertainty with a more or less similar fraction of misaligned BGGs. As with the regular fraction, we also conducted a moving average test of misalignment fraction versus stellar mass (see inset Figure in the middle left panel ($b_{12}$)). This ensures equal numbers (5 galaxies) of galaxies in each data point. It can be seen that the trend of misaligned BGG fractions in unrelaxed groups tends to be systematically above the BGG in relaxed, even though the scatter is large. Once again, we obtain similar results overall, that are consistent with our conclusions on the impact of the group dynamical state on the misalignment of gaseous and stellar components, and also, confirm the significance of the differences between relaxed and unrelaxed groups seen in Figure \ref{fig:SAMI-Hist-align_hist}.
	In summary, at a given stellar mass a small but statistically significant difference has been shown for the BGGs hosted by groups with different dynamical states. In turn, their dynamical state is linked to the formation epoch as shown in our earlier study using cosmological simulations \citep{Raouf2016}.
	
	In the bottom panel of Figure \ref{fig:SAMI-Hist-align_hist} ($c_{1}$), we repeat this process but this time plotting against the Sersic index, $n_{Sersic}$ instead of the stellar mass. 
	Again, we see there is a higher fraction of misaligned BGGs in the unrelaxed groups compare to the relaxed systems.
	
	Furthermore, the corresponding low/high gap (panel $c_{2}$) and high/low offset (panel $c_{3}$) groups show similar differences in misalignment fraction as a function of Sersic index, as shown in the bottom right panels, again verifying the role of dynamical state on the BGG gas-star $\delta$PA independent of Sersic index. Also, thanks to the higher number statistics of BGGs in the low and high offset groups ($\sim$ 2 times bigger than other sub-samples) we see a slight increase in the significance of the difference between them, compared to the other sub-samples. In the other word, the late formed systems (unrelaxed systems) tend to have a higher fraction of misaligned BGGs at fixed Sersic index.
	
	In summary for Figure \ref{fig:SAMI-Hist-align_hist}, it is clear that the difference between the relaxed and unrelaxed groups is marginal in the distribution of $\delta$PA. However, when we split by stellar mass and Sersic index, we find there are consistent offsets across most of the data points in the BGGs of relaxed and unrelaxed systems. The same offsets are also seen for the BGG offset and luminosity gap distribution where the sample size is increased by a factor of $\sim$ 2. The BGG offset and luminosity gap seem to cause roughly equally offsets for a fixed galaxy stellar mass or Sersic index. Although, each individual offset is quite low significance, it is the combination of these results that gives us more confidence in our overall conclusions. Although, we note that \textit{future larger samples will be crucial to enhance the significance of our result}.

	Finally, in Figure \ref{fig:SAMI-Frac-MA} we show the fraction of BGGs that have misaligned gas-star PAs, hosted by the groups with different luminosity gap and BGG offset. The trends show three different stellar mass bins with roughly the same number of BGGs. All $\Delta M_{12}$ and $D_{offset}$ bins include reasonable number statistics (N $>$ 3) except for the very high gap and offset groups in which the statistics are poor, as can be seen by the larger errors in those bins.
	Errors are given by the one-$\sigma$ confidence intervals on the fractions, calculated using the bootstrap method. As can be seen, a higher fraction of misaligned BGGs belongs to the stellar mass range over $10^{11.1} M_{\odot}$ (as already shown by the dashed line in Figure \ref{fig:SAMI-Hist-align_hist}). The misaligned fraction overall tends to increase with BGG offset but slightly decreases with increasing luminosity gap. We note that, although both the gap and offset trend are quite noisy, they both have a slight trend in one direction.
	
	\begin{table}
		\centering
		\small
		\caption{The KS test results (S: Statistic, P: P-value) for comparing the distribution of kinematic $\delta$PA and kinematic asymmetry ($k_5/k_1$) between the different sub-samples of groups. }
		\label{tab:ks}
		
		\begin{tabular}{lcc}  
			Sub-sample & 	$\delta$PA & $k_5/k_1$ \\
			\hline \hline
			relaxed/unrelaxed & (S: 0.19, P: 0.78) &(S: 0.17, P: 0.68) \\
			High/Low gap &      (S: 0.25, P: 0.33) & (S: 0.14, P: 0.82)\\
			Low/High offset &  (S: 0.11, P: 0.83) & (S: 0.20, P: 0.07)\\
			\hline \\
		\end{tabular}
	\end{table}
	
	\section{Discussion and conclusion}
	
	The focus of this study is the connection between the dynamical state of groups and the stellar and gas dynamics of BGGs using a sample of galaxy groups from the GAMA galaxy survey, where the internal kinematic properties of the BGG are measured in the SAMI survey. Our BGGs all have stellar masses over $10^{10.6} M_{\odot}$ with a similar host halo mass distribution. This choice of mass cut causes them to be typically early-type.
	
	To probe the dynamical state of the group halo, we use a combination of two independent optical indicators; the luminosity gap between the two brightest galaxies, $\Delta M_{12}$ (within half a virial radius of the group), and the offset between the position of the BGG and group's luminosity weighted center, $D_{off-set}$. Cosmological simulations show such a selection indirectly leads us to identify BGGs with different merger histories \citep[and different amounts of time since their last major merger; e.g., see fig. 2;][]{Raouf2018}.
	
	For the measurement of the internal kinematics of the BGGs, in the stellar component, we use a Fourier analysis on the regularity of the rotation field \citep[$k_5/k_1$ parameters;][]{Krajnovic2008}, and also consider their location in the spin-ellipticity plane \citep[e.g;][]{Cappellari2016,Emsellem2011}. For the gas kinematics, we measure the position angle of rotation in their H-alpha emission map and compare it with the position angle from their stellar dynamics map \citep[similar to in, ][]{Bryant2019}. The difference between these two position angles is known as the misalignment angle.
	
	We demonstrate that there is a small but statistically significant difference (at the one-$\sigma$ level) in the gas-star kinematics misalignment position angle ($\delta$PA) and, separately, the regularity of the stellar rotation of BGGs, when comparing groups with differing dynamical states of relaxedness.
	We also present the results as a function of stellar mass and Sersic index (in particular for $\delta$PA).
	We find a stronger correlation between gas-star misalignment and Sersic index than with stellar mass, in agreement with \cite{Bryant2019}.
	
	The main results are as follows.
	
	\begin{enumerate}
		\item We find that the BGGs in unrelaxed groups tend to have stellar kinematics with less regular rotation fields than those in relaxed groups. The statistical significance is low (of order one-$\sigma$), perhaps in part due to poor statistics. When we split the sample into high/low offset and high/low luminosity gap, the results remain intact, and the larger samples give improved statistical significance.
		
		\item Placing our sample in the spin parameter-ellipticity plane, we find that more than 50\% of our sample of BGGs have a stellar component that is considered a fast rotator. There is a slightly higher fraction ($\gtrsim$ 10\%) of slow rotators in relaxed groups with a low significance compared to the unrelaxed group. Thus, there is a weak indication that relaxed groups are more likely to host slow rotator BGGs (perhaps due to a slightly higher fraction of massive BGGs in relaxed groups).
		
		\item Now considering the internal gas kinematics as well, we show that the BGGs in unrelaxed groups tend to have higher $\delta$PA compared to relaxed groups. Note that there is a possible trend between the groups dynamics and the kinematics of BGGs, but that the result is not significant considering the errors calculated using the bootstrap method.
		
		\item Splitting the sample by stellar mass, or Sersic index, we find a clear correlation with $\delta$PA. We consistently find a higher fraction of misaligned BGGs hosted in unrelaxed groups. 
		However, given the small sample size, the statistical significance is low. But,  
		when we split the sample into high/low offset and high/low luminosity gap sub-samples our statistics are improved slightly (by a factor of 2) and we see a consistent dependency on the dynamical state of the group, but now with greater significance.
		
		\item In general, the luminosity gap and offset parameters seem to roughly equally contribute to our ability to differentiate relaxed and unrelaxed groups, within the range of parameters we consider in this study. 
		
	\end{enumerate}
	
	Although the statistical significance of any individual result is quite weak (perhaps due to our small sample size), they are all consistent with each other in terms of their dependency on the relaxation state of their group which gives us added confidence that there is a true dependency. We note that, although our sample size is somewhat limited (64 galaxies in total from 154 galaxies), it is still the largest sample that could be collected from currently available surveys. The only way to fully confirm our results with greater significance would involve gathering a significantly larger sample.
	
	We suggest that it takes time for galaxies to become regular rotators and/or their gas and stars PA to align and the time scale for realignment is similar to the time it takes for a group to become dynamically relaxed. Recently, it has been shown that the relaxation time-scale strongly depends on the mass assembly history of the host halo \citep{Haghighi2020}. On the other hand, \cite{Raouf2018} showed that there is a 2 Gyr difference in the time-scale of mass assembly history (on average) of the BGG in relaxed and unrelaxed groups. Furthermore, they showed that the time since the last major merger is statistically lower (typically in the last gigayear) for BGGs in unrelaxed groups compared to BGG in relaxed groups. Simulation studies of the formation of massive early-type galaxies suggest that the relaxation time-scale \citep{Lake1983} for dynamical settling of the gaseous component into the stellar discs (gas-star realignment) after the merger is around 2 Gyrs depending on the extension of gas accretion \citep{vandeVoort2015}.    
		On this foundation, we claim that the group dynamical state leaves a traceable impact on the internal kinematics of both the stars and ionised gas in BGGs. 
	
	The emerging picture from this study is consistent with a number of previous studies that have focused on other aspects of the halo impact on the BGGs. 
	For example, previous observations found that relaxed groups tend to have experienced gas-rich galaxy mergers  \citep{Khosroshahi2006,Brough2006a,Brough2006b} in their evolutionary history compared to the BGGs in unrelaxed systems, but with equal stellar mass. 	
	In our recent studies, we showed that the luminosity gap parameter plays an important role in differences between the BGG stellar population properties such as metallicity, star formation rate, dust, and $NUV-r$ color\citep{Raouf2019b}. We found the BGG in relaxed groups typically have an earlier peak in their merger rate, $\sim$2 Gyr, and had not suffered a recent major merger \citep[$<$1 Gyr][]{Raouf2018}. 
	Generally speaking, the BGGs in relaxed groups have a lack of recent major mergers and the more frequent misalignment of gas rotation with respect to their stars in the BGGs of unrelaxed groups could be the results of new externally accreted gas through their more recent last (major) mergers. As mentioned earlier, galaxies in dense environments show less regularity than those in less dense environments \citep{Krajnovic2011,Lee2018}. Indeed, our result pushes the existing results to a new level by moving beyond studying the role of the local density on stellar kinematic to also considering the role of the group dynamics.
	
	This paper suggests that the accreted gas, following a galaxy merger, would settle (i.e the gas relaxation) in a way that its rotation axis is aligned with that of the stellar component (co-rotating or counter-rotating) when the time scale is equal or shorter than the time since the BGG last major merger, otherwise the gas rotation axis would deviate from that of the stars. 
	This is more likely if the major merger had occurred in an earlier epoch during the evolution of the brightest group galaxies, as there will be more time available for the gas to settle in and around the stellar component.  Therefore, it conceivable to find more counter-rotating (very low statistical significance in this work, around 10\%) and aligned BGGs for the groups with an earlier last major merger (e.g BGGs in relaxed systems in this study).
	In the formation of an unrelaxed group, the time between the mergers is short enough that the gas is still misaligned from the last merger, while the group is still classified as unrelaxed due to its dense core. In such an environment continuous accretion of gas as a result of more frequent mergers compare to the relaxed groups, leaves no time for gas to align with the stellar component.  

	Note that the fraction of new accreted gas and to the pre-existing gas is also expected to be important for the overall gas relaxation time scale. An existing gas disk would relax much more quickly after a merger unless it was very massive and thus the gas richness might have an impact on the gas settling time scale \citep[e.g][for early-type galaxies]{Lagos2015}.
	We also find unrelaxed groups with aligned, and relaxed groups with misaligned BGGs. One possibility is that some of our low Sersic index objects might have contained gas prior to the merger, which could be consistent with the fact that we clearly see a higher aligned fraction for those objects. Another possibility is the fact that our optically measurement of the group dynamical state is not expected to be 100\% correct on a one-to-one basis. 

	Our future study will focus on the source of external accretion, cooling the halo, and stripping or distortion of gas due to group processes by considering the location of galaxies in the cosmic web. 
	
	\section*{Acknowledgments} \label{sec:acknow}
	The SAMI Galaxy Survey is based on observations made at the Anglo-Australian Telescope. The SydneyAAO Multi-object Integral field spectrograph (SAMI) was developed jointly by the University of Sydney and the Australian Astronomical Observatory.
	For the masses, redshift and the rest:  GAMA is a joint European-Australasian project based around a spectroscopic campaign using the Anglo-Australian Telescope. The GAMA input catalog is based on data taken from the Sloan Digital Sky Survey and the UKIRT Infrared Deep Sky Survey. Complementary imaging of the GAMA regions is being obtained by several independent survey programs including GALEX MIS, VST KiDS, VISTA VIKING, WISE, Herschel-ATLAS, GMRT, and ASKAP providing UV to radio coverage. GAMA is funded by the STFC (UK), the ARC (Australia), the AAO, and the participating institutions. The GAMA website is http:\\www.gama-survey.org/. We would like to acknowledge financial support from ICRAR, AAO, ARC, STFC, RS, and ERS for GAMA Panchromatic Swarp Imager (PSI). LC is the recipient of an Australian Research Council Future Fellowship (FT180100066) funded by the Australian Government. Parts of this research were conducted by the Australian Research Council Centre of Excellence for All-Sky Astrophysics in 3 Dimensions (ASTRO 3D), through project number CE170100013. 
	JBH is supported by an ARC Laureate Fellowship that funds Jesse van de Sande and an ARC Federation Fellowship that funded the SAMI prototype. JJB acknowledges the support of an Australian Research Council Future Fellowship (FT180100231). Jvds acknowledges support of an Australian Research Council Discovery Early Career Research Award (project number DE200100461) funded by the Australian Government, and
	this research made use of Astropy, http://www.astropy.org a community-developed core Python package for Astronomy \citep{astropy:2013, astropy:2018}.
	
	\appendix
	\section{Smoothing map for gas PA} \label{Sec:Apendix1}
\begin{figure*}[h]
		\centering
		\caption{(left) The stellar kinematics (SK), and (centre) H-alpha gas kinematics (GK) velocity map. There are two galaxies per row. Labels indicate the PA with error and signal to noise (S/N) within one effective radius. The circle indicates the 15" size of the SAMI hexabundle. (right) The smoothed  map of the GK including labels with the GAMA ID, $\delta$PA, sSFR and the group magnitude gap and BGG offset. The dashed circle indicates the effective radius (Re). The bottom-right corner shows an inset image of a postage stamp in $r$-band from the GAMA Panchromatic Swarp Imager (PSI) with 20" side length (note that inset image is on a very different scale as the SAMI images). Red and blue dashed lines show the major axes of rotation for stellar and gas kinematics, respectively.}
		\includegraphics[width=0.45\linewidth]{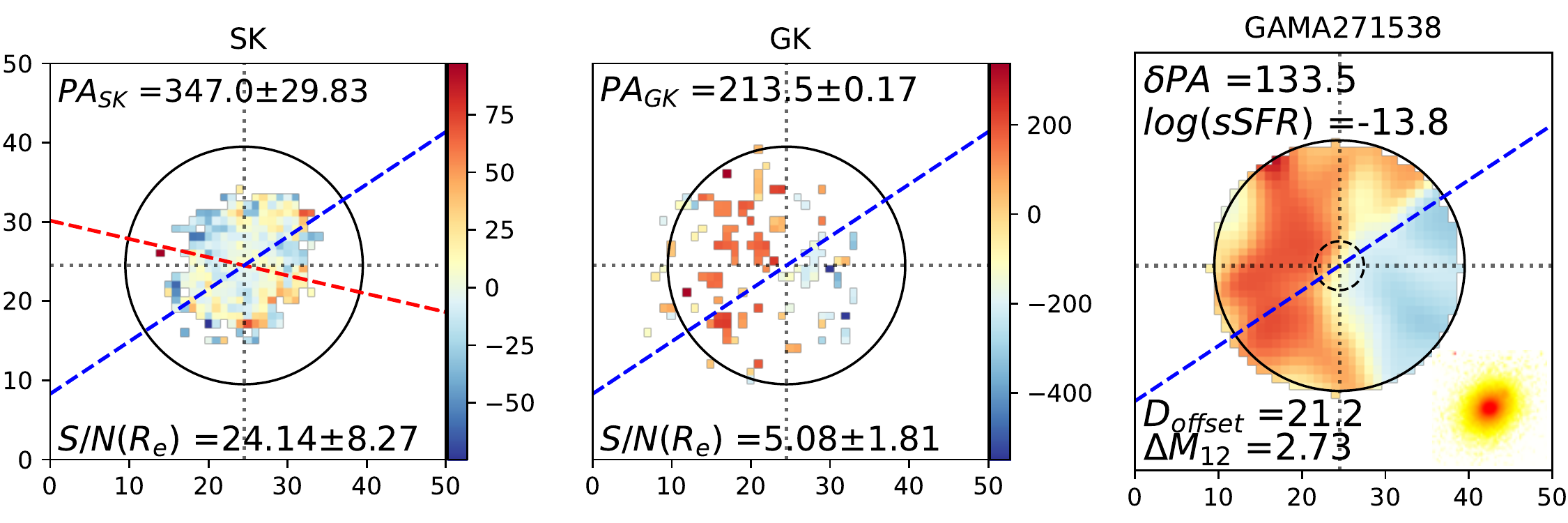} 
		\includegraphics[width=0.45\linewidth]{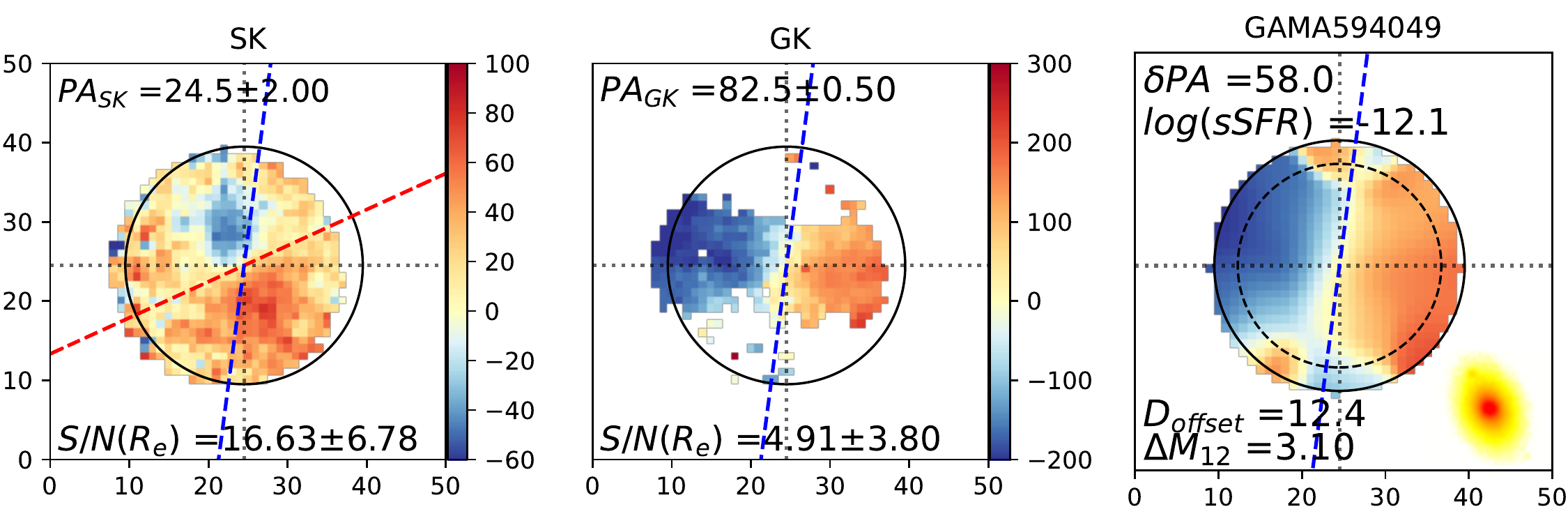} 
		\includegraphics[width=0.45\linewidth]{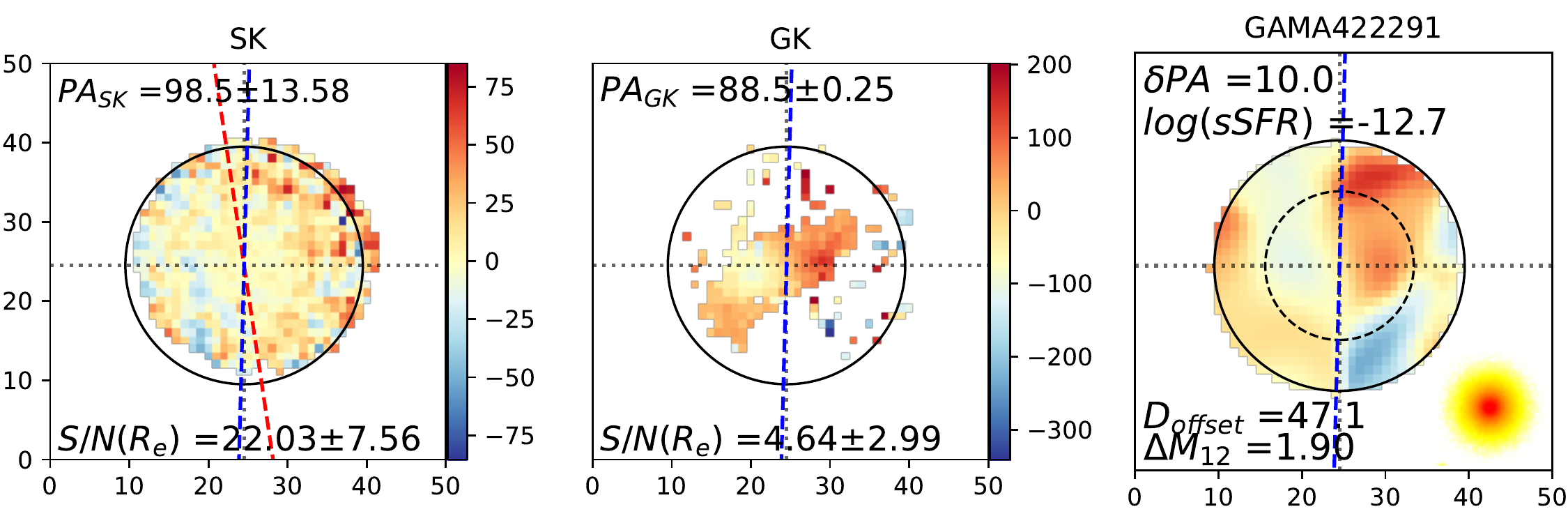} 
		\includegraphics[width=0.45\linewidth]{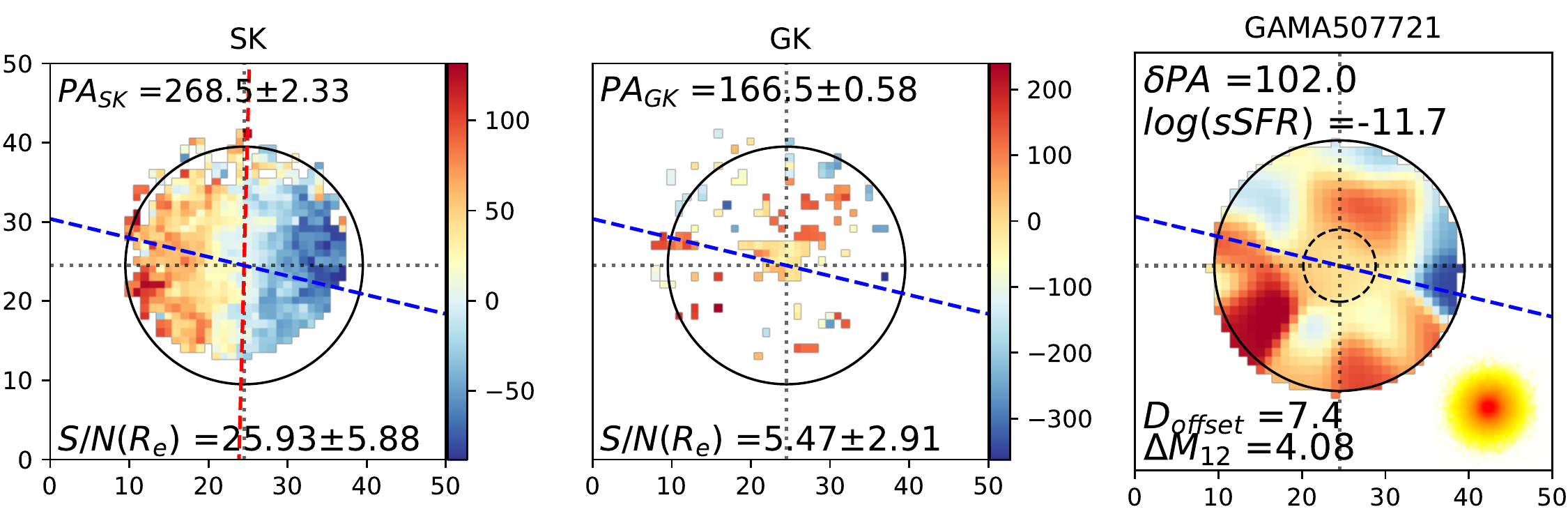} 
		\includegraphics[width=0.45\linewidth]{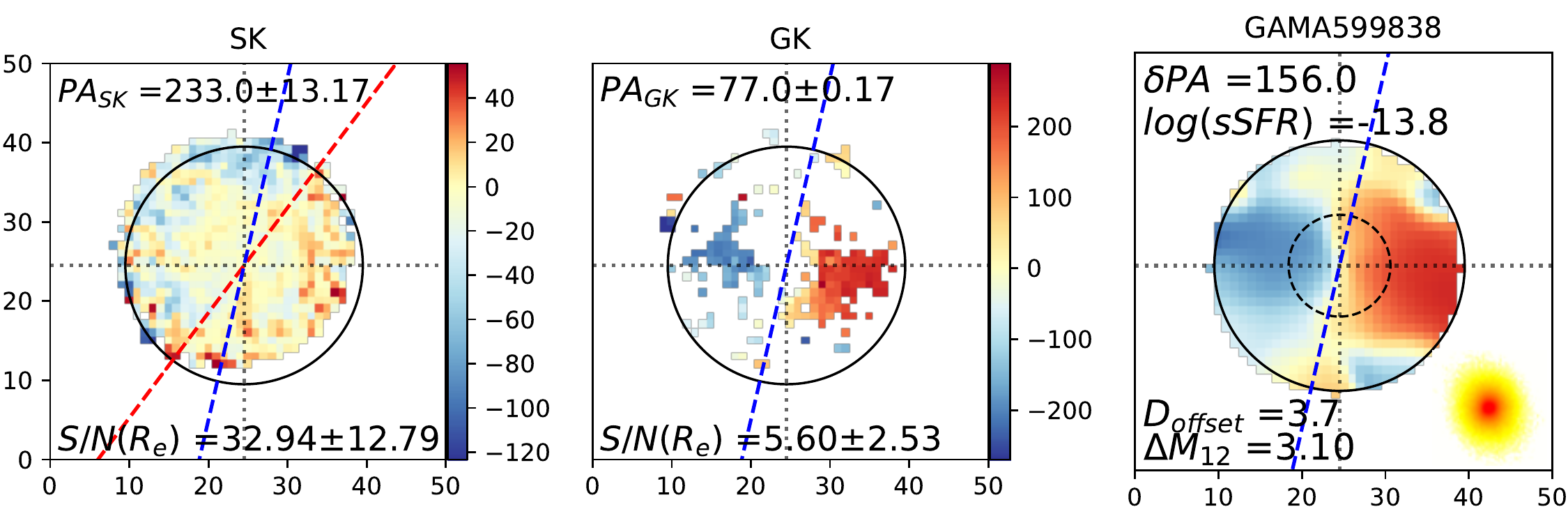} 
		\includegraphics[width=0.45\linewidth]{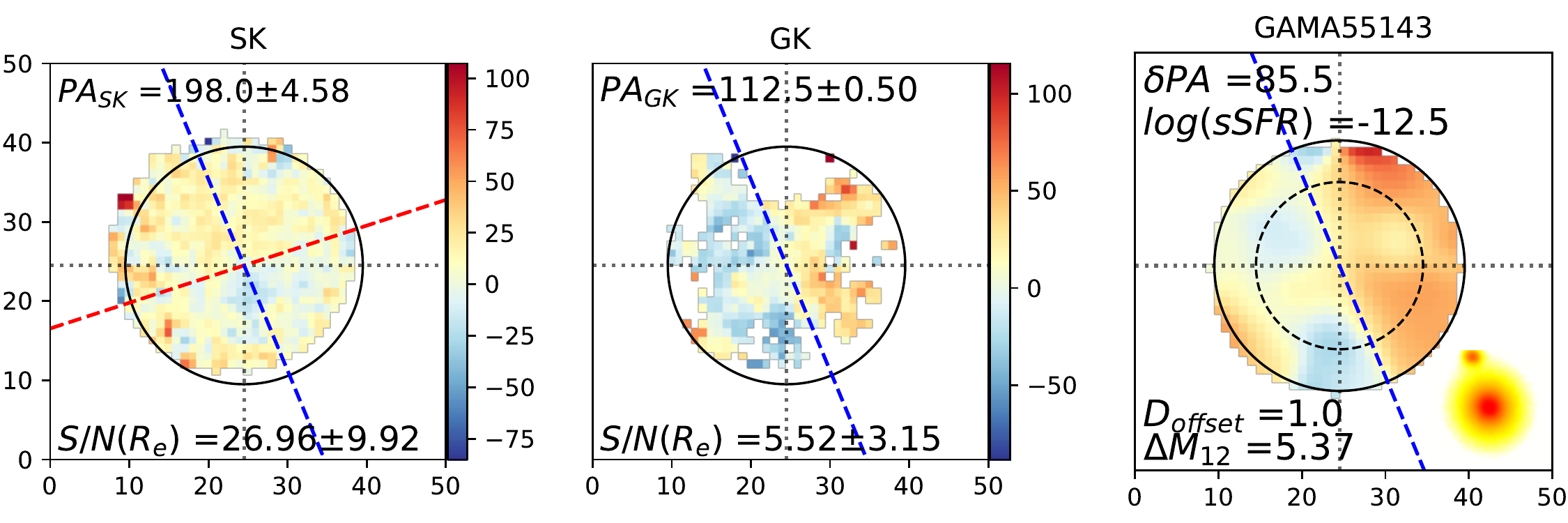}
		\includegraphics[width=0.45\linewidth]{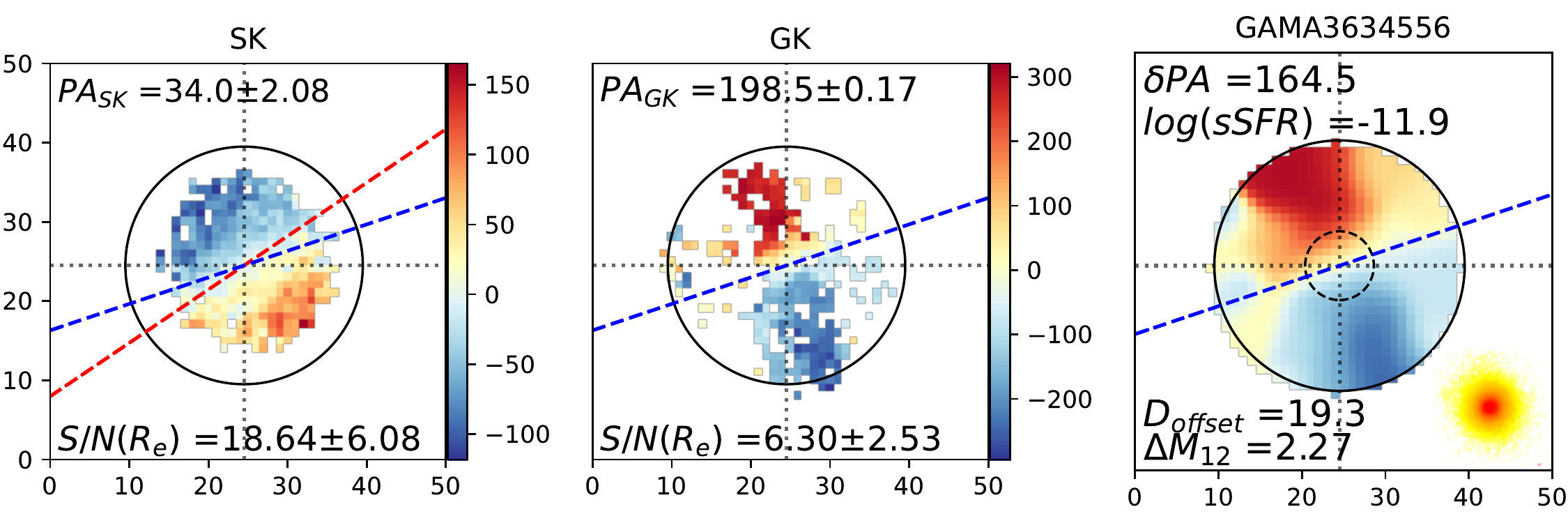}
		\includegraphics[width=0.45\linewidth]{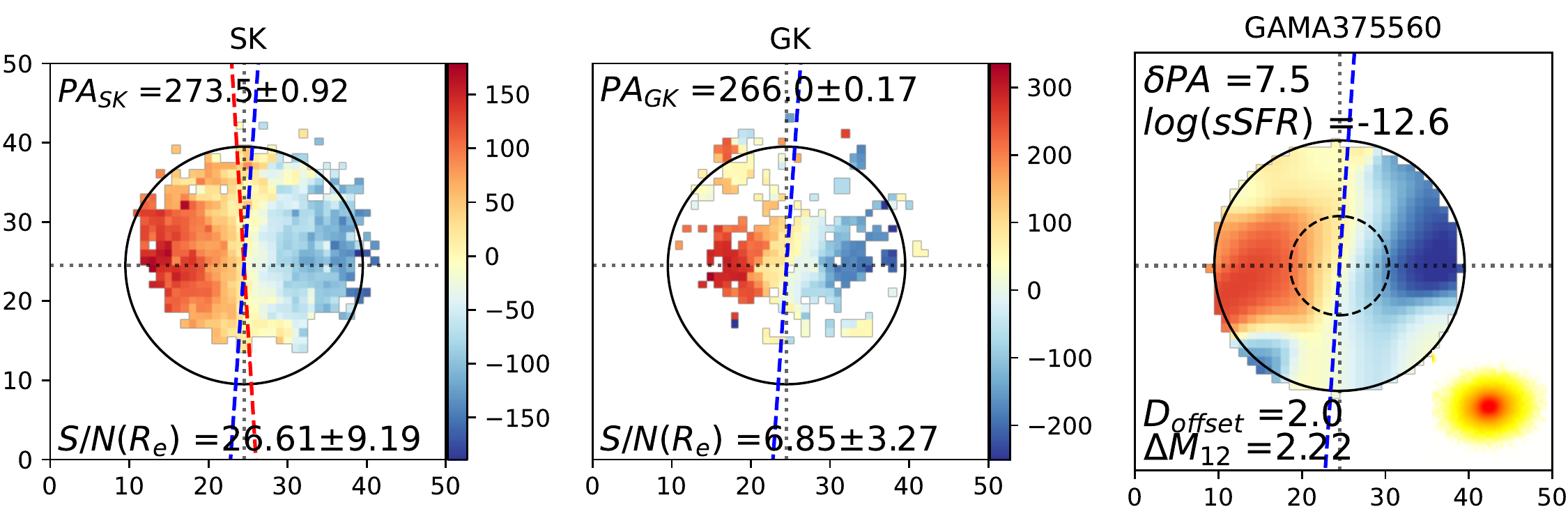} 
		\includegraphics[width=0.45\linewidth]{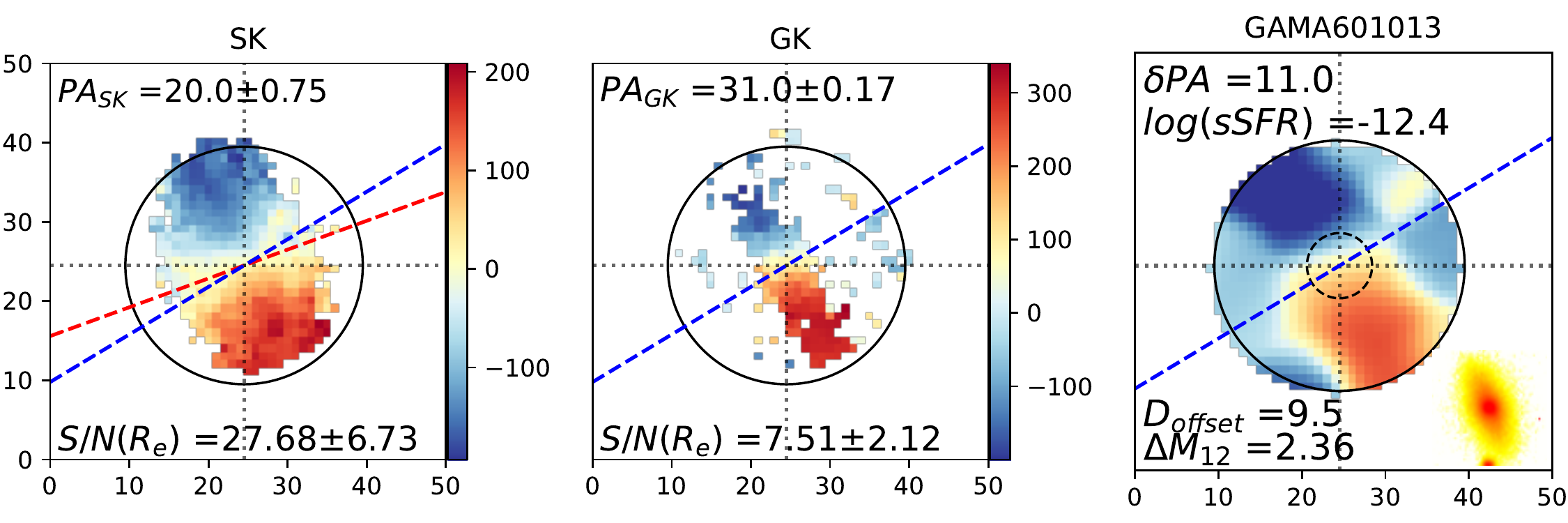} 
		\includegraphics[width=0.45\linewidth]{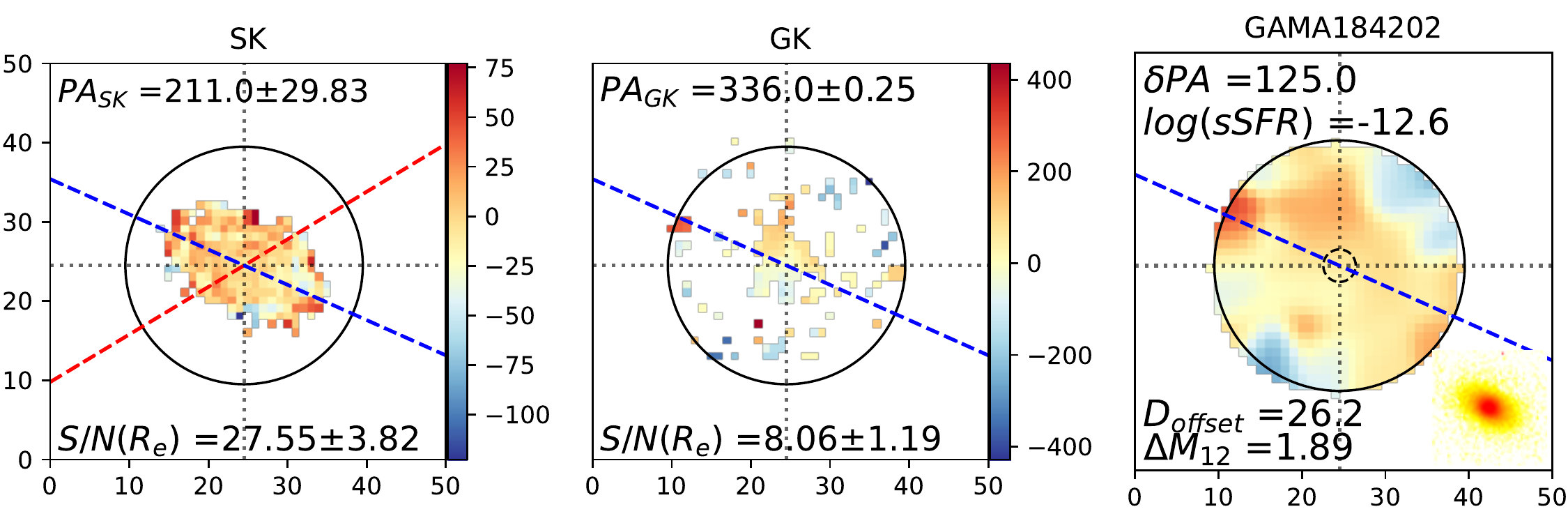} 
		\includegraphics[width=0.45\linewidth]{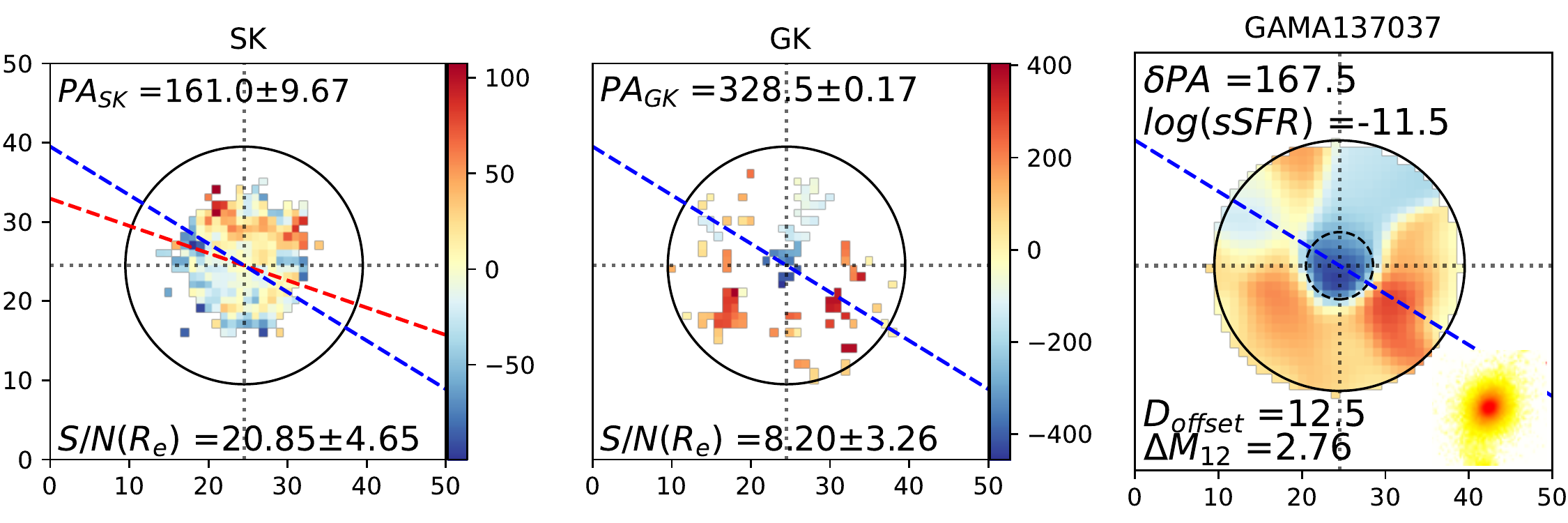} 
		\includegraphics[width=0.45\linewidth]{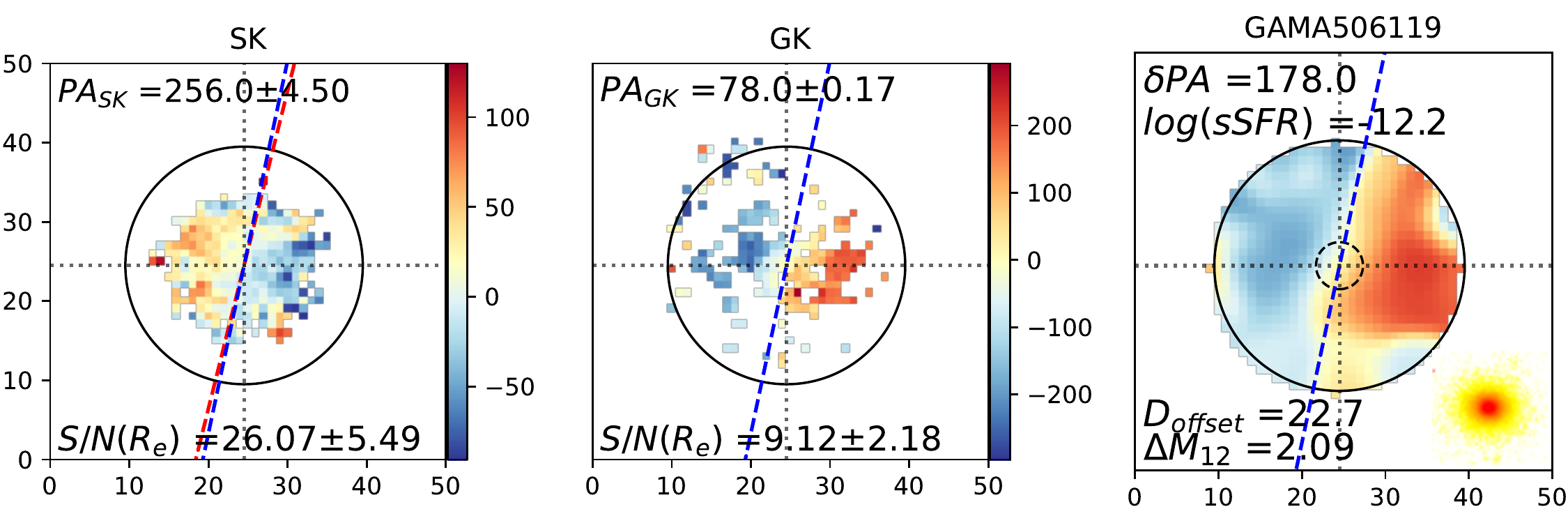} 
		\includegraphics[width=0.45\linewidth]{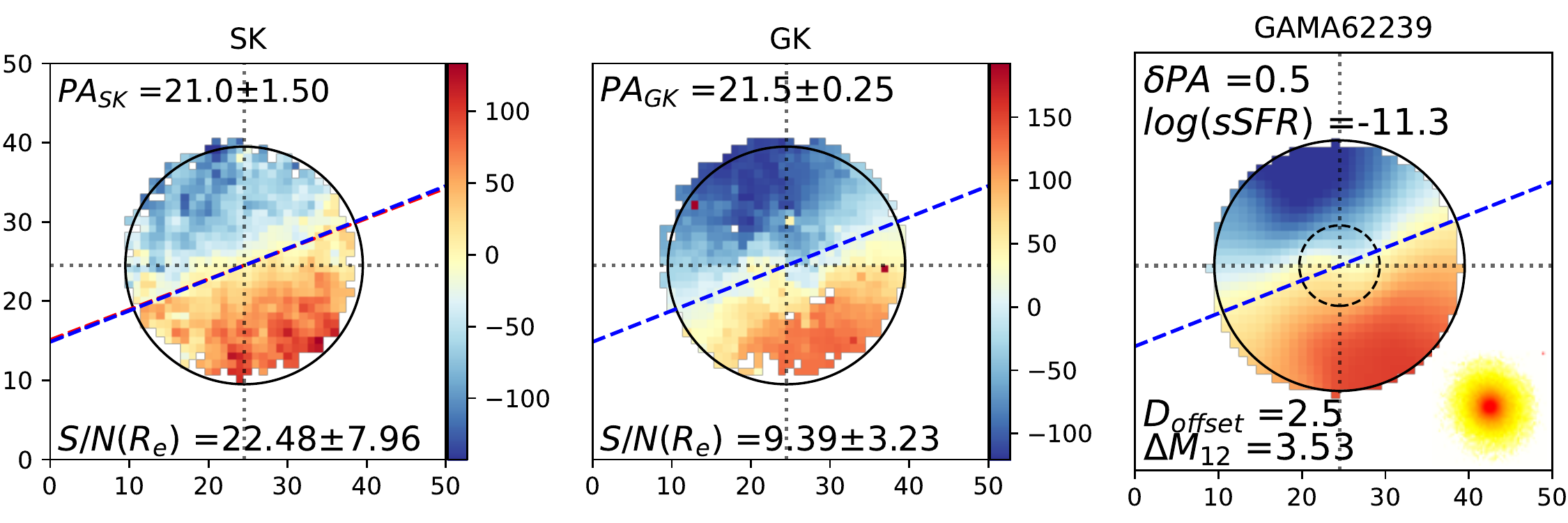}
		\includegraphics[width=0.45\linewidth]{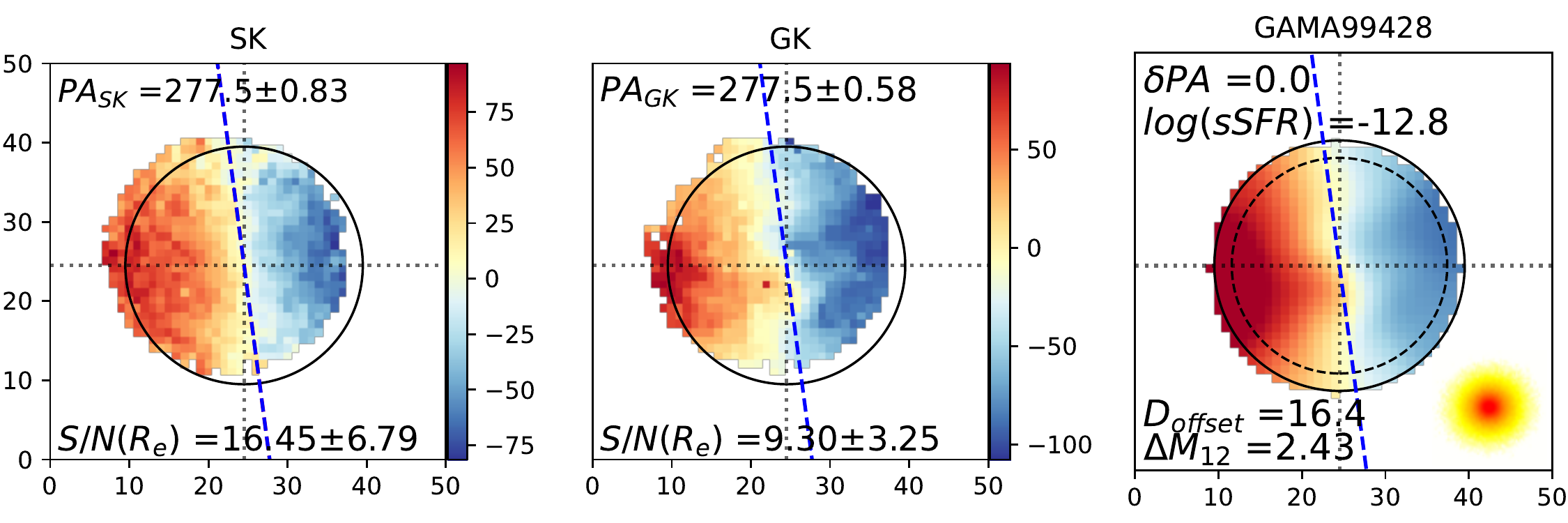}
		\includegraphics[width=0.45\linewidth]{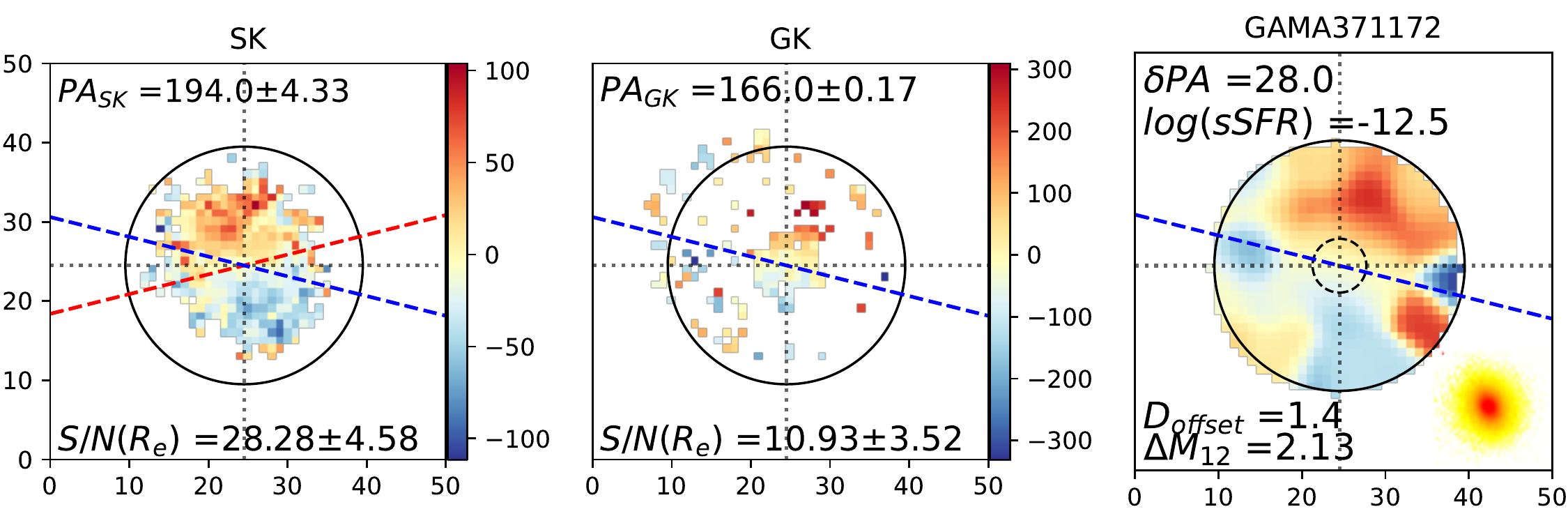} 
		\includegraphics[width=0.45\linewidth]{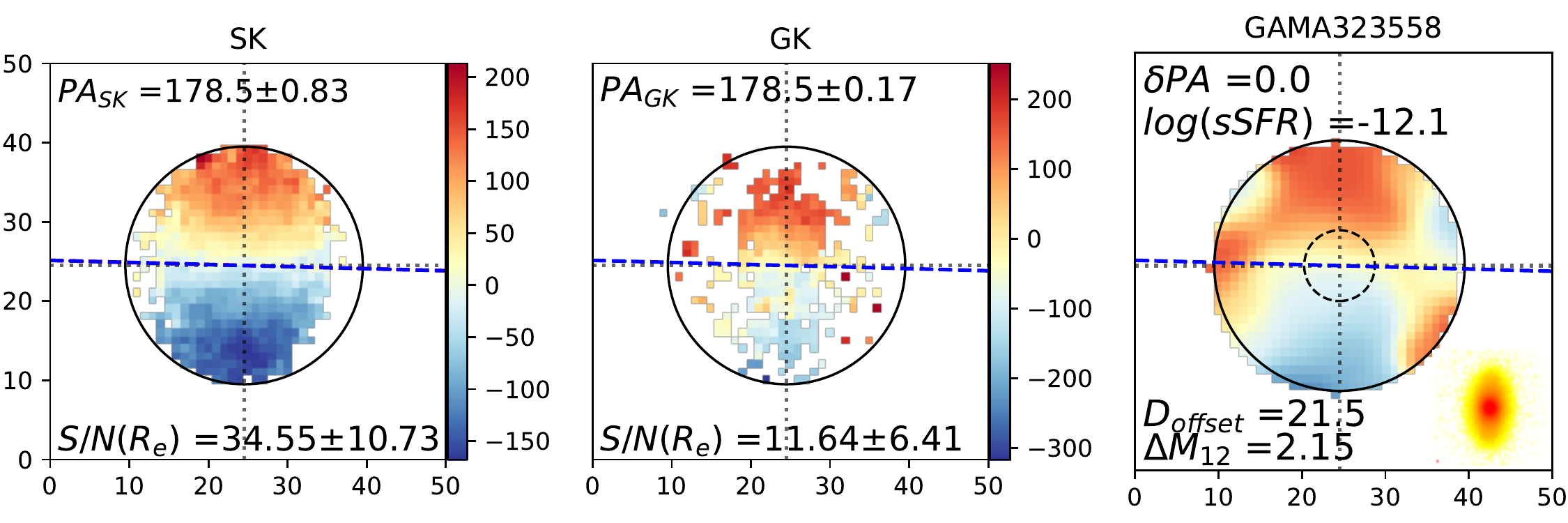} 
		\label{fig:sami-relax}
	\end{figure*}

\begin{figure*}
		\centering
		\caption{Continue relaxed group}
		\includegraphics[width=0.45\linewidth]{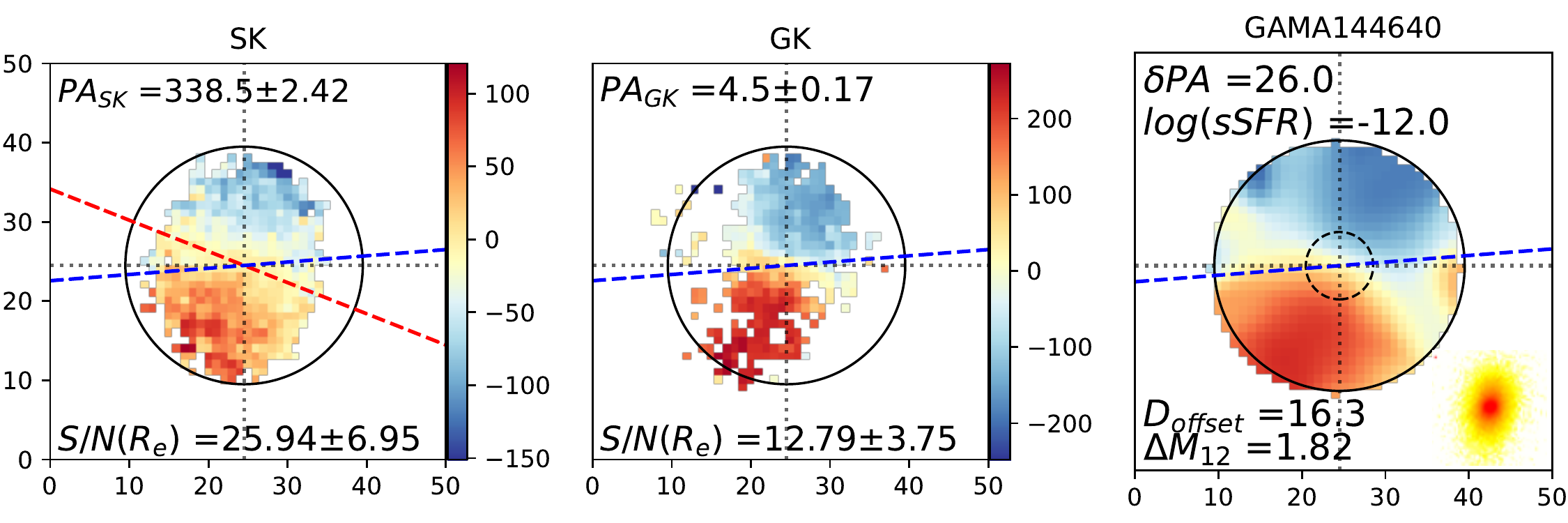} 
		\includegraphics[width=0.45\linewidth]{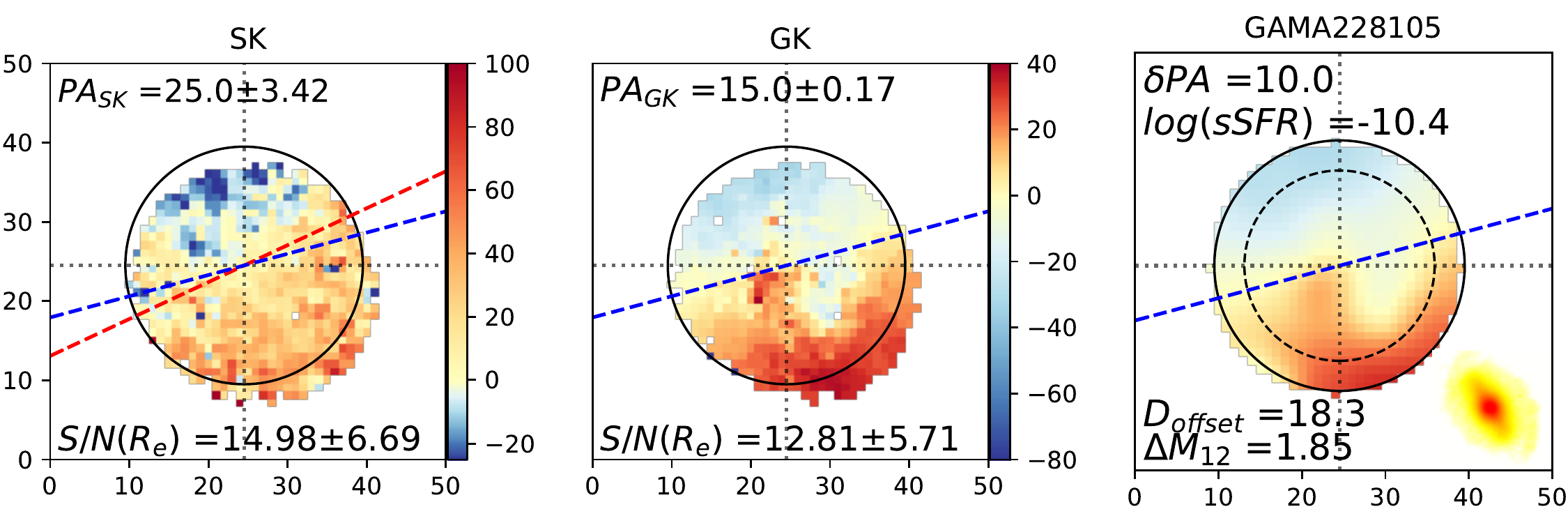} 
		\includegraphics[width=0.45\linewidth]{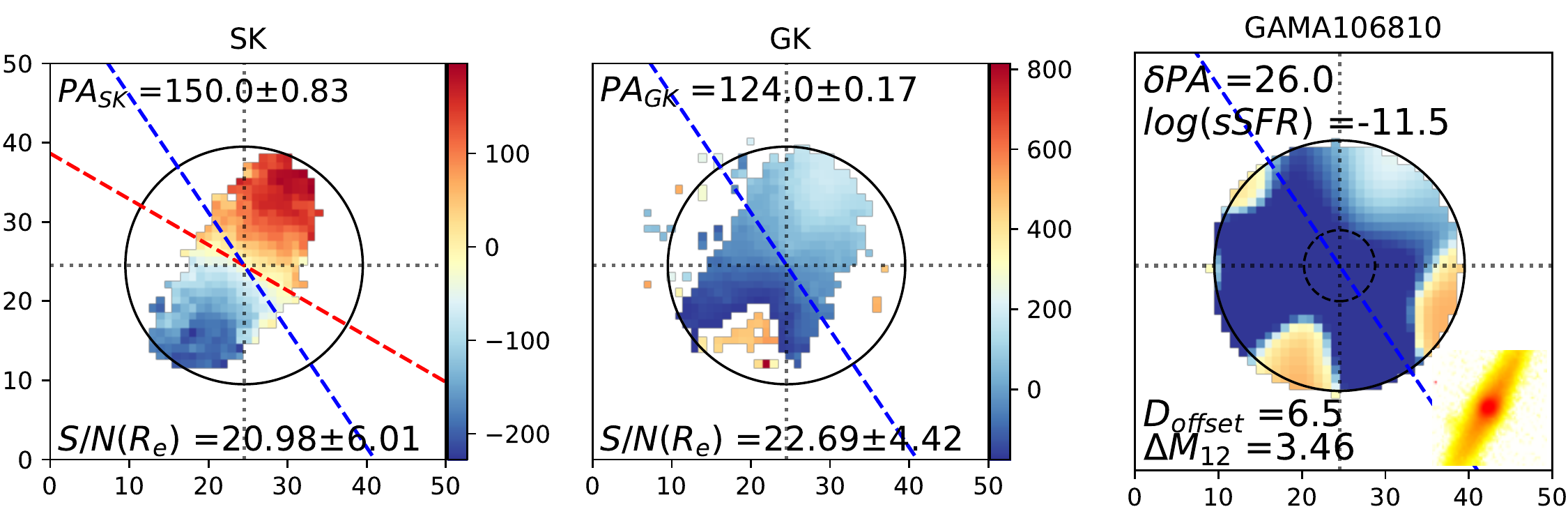} 
		\includegraphics[width=0.45\linewidth]{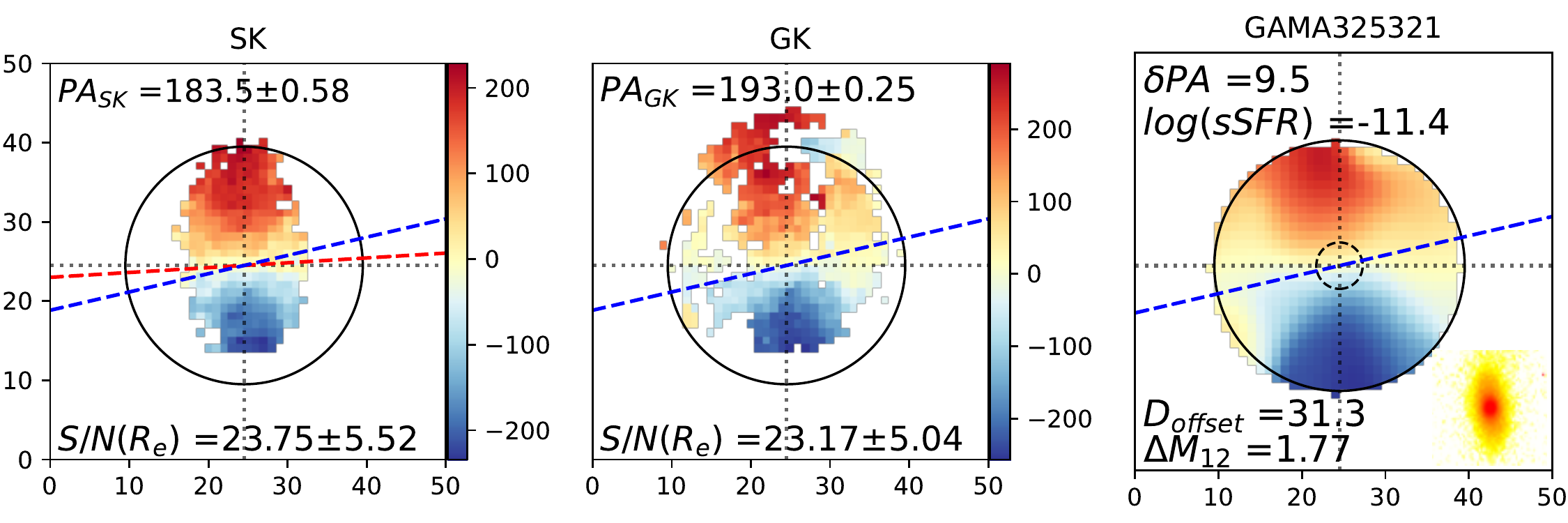} 
		\includegraphics[width=0.45\linewidth]{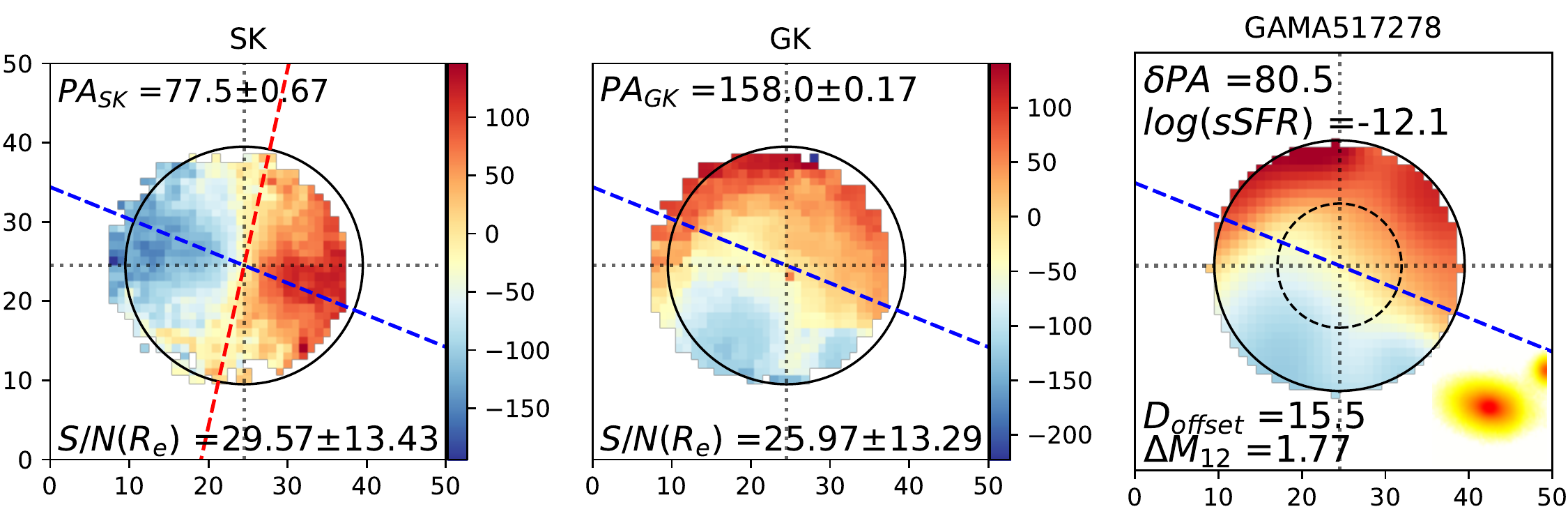} 
		\includegraphics[width=0.45\linewidth]{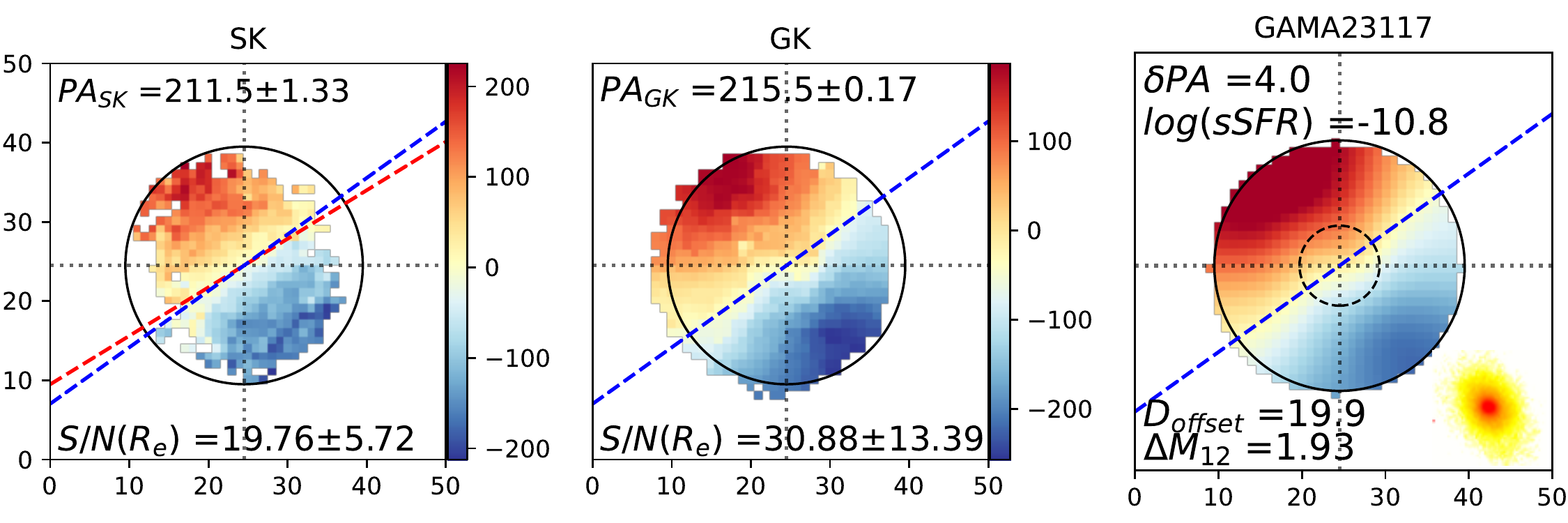}
		\includegraphics[width=0.45\linewidth]{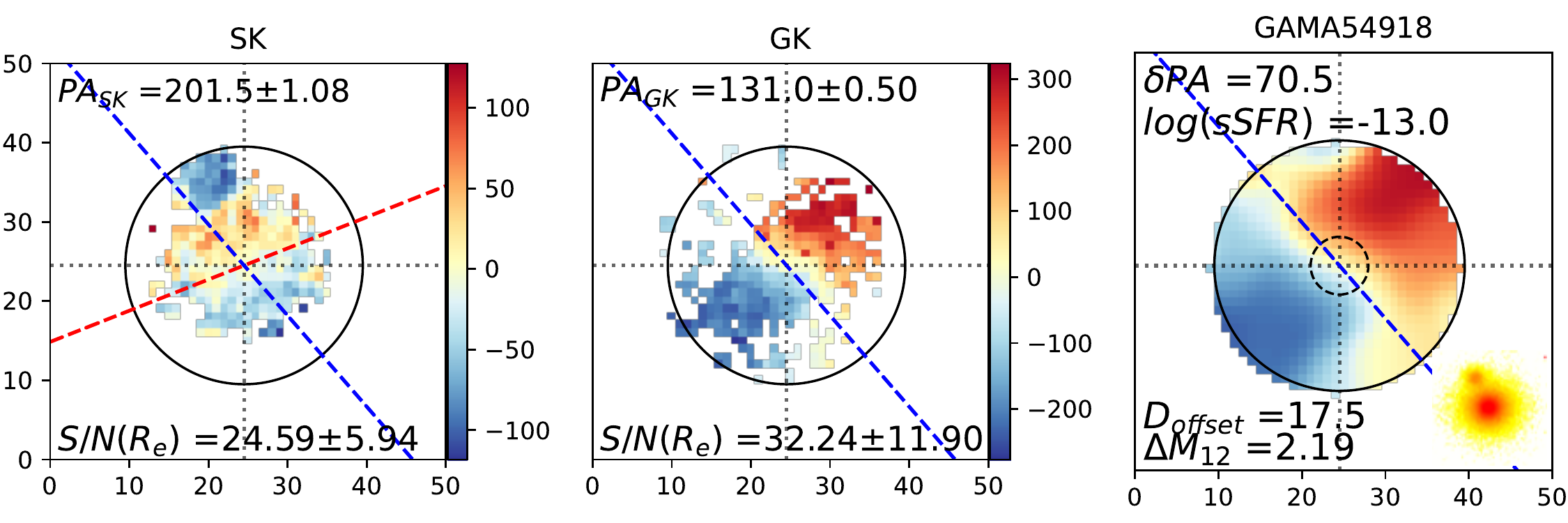}
		\includegraphics[width=0.45\linewidth]{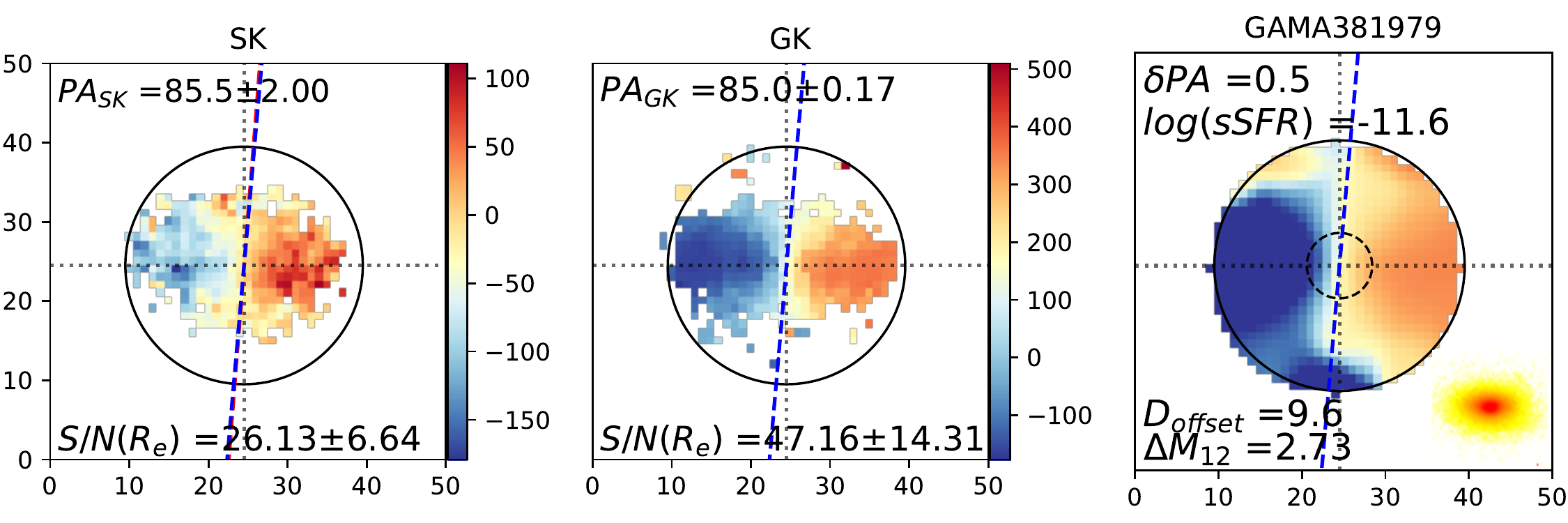} 
		\includegraphics[width=0.45\linewidth]{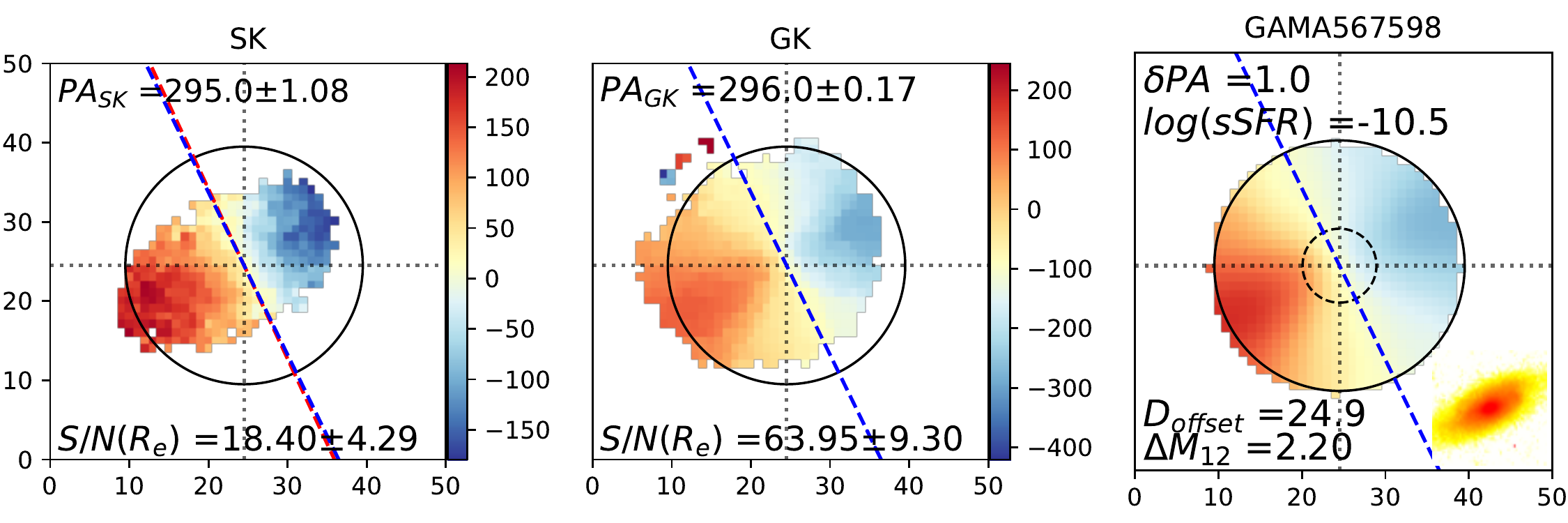} 
		\label{fig:sami-relax2}
	\end{figure*}

\begin{figure*}
		\centering
		\caption{Same as figure \ref{fig:sami-relax} for unelaxed groups}
		\includegraphics[width=0.45\linewidth]{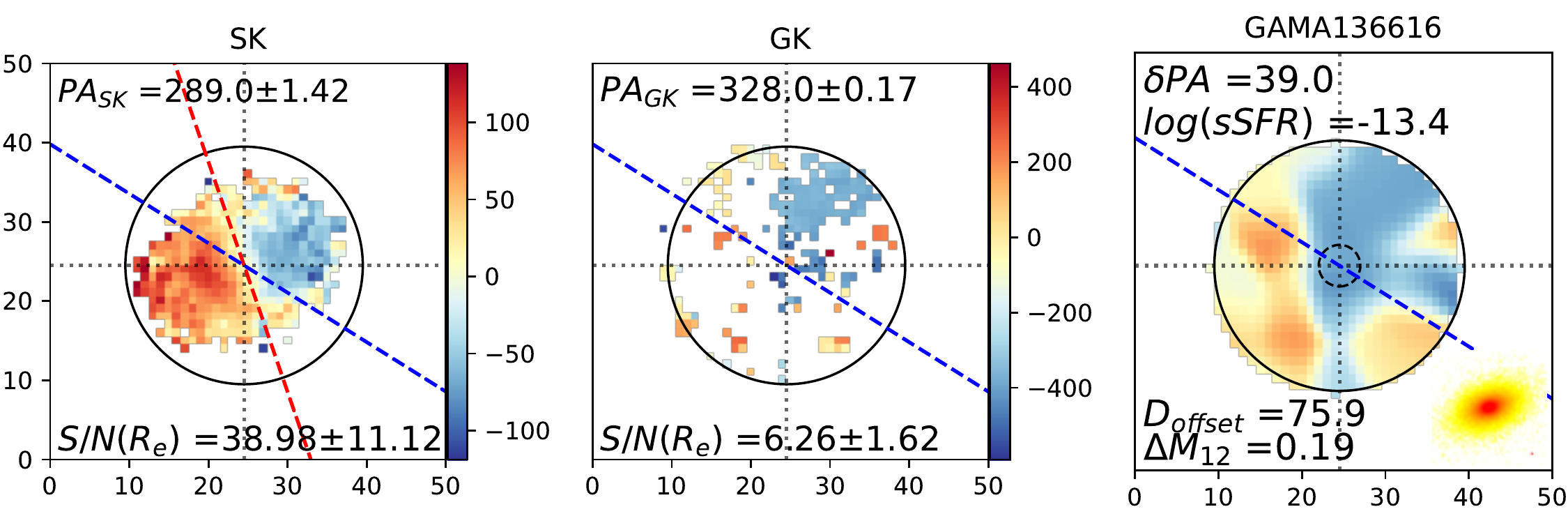}
		\includegraphics[width=0.45\linewidth]{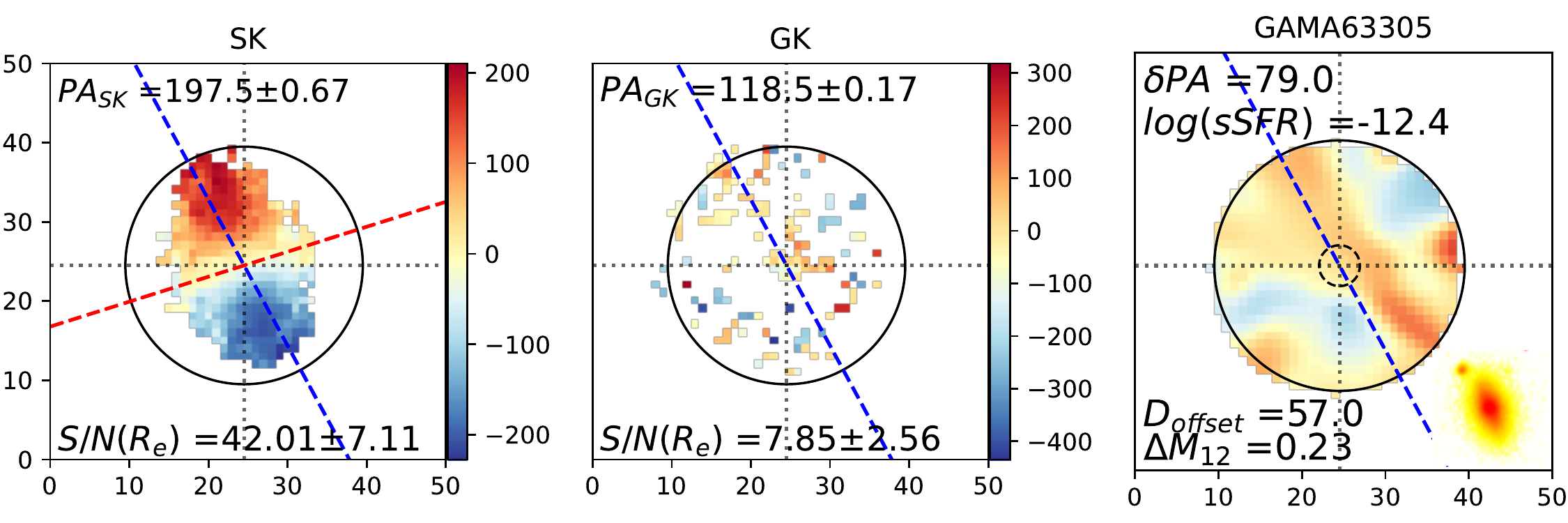} 
		\includegraphics[width=0.45\linewidth]{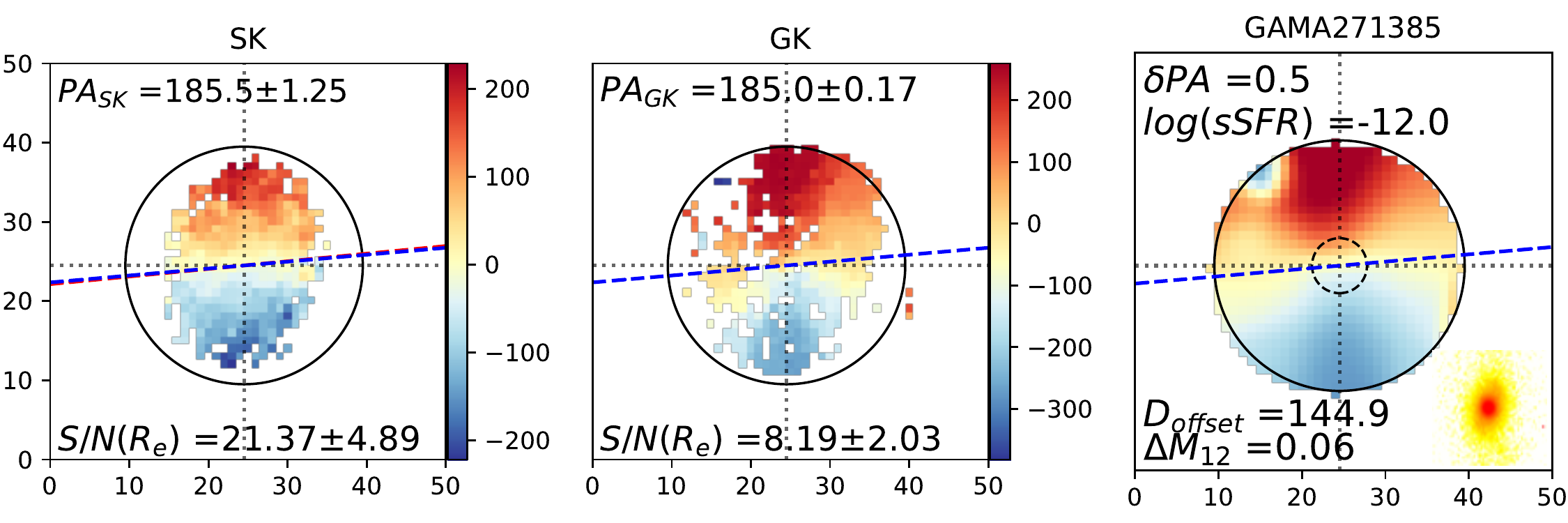}
		\includegraphics[width=0.45\linewidth]{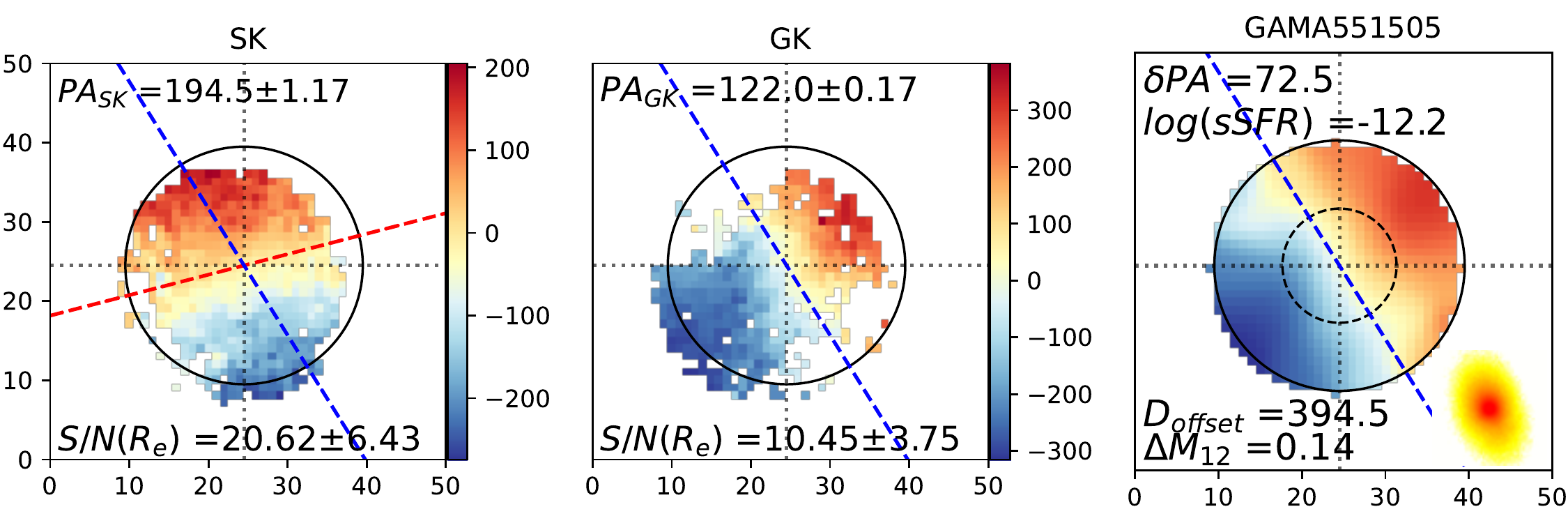}
		\includegraphics[width=0.45\linewidth]{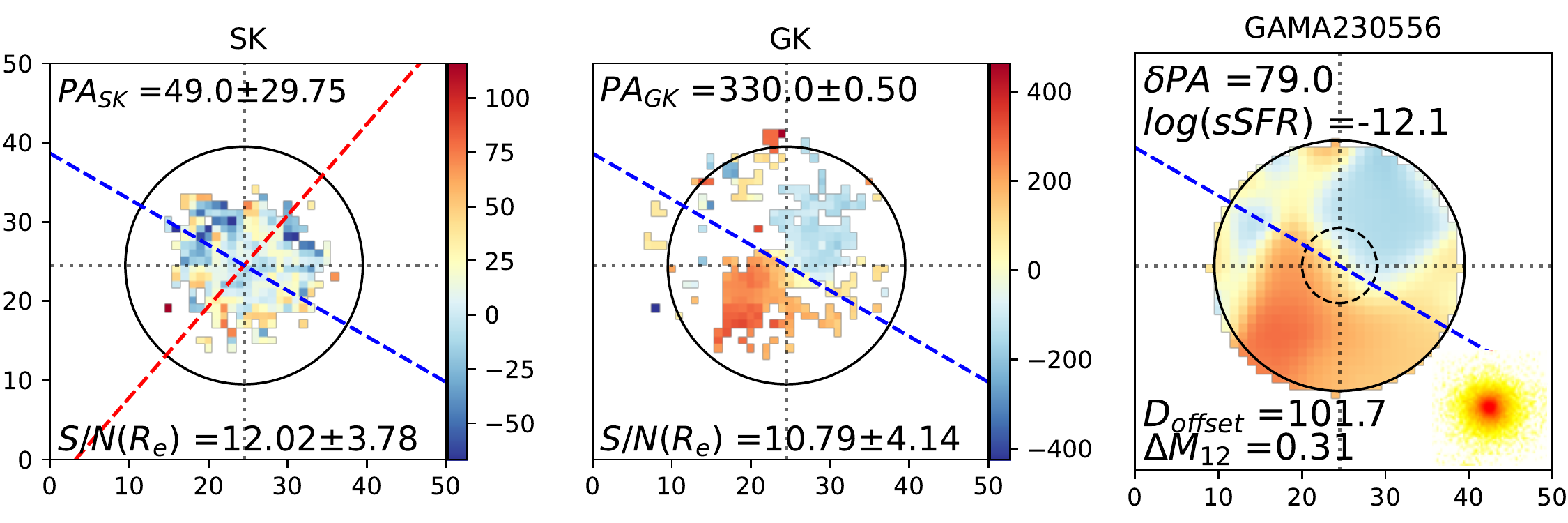}
		\includegraphics[width=0.45\linewidth]{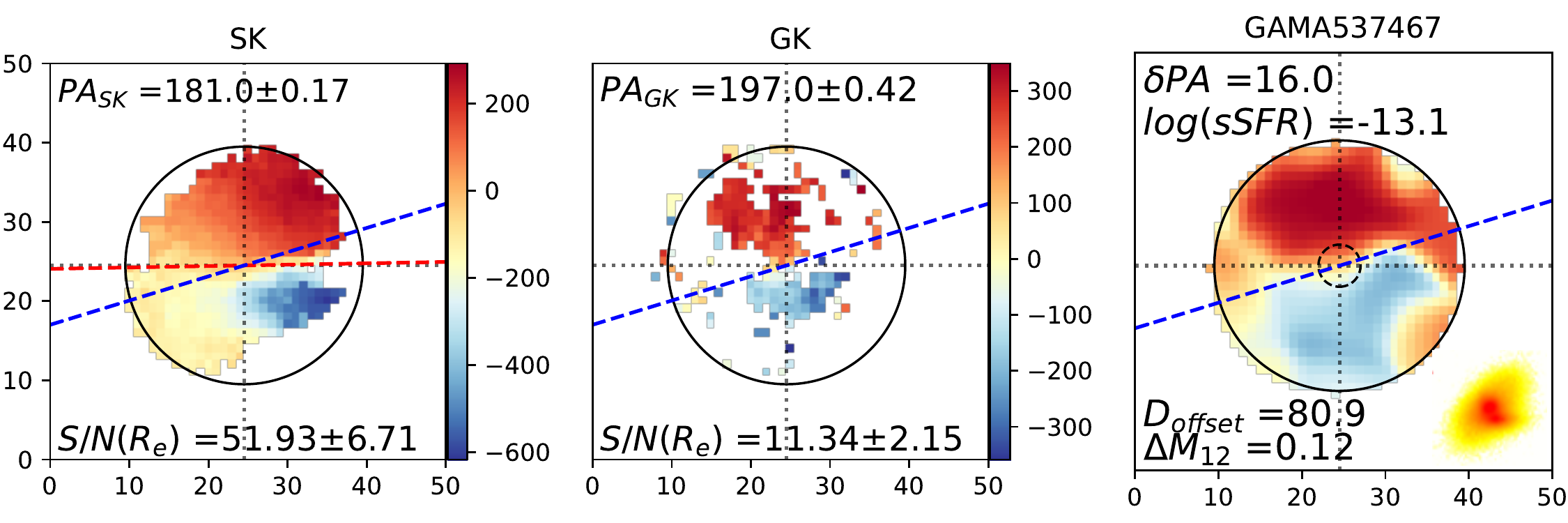}
		
		\label{fig:sami-unrelax}
	\end{figure*}

	To create the smoothed maps of the H-alpha emission maps that are visible in Figure \ref{fig:samiimage}, we use a convolution function with Gaussian kernel and integrated mode. In this way, we try to replace bad spaxels with values interpolated by neighbours (see Figure \ref{fig:sami-relax} - \ref{fig:sami-unrelax}). We note that the smoothed maps are not used for measuring the gas PA. These smoothed maps are only used in combination with the original maps to aid us by eye in evaluating if objects that survive our S/N cut are reliable for measuring the kinematic PA (sec. \ref{Sec:MA}). In each gas velocity map (GK), we only consider the spaxles with H-alpha S/N less than 3 and velocity errors of over 30 $km/s$.
	
	\begin{figure*}
		\centering
		\caption{Continue unelaxed group}
		\includegraphics[width=0.45\linewidth]{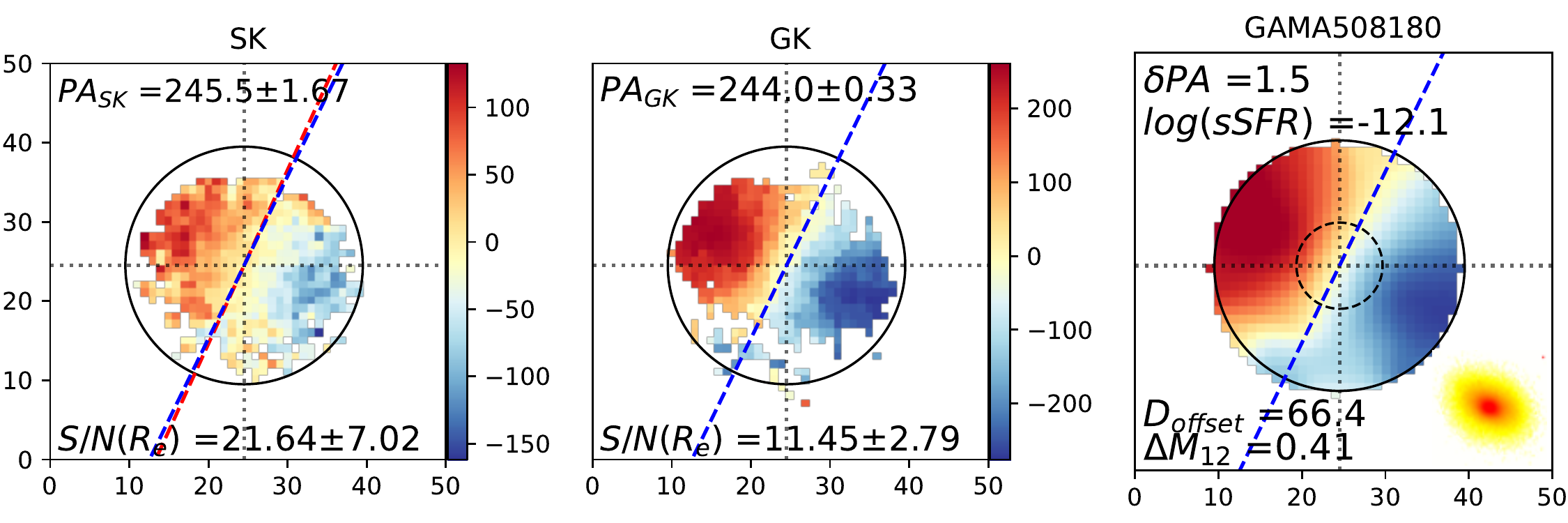}
		\includegraphics[width=0.45\linewidth]{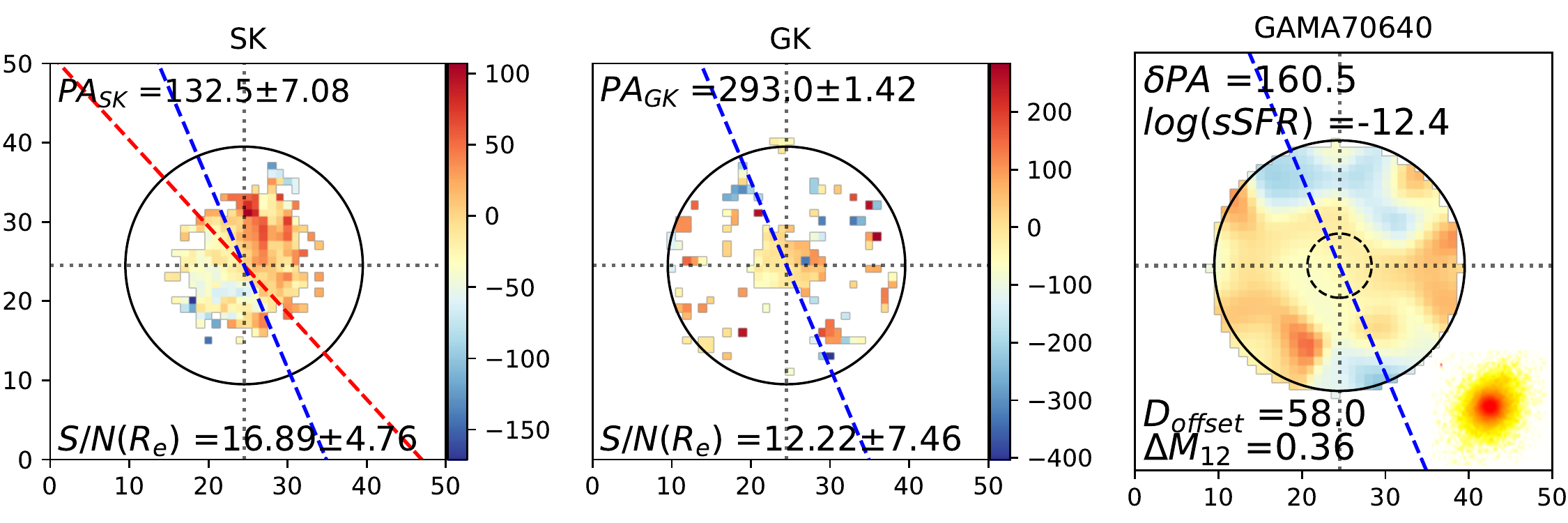} 
		\includegraphics[width=0.45\linewidth]{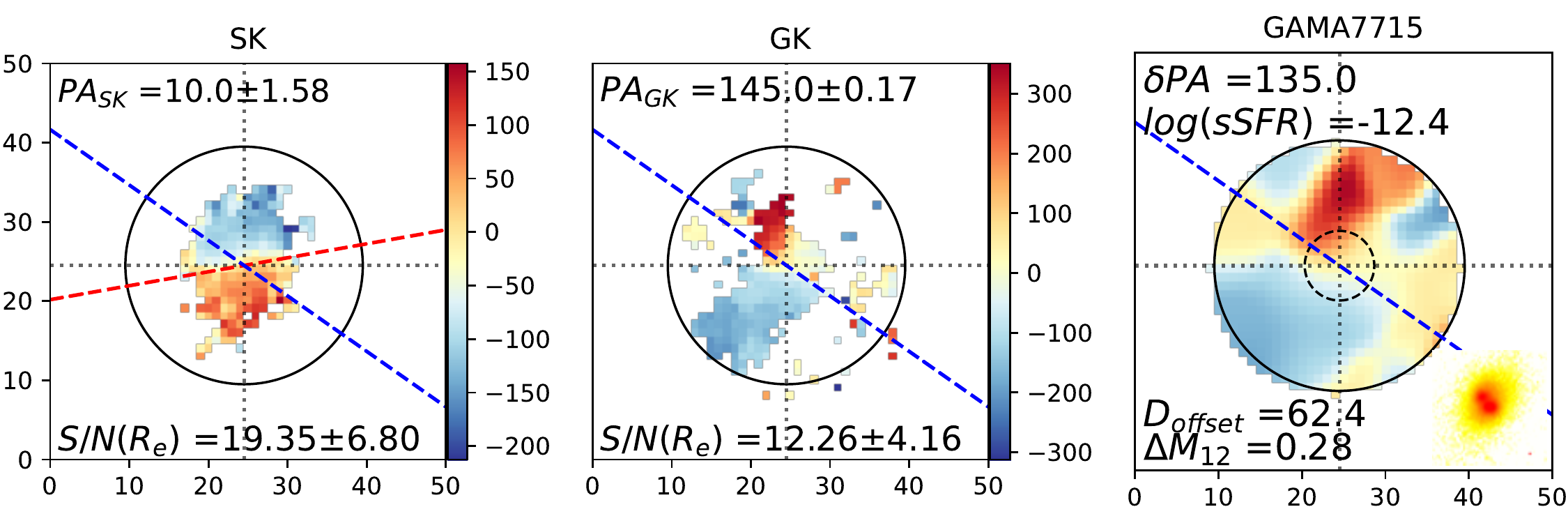}
		\includegraphics[width=0.45\linewidth]{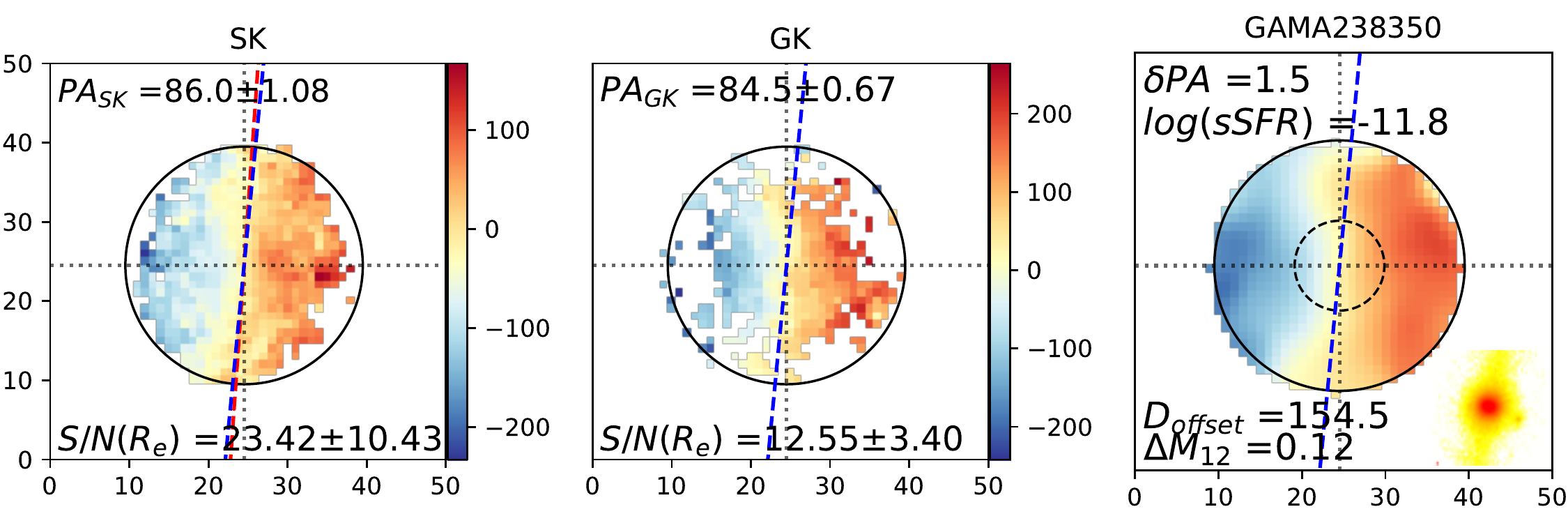}
		\includegraphics[width=0.45\linewidth]{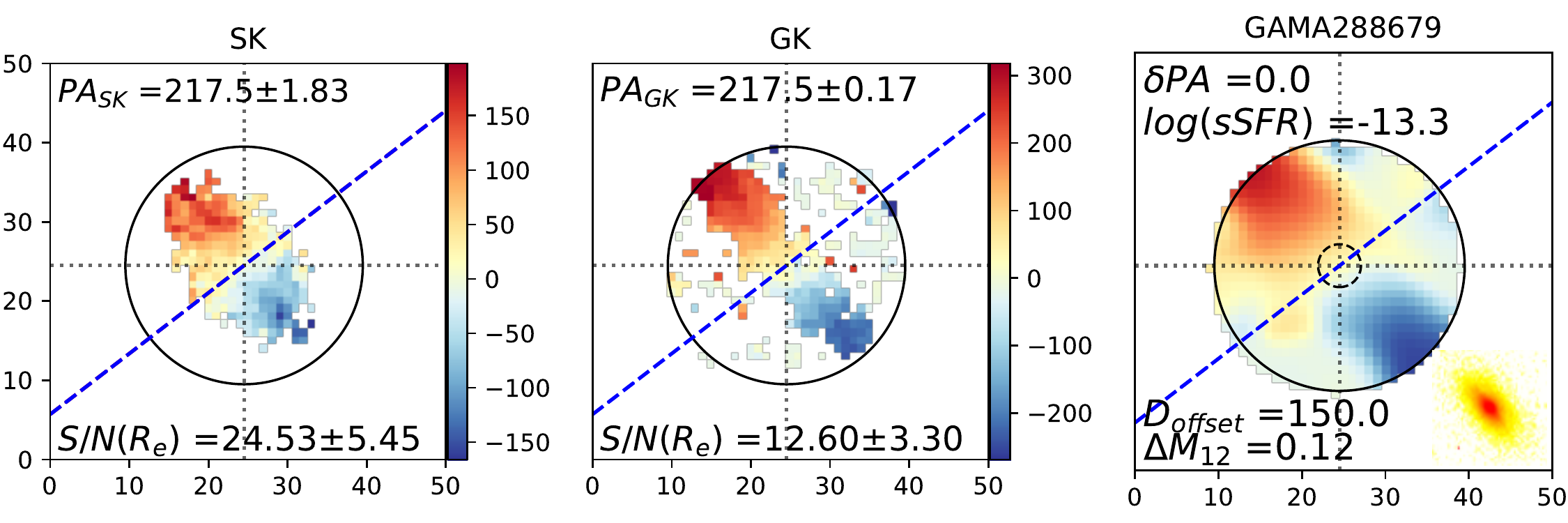}
		\includegraphics[width=0.45\linewidth]{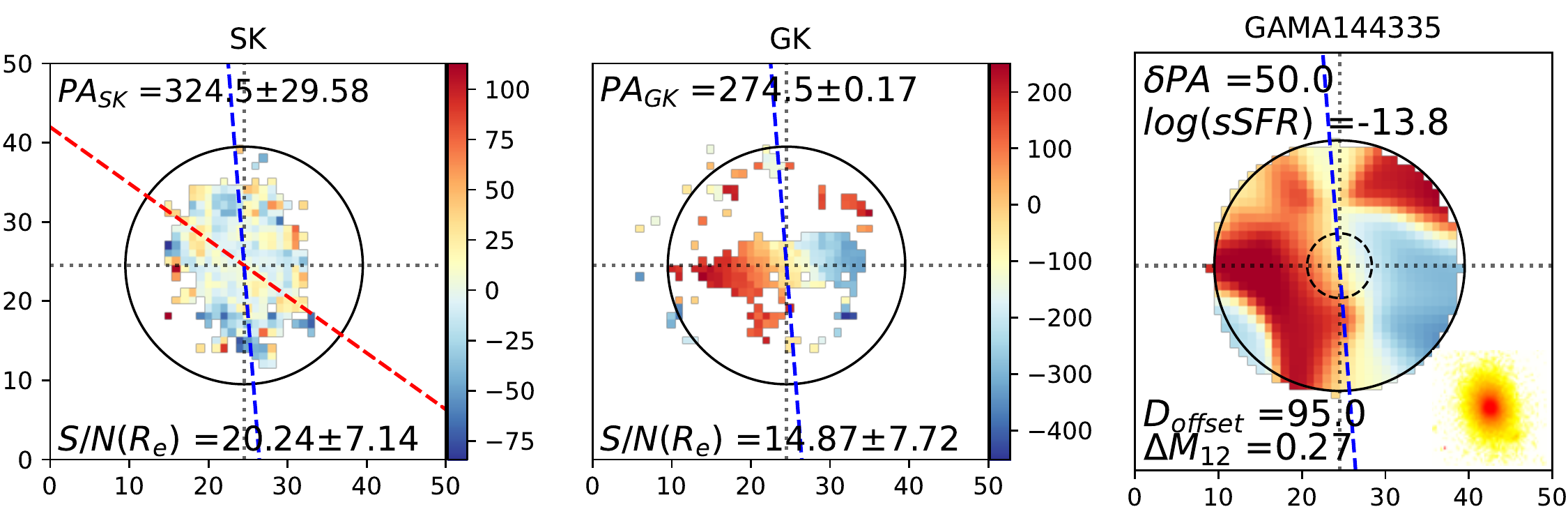}
		\includegraphics[width=0.45\linewidth]{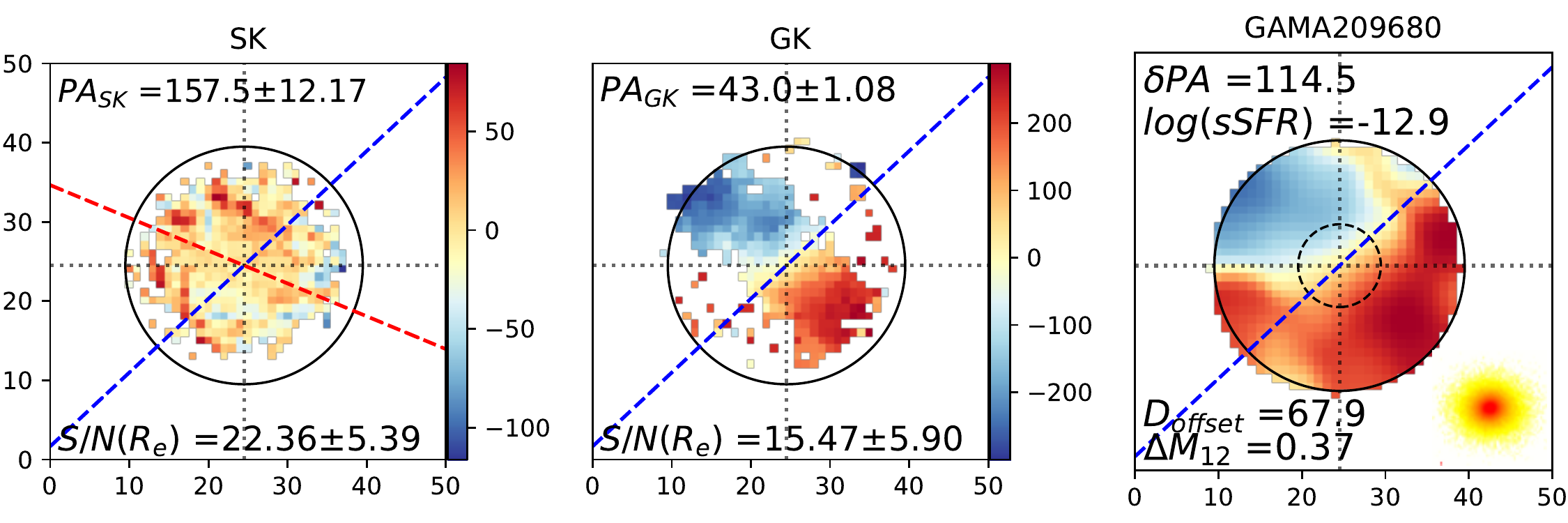}
		\includegraphics[width=0.45\linewidth]{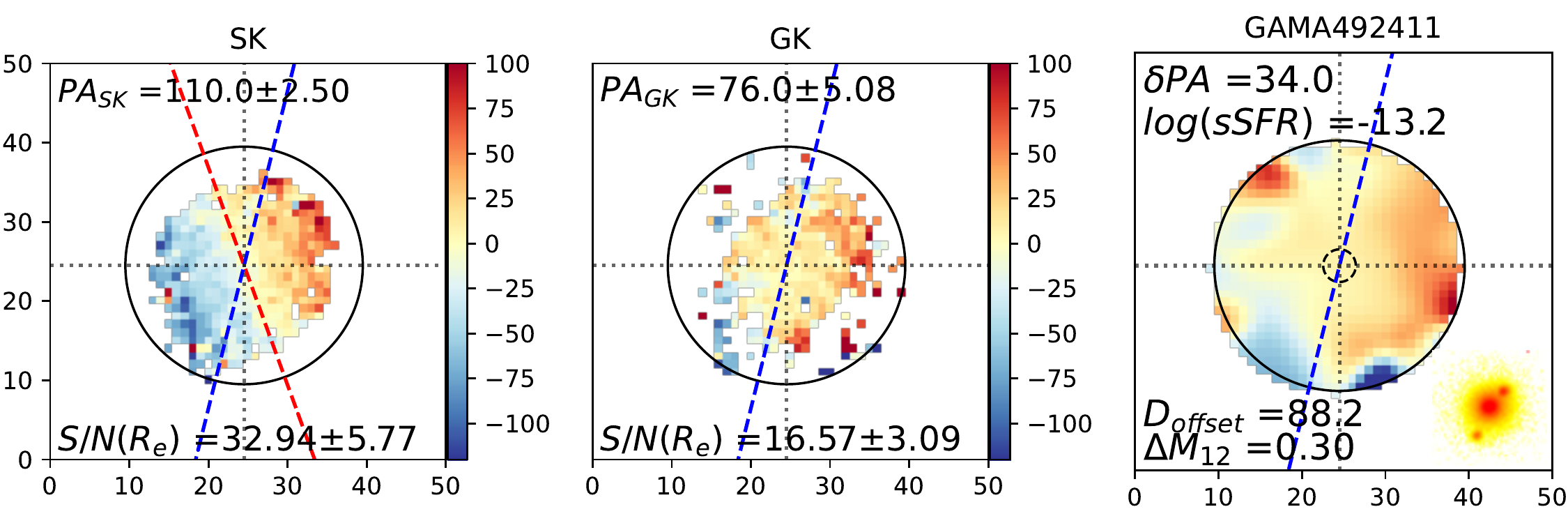}
		\includegraphics[width=0.45\linewidth]{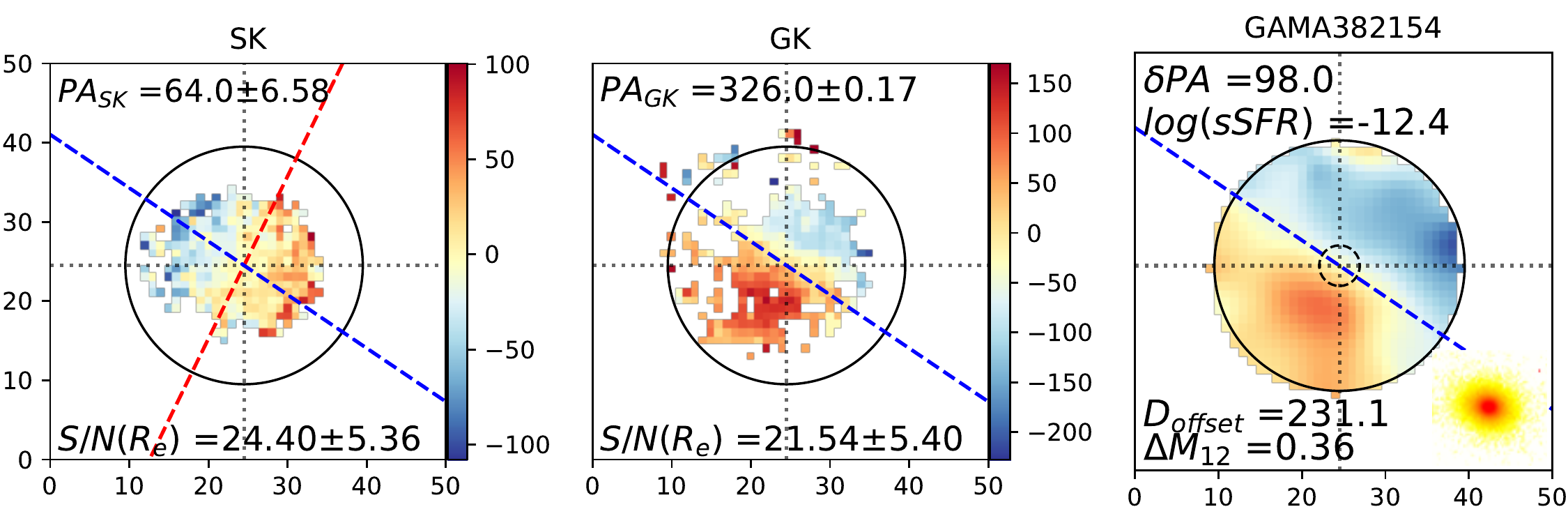}
		\includegraphics[width=0.45\linewidth]{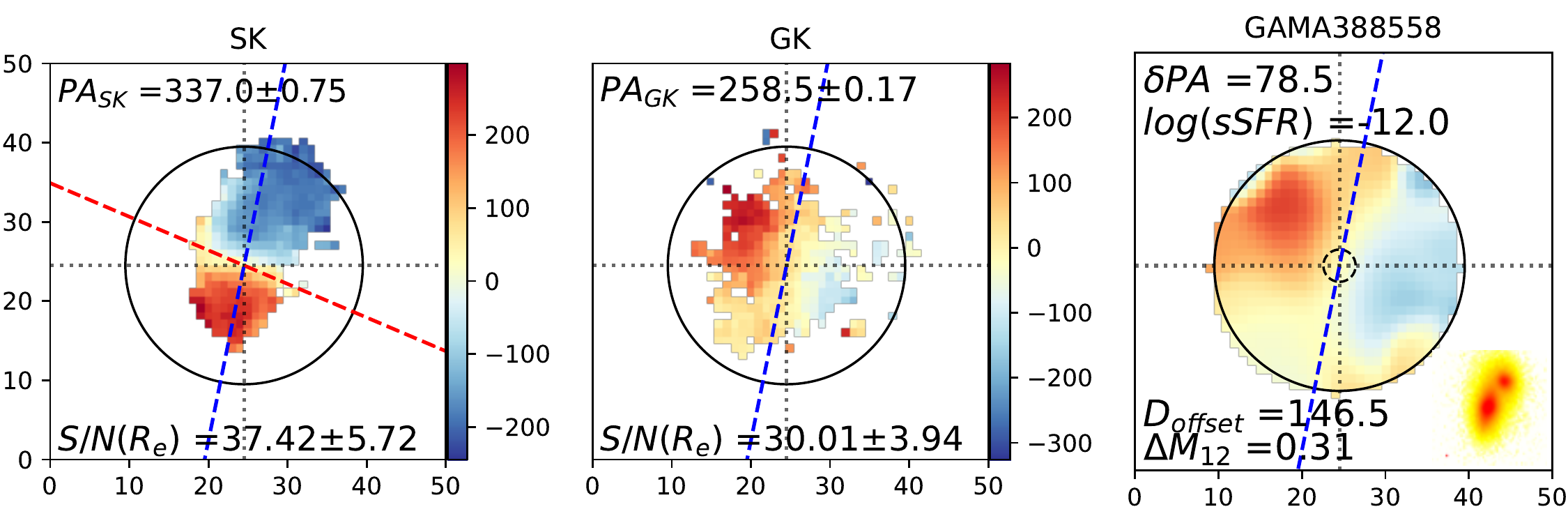}
		\includegraphics[width=0.45\linewidth]{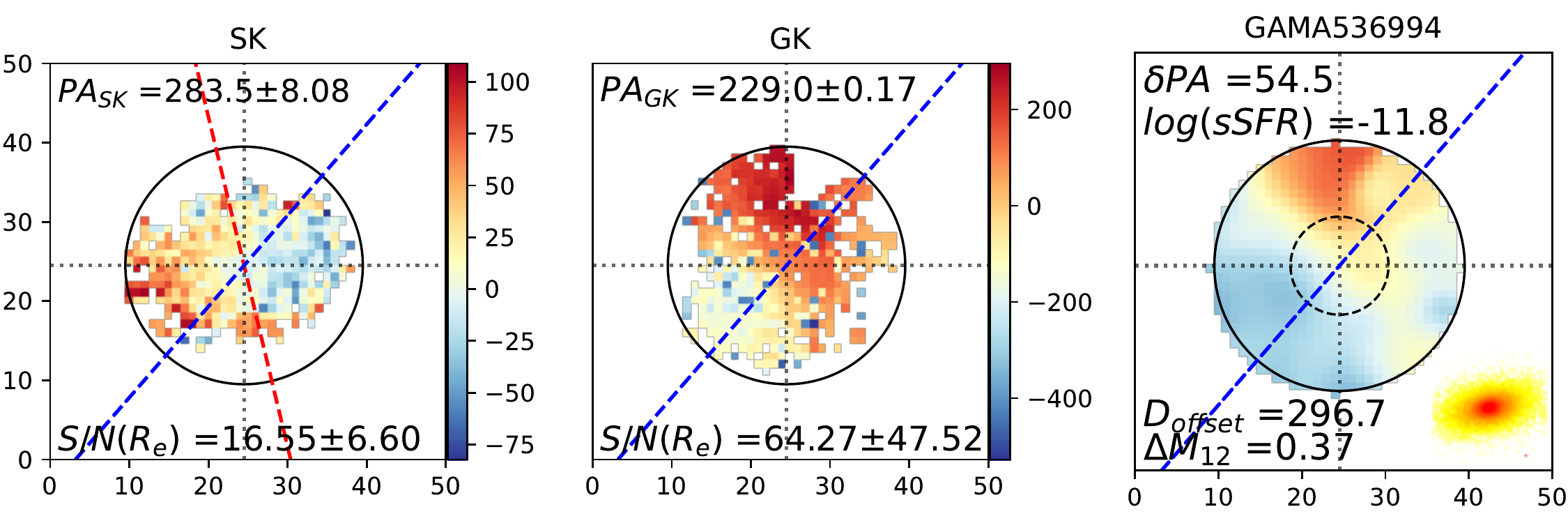}
		\includegraphics[width=0.45\linewidth]{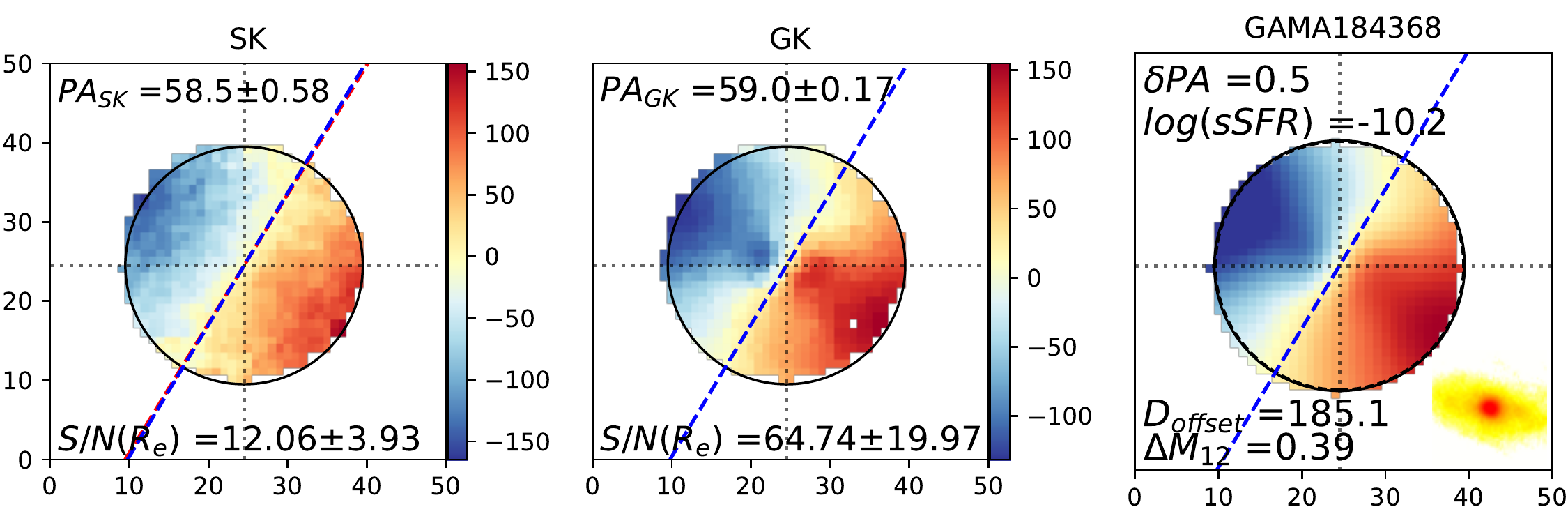}
		\includegraphics[width=0.45\linewidth]{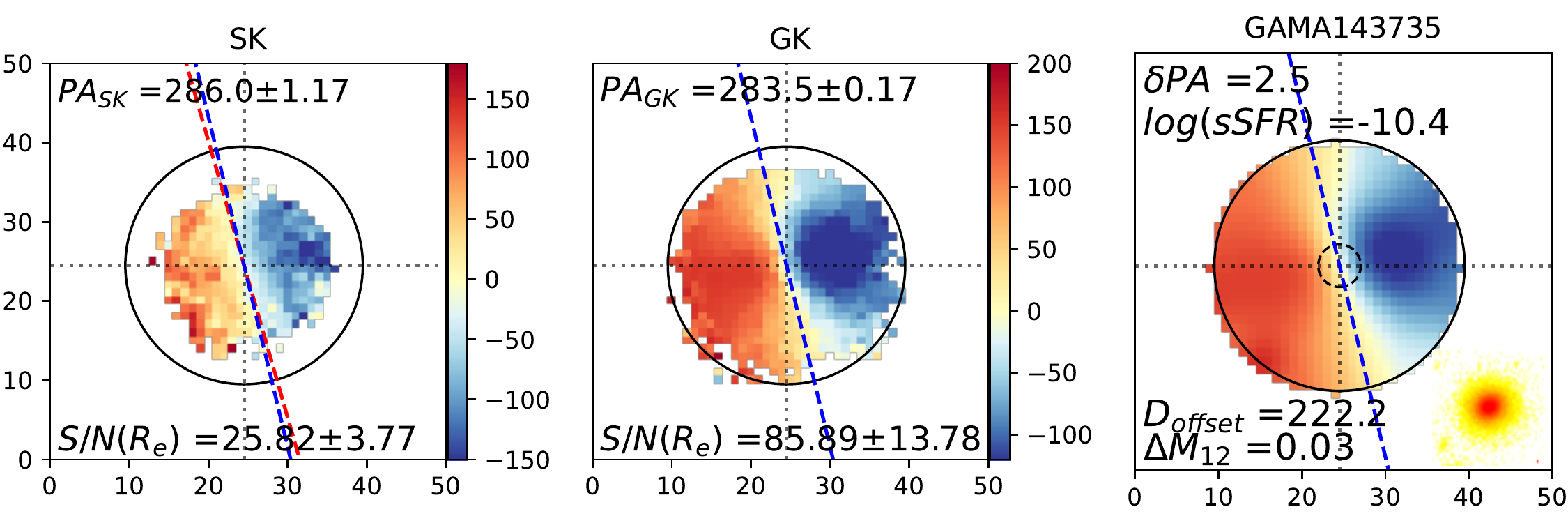}
		\includegraphics[width=0.45\linewidth]{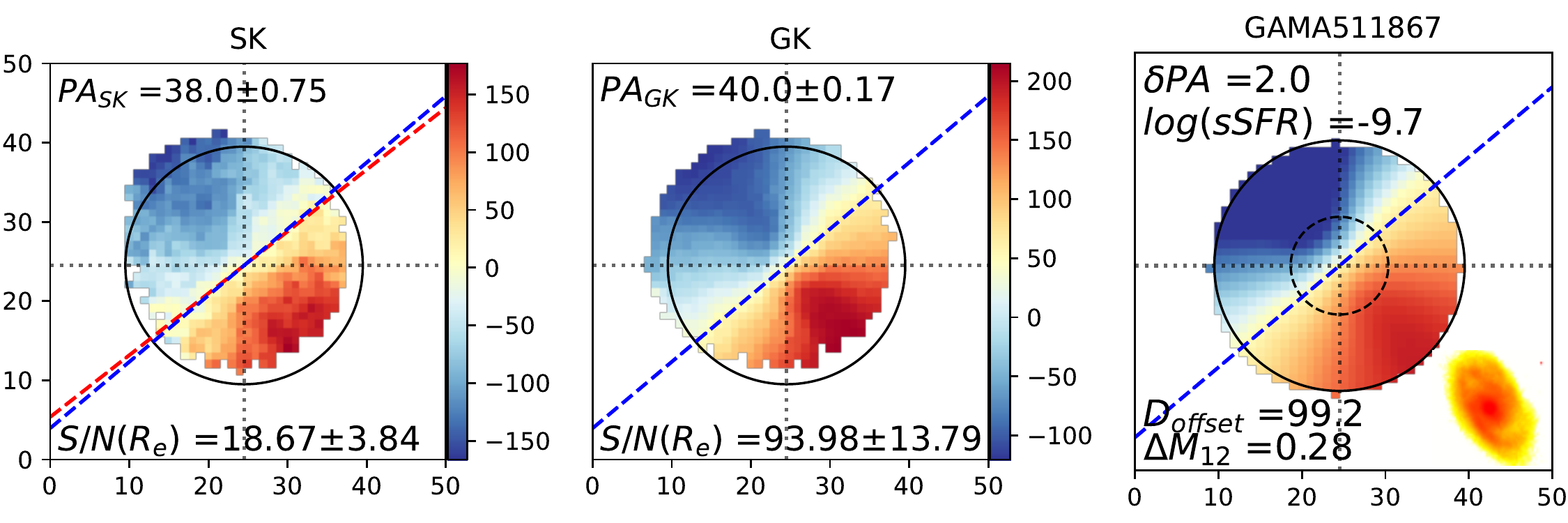}
		\label{fig:sami-unrelax2}
	\end{figure*}

\end{document}